\documentclass[
superscriptaddress,showpacs,preprintnumbers,nofootinbib,amsmath,amssymb, aps,prd,10.5pt]{revtex4-1}
\usepackage{braket}
\usepackage{bm}
\usepackage{dcolumn}
\usepackage{cancel}
\usepackage[dvipdfmx,final]{graphicx}
\input{colordvi.tex}
\usepackage{longtable}
\usepackage[compat=1.1.0]{tikz-feynman}
\usepackage{tabularx}
\usepackage[pdftex,bookmarks,linktocpage,pdfpagelabels,plainpages=false,hyperfigures,linkcolor=blue,citecolor=blue]{hyperref}
\hypersetup{colorlinks=true}
\usepackage{cleveref}
\usepackage{mhchem}
\usepackage{mathtools}
\tikzfeynmanset{every edge={very thick},}
\def\bea{\begin{eqnarray}}
\def\eea{\end{eqnarray}}

\usepackage{geometry}
\usepackage{color}

\begin{document}
\title{Manifesting hidden dynamics of a sub-component dark matter}
\author{Ayuki Kamada}
\email{akamada@fuw.edu.pl}
\affiliation{Center for Theoretical Physics of the Universe, Institute for Basic Science (IBS), Daejeon 34126, Republic of Korea}
\affiliation{Kavli Institute for the Physics and Mathematics of the Universe (WPI), The University of Tokyo Institutes for Advanced Study, The University of Tokyo, Kashiwa 277-8583, Japan}
\affiliation{Institute of Theoretical Physics, Faculty of Physics, University of Warsaw, ul. Pasteura 5, PL–02–093 Warsaw, Poland}
\author{Hee Jung Kim}
\email{heejungkim@ibs.re.kr}
\affiliation{Center for Theoretical Physics of the Universe, Institute for Basic Science (IBS), Daejeon 34126, Republic of Korea}
\affiliation{Department of Physics, KAIST, Daejeon 34141, Korea}
\author{Jong-Chul Park}
\email{jcpark@cnu.ac.kr}
\affiliation{Department of Physics and Institute of Quantum Systems (IQS), Chungnam National University, Daejeon 34134, Republic of Korea}
\author{Seodong Shin}
\email{sshin@jbnu.ac.kr}
\affiliation{Center for Theoretical Physics of the Universe, Institute for Basic Science (IBS), Daejeon 34126, Republic of Korea}
\affiliation{Department of Physics, Jeonbuk National University, Jeonju, Jeonbuk 54896, Republic of Korea}

\date{\today}

\begin{abstract}
We emphasize the distinctive cosmological dynamics in multi-component dark matter scenarios and its impact in probing a sub-dominant component of dark matter.
We find that the thermal evolution of the sub-component dark matter is significantly affected by the sizable self-scattering that is naturally realized for sub-${\rm GeV}$ masses.
The required annihilation cross section for the sub-component sharply increases as we consider a smaller relative abundance fraction among the dark-matter species.
Therefore, contrary to a naive expectation, it can be easier to detect the sub-component with smaller abundance fractions in direct/indirect-detection experiments and cosmological observations.
Combining with the current results of accelerator-based experiments, the abundance fractions smaller than $10\,\%$ are strongly disfavored;
we demonstrate this by taking a dark photon portal scenario as an example.
Nevertheless, for the abundance fraction larger than $10\,\%$, the warm dark matter constraints on the sub-dominant component can be complementary to the parameter space probed by accelerator-based experiments.
\end{abstract}

\maketitle

\section{Introduction}

Evidences for the existence of dark matter (DM) come from observing the gravitational influence of DM alone in various length scales of the Universe.
On the other hand, the particle nature of DM is elusive and our practical viewpoint on DM remains to be a bulk of matter that is dominant in mass.
In the last few decades, there have been extensive efforts to search for non-gravitational interactions of DM with the Standard Model (SM) particles whose mass and interactions are set by the weak scale and the weak interaction of the SM, i.e., the weakly interacting massive particles (WIMP).
Mainly due to the lack of any conclusive experimental signals of non-gravitational interactions of WIMP so far~\cite{Arcadi:2017kky,Roszkowski:2017nbc}, 
many alternative scenarios of dark sector beyond WIMP have been proposed recently. Among them, the scenarios of non-minimal particle contents inside a dark sector have drawn lots of attention because of their abilities resolving various phenomenological issues and providing extra power to many current/future experiments of searching for their signals in new and creative ways.
Examples include the scenarios of inelastic DM~\cite{Tucker-Smith:2001myb}, self-interacting non-minimal dark sector to address small-scale issues~\cite{Loeb:2010gj,Schutz:2014nka,McDermott:2017vyk,Chu:2018nki,Vogelsberger:2018bok,Kamada:2019wjo,Chua:2020svq} and the existence of the supermassive blackholes at high redshifts~\cite{Pollack:2014rja,Choquette:2018lvq,Jo:2020ggs}, and multi-component boosted dark matter (BDM) whose unique signals can be probed in a variety of neutrino and direct-detection experiments~\cite{Agashe:2014yua,Bhattacharya:2014yha,Kong:2014mia,Necib:2016aez,Alhazmi:2016qcs,Kim:2016zjx,Giudice:2017zke,Chatterjee:2018mej,Kim:2018veo,Kim:2019had,Heurtier:2019rkz,Kim:2020ipj,DeRoeck:2020ntj,Alhazmi:2020fju}.
Nevertheless, less attention has been given to exploring the cosmological dynamics of the sub-dominant component of DM and the corresponding impact on their detectability.

A sub-dominant component of DM can play a dominant role in the dynamics of a dark sector.
We already know an example in the SM.
Electrons, a component of matter that is negligible in mass compared to baryon, play an important role in coupling baryons with photons in the early Universe.
The observed baryon acoustic oscillations in the cosmic microwave background (CMB) anisotropies imply that the baryon and the photon bath were tightly-coupled until the recombination epoch.
Without the help of electrons, protons, a dominant component of the baryon, cannot couple to photons until then.
Although electrons have negligible gravitational influence in the point of view from a dark sector, they are actually dominant in interaction and play an important role in the cosmological evolution of the baryon.
This well-known example can be a motivation for paying attention to a sub-component dark matter in a variety of dark-sector scenarios beyond WIMP.

Probes of a sub-dominant component in a dark sector can be promising when it has sizable interactions with the SM particles.
It is well known that a wide range of parameter space of a vanilla model of Higgs portal DM, where the interactions in the thermal freeze-out and the direct-detection experiments are essentially same up to the crossing symmetry,
is strongly constrained even when the DM is a sub-dominant component whose mass is $\gtrsim \mathcal O ({\rm GeV})$~\cite{Cline:2013gha,Athron:2017kgt,Athron:2018hpc,Arcadi:2019lka}.
This is because the large coupling between the DM and the SM particles, which is essential in suppressing its fraction in the total DM, increases the scattering cross section between the DM and the target nucleus in direct-detection experiments. 
Hence, the fraction of the sub-dominant component $\Omega_{\rm sub}/\Omega_{\rm dm,total}$ entering linearly in the direct-detection signal rate is canceled by the large coupling squared in the cross section, allowing the experimental constraints to be applied to the sub-component DM equally.

The strategy to probe a sub-dominant DM component relies on its cosmological evolution, sensitive to the interaction within a dark sector, as stated in the previous paragraph.
In this paper, we study a case where the dynamics within a dark sector affects the detectability of a sub-dominant component.
For a concrete demonstration, we take the minimal two-component DM scenario, where the relic density of two stable DM components are determined by the {\it assisted freeze-out}~\cite{Belanger:2011ww};
the heavier DM particle $\chi_0$, which is dominant in mass, is secluded from SM and directly annihilates only into the lighter DM particle $\chi_1$ while the sub-dominant component $\chi_1$ annihilates into the SM particles.
We show that the dynamics of the assisted freeze-out entails larger annihilation cross sections of the sub-dominant component compared to standard freeze-out scenarios.
This renders the enhanced detectability of $\chi_1$, e.g., in cosmological/astrophysical observations.
We highlight the cosmological evolution of $\chi_1$ by taking into account a large self-scattering cross section of $\chi_1$, i.e., $\sigma_{\rm self}/m\sim 0.1\,{\rm cm^2/g}$, which is naturally realized for sub-{\rm GeV} mass-scale of $\chi_1$ in our reference set-up.
The collaboration between the $\chi_0$-annihilation and the strong self-scattering among $\chi_1$ leads to a distinct thermal evolution of $\chi_1$, which we dub as DM {\it self-heating}~\cite{Kamada:2017gfc,Kamada:2018hte,Chu:2018nki,Kamada:2019wjo}.
The enhanced temperature of $\chi_1$ from DM self-heating affects their velocity-dependent annihilation rate during cosmological epochs sensitive to DM annihilation.
Furthermore, the resultant warmness of $\chi_1$ from DM self-heating affects their gravitational clustering and leaves imprints in matter power spectrum.
In order to guide the attention of readers to their own interests, we devote the rest of the section to providing a scope of our analyses.

\subsection*{Scope of our analyses}

The chemical freeze-out of the sub-dominant component $\chi_1$ has a distinctive feature from the standard freeze-out of single-component DM scenarios.
If the relic abundance of $\chi_1$ is negligible around its freeze-out, i.e., $r_1=\Omega_{\chi_1}/\Omega_{\rm dm, tot}\ll1$, the production of $\chi_1$ from $\chi_0$-annihilation is non-negligible around its freeze-out.
Consequently, the required annihilation cross section of $\chi_1$ is sharply enhanced towards smaller $r_1$.
In the case of $s$-wave ($p$-wave) annihilation, the required annihilation cross section of $\chi_1$ scales as $1/r_1^2$ ($1/r_1^3$), in contrast to a naive expectation, scaling as $1/r_1$.
It is worthwhile to note that considering smaller values of $r_1$ is sometimes referred as a minimal remedy to evade the stringent indirect-detection constraints on sub-${\rm GeV}$ DM annihilations~(for example, see Ref.~\cite{Izaguirre:2013uxa}).
We remark that this is {\it not entirely true} because of the sharp enhancement of the $\chi_1$-annihilation rate towards smaller $r_1$ in our reference scenario.
We provide semi-analytic understandings on the chemical freeze-out of DM in the two-component DM scenario in Section~\ref{section:CFO}.
In order to focus on the impact of the distinct dynamics of the chemical freeze-out, we review the cosmological/experimental bounds on $\chi_1$ while turning-off the self-scattering of $\chi_1$ by hand in Section~\ref{section:noSHcosmology}.

Moreover, we highlight the impact of the self-scattering among the lighter DM particle $\chi_1$ on its cosmological evolution.
After the freeze-out of all the DM particles, residual annihilation of the $\chi_0$ produce $\chi_1$ particles which have enough energy for self-heating due to the mass difference.
This self-heating enhances the temperature of $\chi_1$ and affects its observable signatures such as the suppression of the gravitational clustering in the Galactic scale.
Hence, the constraints for warm dark matter (WDM) enter even for $m_{\chi_1} \gg \mathcal O ({\rm keV})$ and the interpretations of the experimental/observational results from DM direct-detection experiments and the diffuse X-ray/$\gamma$-ray background should be different.
Furthermore, if the annihilation of $\chi_1$ is velocity-suppressed, DM self-heating enhances the annihilation rate of $\chi_1$ during the cosmological epochs sensitive to DM annihilation, e.g., during the photo-dissociation epoch~\cite{Depta:2019lbe} and at the last scattering.
Consequently, the cosmological observations can have more constraining power on the annihilation cross section of $\chi_1$.
The thermal evolution of $\chi_1$ with its self-heating and its impact on cosmological/astrophysical signatures are discussed in Sec.~\ref{section:SHhistory} and Sec.~\ref{section:SHDMconstraint}, respectively.

Although the interesting cosmology for $r_1\ll 1$ provides new possibilities on detecting $\chi_1$ in cosmological observations, we remark that for abundance fractions smaller than $r_1\lesssim 0.1$, the enhanced interaction between $\chi_1$ and SM is usually incompatible with the constraints from terrestrial experiments.
In Section~\ref{section:dphdemon}, we demonstrate this argument for a reference model of two-component singlet scalar DM with a dark photon mediator.
We highlight that the WDM constraints from DM self-heating can be complementary to the parameter space probed by terrestrial experiments.

We conclude in Section~\ref{section:conclusion}.
Further details on the Boltzmann equations of DM, and the temperature evolution of $\chi_1$ in the presence of DM self-heating are collected in Appendix~\ref{appendix:Boltzmanndetail}, \ref{appendix:Boltzmannsolutions}, \ref{section:pwaveann} and \ref{section:Tdmevolapp}.

\section{Cosmology of Two-Component Dark Matter}
\label{section:noSH}

\subsection{Chemical freeze-out}
\label{section:CFO}

In this section, we revisit the processes of the chemical freeze-out of DM particles in a simple reference scenario, two-component DM ($\chi_0$ and $\chi_1$) with mass hierarchy $m_{\chi_0} > m_{\chi_1}$ and the following processes:
\begin{itemize}

\item Annihilation of $\chi_0$: $\chi_0 \, + \, \chi_0 \leftrightarrow \chi_1 \, + \, \chi_1$.

\item Annihilation of $\chi_1$: $\chi_1 \, + \, \chi_1 \leftrightarrow {\rm sm} \, + \, {\rm sm}$, where ``${\rm sm}$" stands for Standard Model particles.

\item Elastic scatterings: $\chi_1 \, + \, \chi_0 \rightarrow \chi_1 \, + \, \chi_0$ and $\chi_1 \, + \, {\rm sm} \rightarrow \chi_1 \, + \, {\rm sm}$.

\end{itemize} 
The DM particles are initially in thermal equilibrium with the SM plasma.
During the decoupling of the annihilations, we assume that DM are in kinetic equilibrium with the SM plasma.
This is justified by the crossing symmetry between DM annihilations and DM elastic scatterings;
the rate of elastic scatterings of a DM particle with some lighter state is typically larger than that of DM annihilations by a factor of $\sim n_{\rm light}/n_{\rm dm}$.
For $\chi_0$, the rate of $\chi_0+\chi_1\leftrightarrow\chi_0+\chi_1$ is larger than the rate of $\chi_0+\chi_0\leftrightarrow\chi_1+\chi_1$ by a factor of $\sim n_{\chi_1}/n_{\chi_0}$ and hence decouples later;
the similar discussion works for $\chi_1$.
For convenience, we introduce the DM yield, $Y_{\chi_i}=n_{\chi_i}/s$, in addition to $x=m_{\chi_1}/T$, where $s=(2\pi^2/45)g_{\ast S} T^3$ and $g_{\ast S}$ is the effective number of relativistic degrees of freedom in entropy density.
Assuming the kinetic equilibrium, the evolution equations for the DM yields are~\cite{Belanger:2011ww}
\begin{equation}
\begin{aligned}
\frac{dY_{\chi_0}}{dx}&=-\frac{\lambda_{\chi_0}(x)}{x}\left[Y_{\chi_0}^{2}-\left(\frac{Y_{\chi_0}^{{\rm eq}}\left(x\right)}{Y_{\chi_{1}}^{{\rm eq}}\left(x\right)}\right)^{2}Y_{\chi_{1}}^{2}\right]\,,\\
\frac{dY_{\chi_{1}}}{dx}&=-\frac{\lambda_{\chi_{1}}(x)}{x}\left[Y_{\chi_{1}}^{2}-\left(Y_{\chi_{1}}^{{\rm eq}}\left(x\right)\right)^{2}\right] + \frac{\lambda_{\chi_0}(x)}{x}\left[Y_{\chi_{0}}^{2}-\left(\frac{Y_{\chi_{0}}^{{\rm eq}}\left(x\right)}{Y_{\chi_{1}}^{{\rm eq}}\left(x\right)}\right)^{2}Y_{\chi_{1}}^{2}\right]\,,
\end{aligned}
\label{eq:yieldevol}
\end{equation}
where we have defined the dimensionless rates $\lambda_{\chi_i}=s\left\langle\sigma_i v_{\rm rel}\right\rangle/H$, and $H^2=g_\ast \pi^2 T^4/(90 m_{\rm pl}^2)$ with $m_{\rm pl}$ being the reduced Planck mass.
The thermally averaged annihilation cross section $\langle \sigma_0 v_{\rm rel} \rangle$ is for the $\chi_0 \chi_0 \to \chi_1 \chi_1$ while $\langle \sigma_1 v_{\rm rel} \rangle$ is $\chi_1 \chi_1$ to SM particles.
In this paper, we explicitly show the velocity dependence as $\langle \sigma_i v_{\rm rel} \rangle \simeq (\sigma_i v_{\rm rel})_s + (\sigma_i v_{\rm rel})_p \langle v_{\rm rel}^2 \rangle$.
For simplicity, we focus on the regime where the annihilation of the heavy component $\chi_0$ decouples first while the lighter component $\chi_1$ remains in thermal equilibrium.
In such a case the chemical freeze-out of $\chi_0$ proceeds like the standard WIMP freeze-out, and the asymptotic value of the yield is 
\begin{equation}
Y_{\chi_0}(\infty)\approx \frac{n_0+1}{\lambda_{\chi_0}(x_{\rm fo,0})}\,,
\label{eq:chi2relic}
\end{equation}
where $n_0=0$ in the case of $s$-wave annihilation of $\chi_0$ and $x_{\rm fo,0}=m_{\chi_1}/T_{\rm fo,0}$ with $T_{\rm fo,0}\sim m_{\chi_0}/20$ being the freeze-out temperature.~\footnote{We determine $T_{\rm fo,0}$ as in the case of freeze-out of WIMP, following Ref.~\cite{Kolb:1990vq}.}
Note that the estimation of $Y_{\chi_0}(\infty)$ can considerably change for mass difference as small as $\delta m=m_{\chi_0}-m_{\chi_1}\lesssim m_{\chi_0}/10$ where the chemical freeze-out processes of $\chi_0$ and $\chi_1$ interfere.
Even for $\delta m\gtrsim m_{\chi_0}/10$, the interference occurs in the case that $\chi_0$ abundance is exponentially suppressed, i.e., $r_0=1-r_1\ll e^{-\delta m/T_{\rm fo,0}}$, since the freeze-out of $\chi_0$ can be delayed and thus interfere with that of $\chi_1$.~\footnote{The chemical freeze-out with small mass differences and exponentially suppressed $r_0$'s are explored in Refs.~\cite{Maity:2019hre,Saez:2021oxl}.}
Hereafter, we will implicitly avoid such regime and focus on the simplest case where $\chi_0$ relic abundance is estimated as Eq.~\eqref{eq:chi2relic}, as our main purpose is to demonstrate the impact of self-heating in a given scenario.

The estimation of the final yield of $\chi_1$ is more involved.
After the $\chi_0$ freeze-out, evolution of $Y_{\chi_1}$ is written as
\begin{equation}
\frac{dY_{\chi_{1}}}{dx}\simeq-\frac{\lambda_{\chi_{1}}(x)}{x}\left[Y_{\chi_{1}}^{2}-\left(Y_{\chi_{1}}^{{\rm eq}}\left(x\right)\right)^{2}-Y_{\rm ast.}^{2}\left(x\right)\right]\,,
\label{eq:chi1evol}
\end{equation}
where $Y_{\rm ast.}$ is defined as
\begin{equation}
Y_{\rm ast.}\left(x\right)=\sqrt{\frac{\left\langle \sigma_{0}v_{\rm rel}\right\rangle}{\left\langle \sigma_{1}v_{\rm rel}\right\rangle}}Y_{\chi_{0}}(x)\,.
\label{eq:YastDef}
\end{equation}
The term proportional to $Y_{\rm ast.}$ represents the light DM production from the heavy DM annihilation, $\chi_0 \chi_0 \to \chi_1 \chi_1$.
If $Y_{\rm ast.}$ is negligible compared to $Y^{\rm eq}_{\chi_1}$ around the standard freeze-out point of $\chi_1$, i.e., $T_{\rm fo,1}\sim m_{\chi_1}/20$, the final relic of $\chi_1$ is estimated as $Y_{\chi_1}(\infty)\approx (n_1+1)/\lambda_{\chi_1}(x_{\rm fo,1})$
where $n_1=0$ in the case of $s$-wave annihilation of $\chi_1$ and $x_{\rm fo,1}=m_{\chi_1}/T_{\rm fo,1}$ with $T_{\rm fo,1}\sim m_{\chi_1}/20$ being the freeze-out temperature.
But as we consider smaller $r_1\ll1$, $x_{\rm fo,1}$ would become larger while $Y^{\rm eq}_{\chi_1}(x_{\rm fo,1})\propto e^{-x_{\rm fo,1}}$ becomes more suppressed;
eventually, the production rate of $\chi_1$ from the $\chi_0$-annihilation becomes non-negligible compared to the annihilation rate of $\chi_1$ into the SM particles where we dub this situation {\it assisted regime}.
In the assisted regime, the final relic would be larger than the estimation in the standard freeze-out regime.
Below, we discuss the estimation of the final yield in the two illustrative cases, i.e., the cases of $s$-wave and $p$-wave annihilation of $\chi_1$ while the $\chi_0$-annihilation is fixed to be $s$-wave for simplicity.
Nevertheless, the analytic estimations we present can be used for general partial-wave annihilations of DM.

\begin{figure}[t!]
\centering
\includegraphics[scale=0.6]{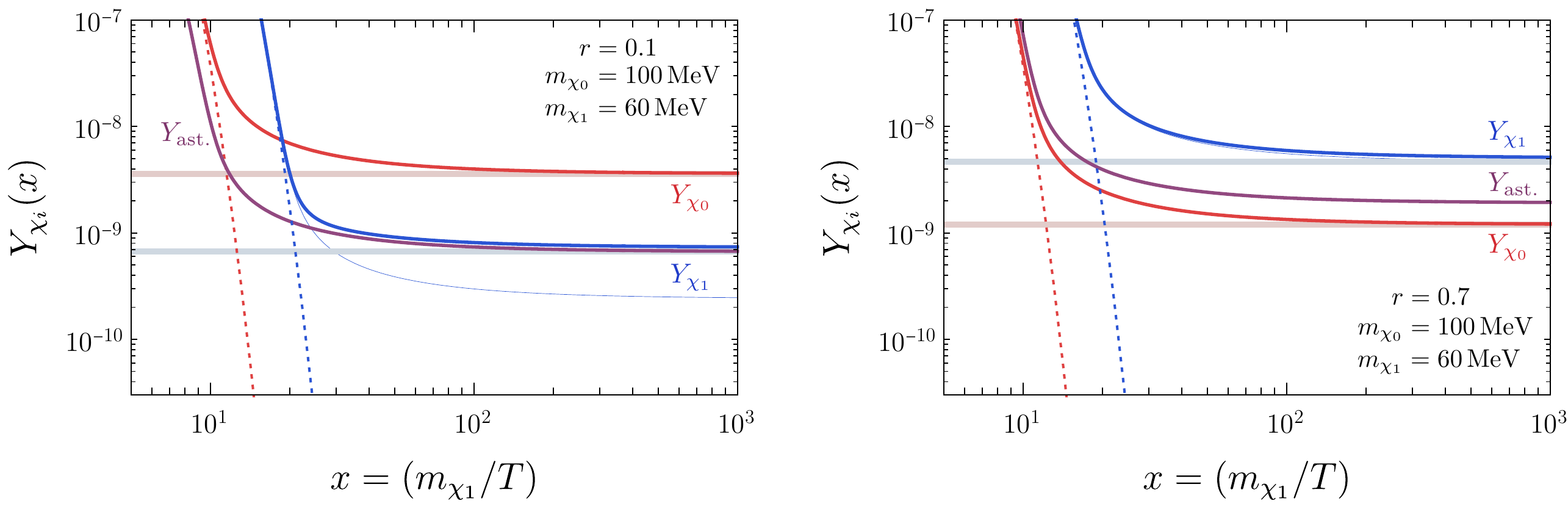}
\caption{Evolution of DM yields (thick solid) in the case of $s$-wave annihilation of $\chi_1$. 
The left (right) panel demonstrates the chemical freeze-out of $\chi_1$ in the assisted (standard) freeze-out regime.
We also present the solutions in the purely standard freeze-out case, e.g., a solution to Eq.~\eqref{eq:chi1evol} while neglecting $Y_{\rm ast.}$ for $Y_{\chi_1}$, as the thin solid curves.
The horizontal lines are the analytic estimations for the final relic abundance of DM, while the dotted curves are the equilibrium abundances.
}
\label{fig:yields}
\end{figure}

\begin{figure}[t!]
\centering
\includegraphics[scale=0.6]{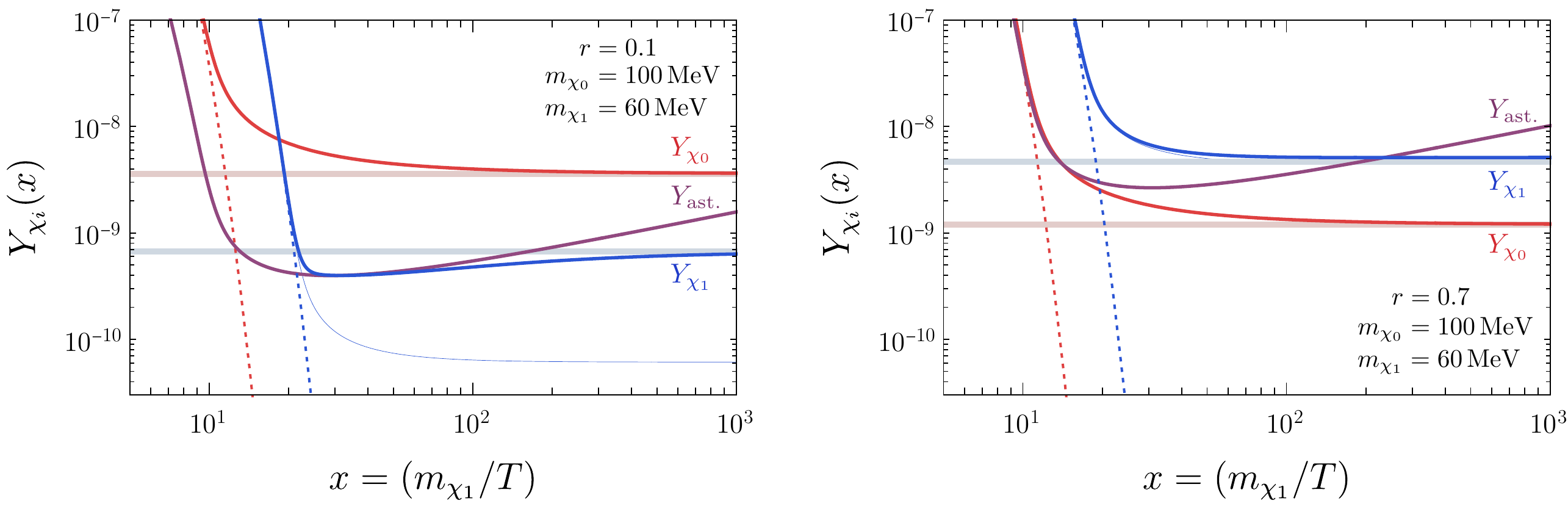}
\caption{Same as in Figure~\ref{fig:yields}, but in the case of $p$-wave annihilation of $\chi_1$. In the left panel, the departure point of $Y_{\chi_1}$ from $Y_{\rm ast.}$ is $x_{\rm fo}^\prime\simeq79$~[Eq.~\eqref{eq:Xfoprime}].
}
\label{fig:yield}
\end{figure}

Figure~\ref{fig:yields} shows the numerical solutions to Eqs.~\eqref{eq:yieldevol} in the case that the $\chi_1$-annihilation is $s$-wave.
The left (right) panel shows the chemical freeze-out in the assisted (standard) freeze-out regime.
We also present the solution in the case of standard freeze-out, i.e., ignoring $Y_{\rm ast.}$ in Eq.~\eqref{eq:chi1evol}, as the thin solid line.
We dub this {\it standard regime} for simplicity.
In the assisted regime, the final yield of $\chi_1$ is significantly enhanced compared to the case of standard regime.
Around $x\sim 30$, instead of following the equilibrium trajectory (dotted) further, $Y_{\chi_1}$ follows the constant $Y_{\rm ast.}$ (purple) asymptotically;
this is because the volumetric production/annihilation rate from $\chi_0$-annihilation/$\chi_1$-annihilation balance there, and hence the yield of $\chi_1$ seizes to decrease down to the yield predicted in the case of standard freeze-out.
The final yield of $\chi_1$ is estimated by the balance condition as $Y_{\chi_1}(\infty) \approx Y_{\rm ast.}(\infty)$.
The detailed analytic arguments for this estimation can be found in Appendix~\ref{appendix:Boltzmanndetail}.
Putting our understandings in the standard/assisted freeze-out regimes together, we estimate the final yield of $\chi_1$ as
\begin{equation}
Y_{\chi_1}(\infty) \approx \max \left[Y_{\rm ast.}(\infty) , \frac{n_1+1}{\lambda_{\chi_1}(x_{\rm fo,1})} \right]\,,
\label{eq:chi1relicassisteds}
\end{equation}
where $Y_{\rm ast.}(\infty)$ is given as
\begin{equation}
Y_{\rm ast.}(\infty) = \sqrt{\frac{(\sigma_0 v_{\rm rel})_s}{(\sigma_1 v_{\rm rel})_s}} Y_{\chi_0}(\infty)\,.
\end{equation}
When the first (second) term inside the maximum determines the final yield of $\chi_1$, the freeze-out of $\chi_1$ is in the assisted (standard) regime.
Note that $(\sigma_i v_{\rm rel})_s$ (and $(\sigma_i v_{\rm rel})_p$ later) is independent of the velocity $v_{\rm rel}$ in our notation.
In the assisted freeze-out regime, the required annihilation cross section of $\chi_1$ for a given $r_1$ is
\begin{equation}
\left(\sigma_{1}v_{{\rm rel}}\right)_s\ensuremath{\simeq4.7\times10^{-24}{\rm cm^{3}/s}\,\left(\frac{0.1}{r_{1}}\right)^{2}\left(\frac{m_{\chi_{1}}/m_{\chi_{0}}}{0.6}\right)^{2}\left(\frac{\sqrt{g_{\ast}}}{g_{\ast S}}\right)_{x_{\rm fo,0}}}\,.
\label{eq:assistedxsections}
\end{equation}
We remark that the annihilation cross section is enhanced towards smaller values of $r_1$ as $(\sigma_1 v_{\rm rel})_s\propto1/r_1^2$.
The $r_1$-dependence of the annihilation cross section in the assisted regime is sharper than that in the standard freeze-out regime where the annihilation cross section scales as $\propto1/r_1$.

Figure~\ref{fig:yield} shows the numerical solutions to Eqs.~\eqref{eq:yieldevol} in the case of $p$-wave annihilation of $\chi_1$ pair into the SM particles.
The left (right) panel shows the chemical freeze-out in the assisted (standard) freeze-out regime.
Again, the final yield of $\chi_1$ in the assisted freeze-out regime is significantly larger than what is expected in the standard freeze-out regime, as clearly seen by comparing the thick and thin blue curves in the left panel of Figure~\ref{fig:yield}.
The difference from the case of $s$-wave annihilation of $\chi_1$ is that $Y_{\chi_1}$ follows $Y_{\rm ast.}$ (purple) only until $x\sim 100$ and gradually reaches a constant value asymptotically since the asymptotic value of $Y_{\rm ast.}(\infty)$ is no longer a (constant) scaled value of $Y_{\chi_0}(\infty)$; the ratio $\langle \sigma_0 v_{\rm rel} \rangle$/$\langle \sigma_1 v_{\rm rel} \rangle$ now increases as the temperature $T$ decreases. 
We denote the SM temperature at the departure point from $Y_{\rm ast.}$ as $T_{\rm fo}^\prime$.
The final yield of $\chi_1$ roughly coincides with $Y_{\rm ast.}(x_{\rm fo}^\prime)$.
More precisely, the final relic abundance $\chi_1$ in the assisted regime can be estimated as $Y_{\chi_1}(\infty) \approx (n_1+1)/\lambda_{\chi_1}(x_{\rm fo}^\prime)$
where the $x_{\rm fo}^\prime=m_{\chi_1}/T_{\rm fo}^\prime$ is defined by the point where the relative deviation of $Y_{\chi_1}$ from $Y_{\rm ast.}$ becomes order unity;
detailed analysis and the accuracy of this estimation are collected in the Appendix~\ref{appendix:Boltzmanndetail}.
We estimate the final yield of $\chi_1$ as
\begin{equation}
Y_{\chi_1}(\infty) \approx \max \left[ \frac{n_1+1}{\lambda_{\chi_1}(x_{\rm fo}^\prime)} , \frac{n_1+1}{\lambda_{\chi_1}(x_{\rm fo,1})} \right]\,.
\label{eq:chi1relicassisted}
\end{equation}
When the first (second) term inside the maximum determines the final yield of $\chi_1$, the freeze-out of $\chi_1$ is in the assisted (standard) regime.
At $x=x_{\rm fo}^\prime$, $(Y_{\rm ast.}-Y_{\chi_1})/Y_{\rm ast.}=c^\prime$ and $c^\prime\simeq0.35$ is a numerical constant to fit the final relic abundance to numerical results.
$x_{\rm fo}^\prime$ is given by
\begin{equation}
x_{{\rm fo}}^{\prime}\simeq47\,\left(\frac{c^\prime}{0.35}\right)^{2/3}\left(\frac{m_{\chi_{1}}/m_{\chi_{0}}}{0.6}\right)^{2/3}\left(\frac{(\sigma_1 v_{\rm rel})_p}{4.5\times10^{-23}\,{\rm cm^3/s}}\right)^{1/3}\left(\frac{g_{\ast S}}{\sqrt{g_{\ast}}}\right)_{x_{\rm fo}^\prime}^{2/3}\left(\frac{\sqrt{g_{\ast}}}{g_{\ast S}}\right)_{x_{\rm fo,0}}^{1/3}\,.
\label{eq:Xfoprime}
\end{equation}
The required annihilation cross section of $\chi_1$ for a given $r_1$ is given as
\begin{equation}
(\sigma_1 v_{\rm rel})_p\simeq4.2\times10^{-24}\,{\rm cm^3/s}\,\left(\frac{c^\prime}{0.35}\right)^4\left(\frac{m_{\chi_{1}}/m_{\chi_{0}}}{0.6}\right)^{4}\left(\frac{0.1}{r_{1}}\right)^{3}\left(\frac{g_{\ast S}}{\sqrt{g_{\ast}}}\right)_{x_{\rm fo}^\prime}^{4}\left(\frac{\sqrt{g_{\ast}}}{g_{\ast S}}\right)_{x_{\rm fo,0}}^{2}\,,
\label{eq:assistedxsection}
\end{equation}
where we define $(\sigma_1 v_{\rm rel})_p$ through the relation $\langle\sigma_1 v_{\rm rel}\rangle=(\sigma_1 v_{\rm rel})_p\langle v_{\rm rel}^2\rangle$ with the thermal average of the squared relative scattering velocity among $\chi_1$, $\langle v_{\rm rel}^2\rangle\simeq 6T_{\chi_1}/m_{\chi_1}$.
One would recover the value of $(\sigma_1 v_{\rm rel})_p$ used in Eq.~\eqref{eq:Xfoprime} by taking $g_\ast = g_{\ast S}=10.75$.
Note that the $r_1$-dependence of the annihilation cross section, i.e., $(\sigma_1 v_{\rm rel})_p\propto1/r_1^3$, is even sharper than in the case of $s$-wave annihilation cross section $(\sigma_1 v_{\rm rel})_s\propto1/r_1^2$.
This is because $Y_{\rm ast.}(x)$ increases with $x$ contrary to the $s$-wave case, due to the velocity dependence of $\langle \sigma_1 v_{\rm rel}\rangle$ for the $p$-wave case~[Eq.~\eqref{eq:YastDef}]. Therefore, $Y_{\chi_1}$ is lifted up more by following $Y_{\rm ast.}$ until $x \sim x'_{\rm fo}$. From Eq.~\eqref{eq:Xfoprime}, the value of $x'_{\rm fo}$ increases for larger values of $(\sigma_1 v_{\rm rel})_p$, keeping the above effect longer.
We remark that, regardless of the DM masses, the assisted regime emerges as we consider $r_1\ll 1$, i.e., the first term inside the maximum dominates over the second term in Eqs.~\eqref{eq:chi1relicassisted} and \eqref{eq:chi1relicassisteds} for $r_1\ll1$.
This is because in the assisted regime, the required annihilation cross section to realize a given $r_1$ exhibits sharper dependence on $r_1$, $(\sigma_1 v_{\rm rel})_{s,p} \propto 1/r_1^{2,3}$, compared to the standard freeze-out regime, $\propto 1/r_1$.
The sharp enhancement of the required cross section towards smaller $r_1$ generally makes the two-component DM scenario tightly constrained by direct-detection experiments and cosmological observations compared to the single component DM case, as will be discussed in the next section.

\subsection{Scenario without DM self-heating}
\label{section:noSHcosmology}

After the chemical freeze-out of DM, residual $\chi_1$-annihilations can produce significant flux of energetic SM particles that can be probed through cosmological/astrophysical observations.
The non-observation of such signatures provides bounds on DM annihilation cross sections.
Meanwhile, if $\chi_1$ exhibit sizable self-scattering, the temperature evolution of $\chi_1$ could be sensitive to $\chi_0$-annihilations;
the residual $\chi_0$-annihilations may lead to DM self-heating.
The modifications on the thermal history of $\chi_1$ could directly affect the bounds on $\chi_1$-annihilation if the annihilation cross section of $\chi_1$ depends on $T_{\chi_1}$.

In this section, in order to focus on the impacts of introducing the assisted regime, we first review the thermal history and the cosmological/experimental bounds on $\chi_1$ while turning-off the self-scattering of $\chi_1$ by hand.
Note that it is actually more natural to expect sizable self-scattering among $\chi_1$ particles for our reference mass range of $m_{\chi_1} < \mathcal O (0.1\,{\rm GeV})$ in many multi-component dark matter scenarios, which will be discussed in later sections.

\subsubsection{$s$-wave annihilation of $\chi_1$}

The cosmological/astrophysical bounds on DM annihilations are very stringent for sub-${\rm GeV}$ DM due to their enhanced number density.
If DM dominantly annihilates into electromagnetic particles, the bounds on sub-${\rm GeV}$ DM annihilations disfavor the standard single-component thermal DM in the case of $s$-wave annihilation.
The two-component DM scenario is sometimes considered to be a minimal remedy to be consistent with the stringent bounds on DM annihilations~\cite{Izaguirre:2013uxa};
the sub-dominant DM component $\chi_1$ with abundance fraction $r_1\ll 1$ annihilates into SM with the annihilation cross section enhanced as $(\sigma_1 v_{\rm rel})_s \propto1/r_1$ and the volumetric annihilation rate is suppressed towards smaller $r_1$ as $n_{\chi_1}^2 (\sigma_1 v_{\rm rel})_s\propto r_1$.
Since the bounds on DM annihilations are basically given in terms of the quantity proportional to the volumetric rate, considering $r_1\ll1$ seems to be a viable possibility at the first sight.
However, this is {\it not entirely true} since, as we have seen in Section~\ref{section:CFO}, the relic abundance of $\chi_1$ is determined in the assisted regime where the required annihilation cross section scales as $(\sigma_1 v_{\rm rel})_s \propto1/r_1^2$ and thus the volumetric annihilation rate is virtually independent of $r_1$.
Therefore, considering smaller $r_1$ may not relax the bounds on $\chi_1$ annihilation as naively expected.
In the rest of this section, we review the possible indirect-detection constraints on $s$-wave $\chi_1$-annihilation for a vast range of the abundance ratio $r_1$, keeping in mind the caveat on the required annihilation cross section in the assisted regime.
Since the constraints on $s$-wave annihilation do not depend on the temperature evolution of $\chi_1$, we leave the discussion on the temperature evolution for the next section where we discuss the case of $p$-wave annihilation of $\chi_1$.
We first summarize the considered list of constraints below.

\begin{itemize}

\item {\bf Bounds on {\rm MeV}-scale freeze-out of DM}~\cite{Sabti:2019mhn}:
Light DM particles that are in thermal equilibrium exclusively with the baryon-photon plasma or neutrinos, beyond the neutrino decoupling, i.e., $T\lesssim T_{\nu,{\rm dec}}\sim2\,{\rm MeV}$, are constrained by the BBN and CMB observations.
Around $T\sim1\,{\rm MeV}$, DM energy density may considerably contribute to the expansion rate of the Universe, and DM annihilations may release significant amount entropy exclusively into the baryon-photon plasma or neutrinos.
As a consequence, the temperature ratio between neutrinos and photons after the neutrino decoupling and the synthesis of the primordial elements during BBN may be considerably affected.
Cosmological observables such as $N_{\rm eff}$ from the CMB observations and the observations on the primordial abundances of light elements (e.g., helium and deuterium) from BBN will thus provide constraints on the DM mass.
DM with masses greater than $m_{\rm dm}\gtrsim 40\,{\rm MeV}$ will not be constrained since they freeze-out before the neutrino decoupling and their energy density is negligible around $T\sim1\,{\rm MeV}$.
For electrophilic thermal DM, DM annihilations will raise the photon temperature and thus lead to $N_{\rm eff}$ smaller than the SM prediction.
We will adopt the constraint coming from the {\it Planck} data alone~\cite{Aghanim:2018eyx}, rather than a joint analysis with both the local measurements and BBN observations.~\footnote{Ref.~\cite{Sabti:2019mhn} provides limits on DM mass through joint analyses by combining the {\it Planck} data with the local measurement of $H_0$~\cite{Riess:2019cxk} ({\it Planck}$+H_0$), or with the measurements of the primordial abundances of light nuclei~\cite{Tanabashi:2018oca} ({\it Planck}$+{\rm BBN}$). Each joint analysis prefers larger values of $N_{\rm eff}$ compared to the analysis of the {\it Planck} data alone for non-annihilating DM.
This is because of the apparent tension on the determination of $H_0$ from local measurements and {\it Planck} data, and the slight $\sim 0.9\,\sigma$ tension on the inferred $\Omega_b h^2$ between BBN and CMB observations~\cite{Pitrou:2018cgg}.
Since electrophilic DM lowers $N_{\rm eff}$, joint analyses provide stronger limits on the masses of electrophilic DM;
for complex scalar DM, the limits are $m_{\rm dm}\gtrsim 9.2\,{\rm MeV}$ for {\it Planck}$+H_0$ and $m_{\rm dm}\gtrsim 8.1\,{\rm MeV}$ for {\it Planck}$+{\rm BBN}$.
To be conservative, we take the limit from {\it Planck} data alone.}
For a complex scalar DM, which will be the illustrative case in Section~\ref{section:dphdemon}, {\it Planck} data alone constrains DM mass to be $m_{\rm dm}\gtrsim 4.6\,{\rm MeV}$ at $95.4\,\%$ CL~\cite{Sabti:2019mhn};
remark that the constraints apply irrespective of $r_1$.

\item {\bf Photo-dissociation constraints on DM annihilation}~\cite{Depta:2019lbe}:
The residual annihilation of DM after the freeze-out could affect the abundances of light elements through the process of photo-dissociation.
We first briefly review the case of DM mass larger than a few ${\rm GeV}$~\cite{Kawasaki:2015yya}, and discuss the caveats of sub-${\rm GeV}$ DM annihilations.
When DM annihilate into electromagnetic components of SM, e.g., $e^+e^-$ or $\gamma\gamma$, the energetic final state particles initiate the electromagnetic cascade, e.g., by scattering with background photons, thermal electrons, and nuclei.
The cascade process redistributes the injected energy from DM annihilations among the electromagnetic particles and produces an energetic photon spectrum.
The photon spectrum is exponentially suppressed above a high-energy cutoff, $E\sim m_e^2/22T$;
photons above the cutoff are efficiently degraded through the pair annihilation process $\gamma\gamma_{\rm b}\rightarrow e^+ e^-$ ($\gamma_b$ denotes the background photon)~\cite{Protheroe:1994dt,Kawasaki:1994af,Cyburt:2002uv}.
When the cutoff is larger than the thresholds of the photo-dissociation processes of light nuclei, e.g., $\ce{D}$, $\ce{^{3}He}$ and $\ce{^{4}He}$, the processes are triggered.
The triggered photo-dissociation processes modify the abundance ratios among the light nuclei.
The predicted abundance ratios are compared with the observed values to give upper bounds on DM annihilation cross section.
The photo-dissociation processes from DM annihilation are relevant at temperatures long after the BBN, $100\,{\rm eV} \lesssim T\lesssim 10\,{\rm keV}$~\cite{Hufnagel:2018bjp,Depta:2019lbe};
for $T\lesssim 10\,{\rm keV}$, the high-energy cutoff of the resultant photon spectrum become larger than the dissociation thresholds of light nuclei;
for $T\lesssim 100\,{\rm eV}$, the high-energy cutoff is larger than the dissociation thresholds (of $\ce{^{4}He}$) while the energy injection rate redshifts towards lower temperatures.

For the annihilation of sub-${\rm GeV}$ dark matter, small $m_{\rm dm}$ renders the high-energy cutoff at $E\sim\min\left[m_e^2/22T,m_{\rm dm}\right]$;
this is because photons of $E\gtrsim m_{\rm dm}$ are limited by the initial energy injection spectrum from DM annihilation.~\footnote{See Refs.~\cite{Poulin:2015woa,Hufnagel:2018bjp,Depta:2019lbe} for the dedicated analyses and discussions on the resultant photon spectrum when the high-energy cutoff is limited by the DM mass.}
For example, for $m_{\rm dm}\lesssim2\,{\rm MeV}$, the cutoff is smaller than the threshold energy of $\ce{D}$ and the photo-dissociation constraint disappears.
We employ the photo-dissociation constraints on sub-${\rm GeV}$ DM annihilations presented in Ref.~\cite{Depta:2019lbe}.
For $s$-wave annihilating $\chi_1$, we simply rescale the constraints with respect to the factor $r_1^{-2}$.

\item {\bf CMB bounds on DM annihilation}~\cite{Aghanim:2018eyx}:
After the freeze-out of $\chi_1$, residual annihilation of $\chi_1$ into SM particles continues all the way down to the recombination epoch.
Although their annihilation rate per volume, $\sim n_{\chi_1}^2\langle\sigma_{\rm ann} v_{\rm rel}\rangle$, is tiny, it can be significant enough to affect the CMB through energy injection into the SM plasma.
The energy injection from DM annihilation ionizes the neutral hydrogen and modifies the ionization history between the recombination and the reionization.
The additional free electrons scatter with CMB photons and make the last scattering surface thicker.
The broadening of the last scattering surface affects the CMB temperature power spectrum~\cite{Padmanabhan:2005es,Green:2018pmd}.
The temperature power spectrum on scales smaller than the acoustic horizon at the recombination ($l\gtrsim200$) is relatively suppressed from the enhanced Landau damping (not Silk damping) of the CMB photons.
On the other hand, the polarization power spectrum on scales larger than the acoustic horizon at the recombination ($20 \lesssim l\lesssim200$) is enhanced because of the increased probability of the Thomson scattering of CMB photons between the recombination and the reionization.
The quantity constrained from CMB observations is the energy injection rate per volume given as
\begin{equation}
\frac{dE}{dtdV}=f_{\rm eff}\times \Delta E \times n_{\chi_1}^2\langle\sigma_{\rm ann} v_{\rm rel}\rangle\,,
\label{eq:Einjrate}
\end{equation}
where $\Delta E\sim m_{\rm dm}$ is the injected energy per annihilation, and $f_{\rm eff}$ is the efficiency of energy deposition which is typically an order unity number depending on the annihilation product.
Assuming that the component annihilating into SM accounts for the total observed DM density, the recent data from {\it Planck}~\cite{Aghanim:2018eyx} constrains DM annihilation as
\begin{equation}
f_{{\rm eff}}\frac{\left\langle \sigma_1 v_{{\rm rel}}\right\rangle_{{\rm rec}}}{m_{\chi_1}}\lesssim3.2\times10^{-28}\,{\rm cm^{3} \, s^{-1} \, GeV^{-1}}\cdot\left(1/r_1\right)^2\,,
\label{eq:CMB}
\end{equation}
where we scale the constraint with respect to $r_1$, since both $\chi_0$ and $\chi_1$ contribute to the DM density but only $\chi_1$ annihilates into SM particles.
Hereafter, we take $f_{\rm eff}=1$.
For simplicity, we assume neither resonances nor non-perturbative enhancements of the annihilation cross section.

\item {\bf DM annihilations in the Milky Way}~\cite{Essig:2013goa,Cirelli:2020bpc}:
DM annihilations in the Milky Way halo could produce significant flux of diffuse X-ray and $\gamma$-ray photons.
Therefore, the measured photon flux from the satellite observations sets upper bounds on the annihilation cross section of DM.
We employ the bounds presented in Refs.~\cite{Essig:2013goa,Cirelli:2020bpc};
assuming that DM dominantly annihilates into $e^+e^-$, the photon flux from final state radiation off the DM annihilations and the inverse Compton scattering of the produced $e^+ /e^-$ with low energy photons (CMB, infrared light and starlight) should be smaller than the observed one.
In the case of $s$-wave annihilation of sub-${\rm GeV}$ DM, the CMB bounds on DM annihilation is roughly a few orders of magnitude stronger than the one from the DM annihilations in the Milky Way halo.
Since the upper bounds on DM annihilation are basically given in term of the rate $n_{\chi_1}^2 (\sigma_1 v_{\rm rel})_s$ at the Galactic velocity scales, we rescale the bounds on DM annihilation cross section with respect to the factor $r_1^{-2}$.

\end{itemize}

\begin{figure}[t!]
\centering
\includegraphics[scale=0.61]{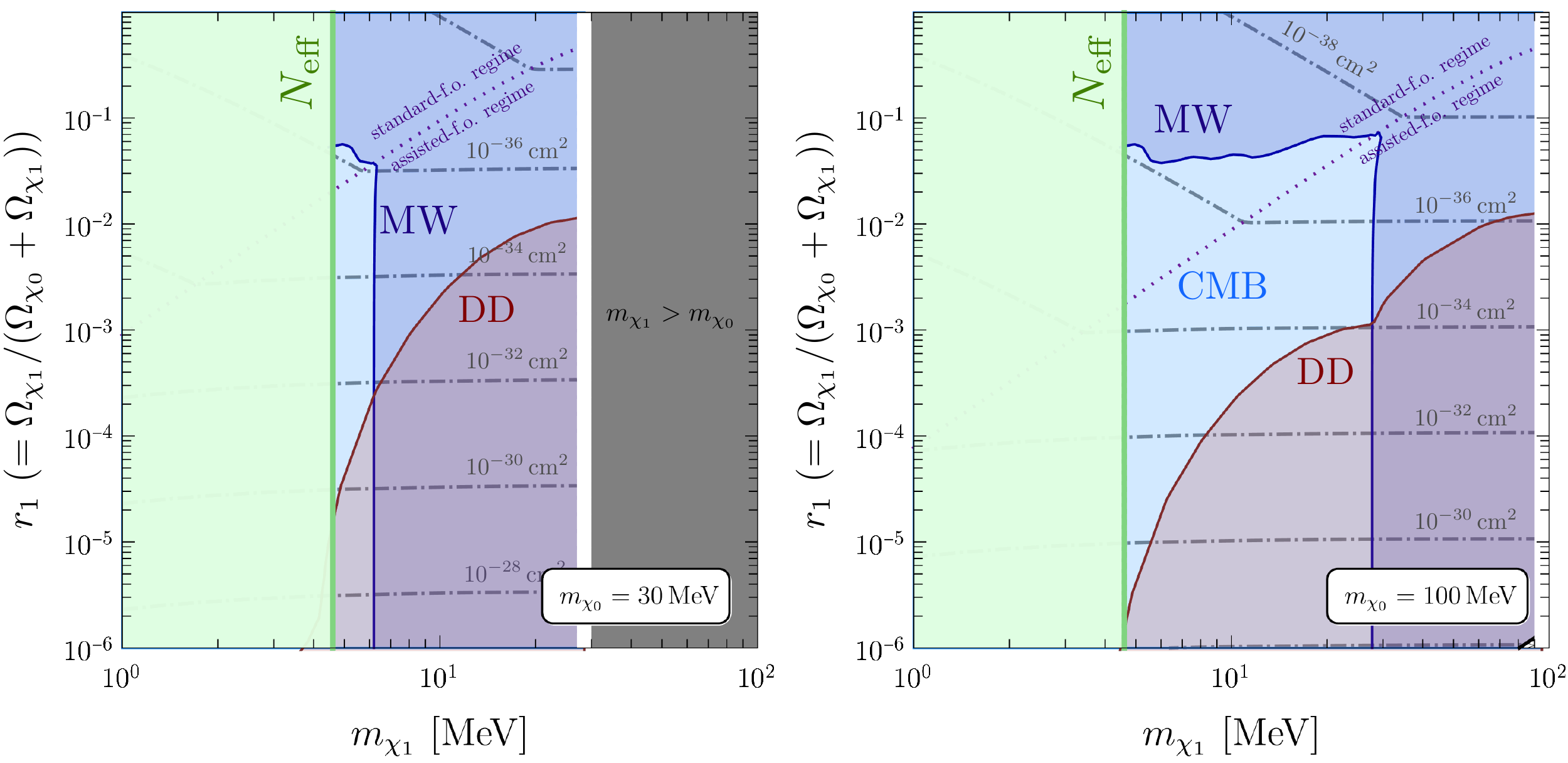}
\caption{
Summary of constraints on $s$-wave annihilating $\chi_1$ in the absence of self-heating.
In the assisted regime (below the dotted curve), the annihilation cross section is sharply enhanced towards smaller $r_1$, i.e., $(\sigma_1 v_{\rm rel})_s\propto1/r_1^2$.
Since the volumetric annihilation rate $n_{\chi_1}^2(\sigma_1 v_{\rm rel})_s$ is virtually independent of $r_1$, the constraints on $\chi_1$-annihilation is independent of $r_1$ in the assisted regime.
In the region that is not constrained by the ${\rm MeV}$-scale freeze-out~\cite{Sabti:2019mhn} (green), i.e., $m_{\chi_1}\gtrsim 4.6\,{\rm MeV}$, the strongest constraint are the bounds on DM annihilations from observations on CMB~\cite{Aghanim:2018eyx} (sky-blue) and Galactic diffuse X-ray and $\gamma$-ray photons~\cite{Essig:2013goa} (deep-blue);
the constraints rule out most of the parameter space of the sub-${\rm GeV}$ two-component DM scenario.
The constraints from photo-dissociation of light nuclei does not appear in the presented parameter space.
For reference, we also display the contours for the minimal contribution to $\sigma_{\chi_1e}$ in the heavy mediator limit~[Eq.~\eqref{eq:sigmachi1e}] (dot-dashed), and the corresponding direct-detection limits~\cite{Angle:2011th,Aprile:2016wwo,Essig:2017kqs,Agnes:2018oej} (brown).
}
\label{fig:paramnoSH_s}
\end{figure}

We summarize the aforementioned indirect-detection constraints on DM annihilations in Figure~\ref{fig:paramnoSH_s} in the $m_{\chi_1}$ versus $r_1$ plane for a given $m_{\chi_0}$.
At each point in the plane, we determine $(\sigma_1v_{\rm rel})_s$ according to Eq.~\eqref{eq:chi1relicassisteds}.
The dotted curve in Figure~\ref{fig:paramnoSH_s} separates the two regimes, i.e., the standard and assisted regimes, for the chemical freeze-out of $\chi_1$.
As a reference, we plot the contours for the minimal contribution to the elastic scattering cross section between $\chi_1$ and electron in the heavy mediator limit $\sigma_{\chi_1 e}$ (dot-dashed) given by
\begin{equation}
\sigma_{\chi_1{\rm sm}}\sim (\sigma_1v_{\rm rel})_s\times \left( \frac{\mu_{\chi_1{\rm sm}}}{m_{\chi_1}/2} \right)^2
\label{eq:sigmachi1e}
\end{equation}
where $\mu_{\chi_1{\rm sm}}$ is the reduced mass of the $\chi_1$-${\rm sm}$ system.
We also plot the direct-detection constraints based on this minimal contribution (in brown):
\begin{itemize}

\item {\bf Direct-detection constraints on $\chi_1$-SM interaction}:
The elastic scattering cross section of $\chi_1$ with SM can receive a minimal contribution given by Eq.~\eqref{eq:sigmachi1e};
for concreteness, we assume the heavy mediator limit.
We present the direct-detection constraints on $\chi_1$-$e$ scattering cross section $\sigma_{\chi_1e}$ in Figure~\ref{fig:paramnoSH_s} as a reference.
We employ the direct-detection constraints on sub-${\rm GeV}$ DM from the following experiments (with the mass range where they are most sensitive):
SuperCDMS~\cite{Agnese:2018col} and SENSEI~\cite{Crisler:2018gci,Abramoff:2019dfb} ($m_{\rm dm}\lesssim 4\,{\rm MeV}$);
XENON10~\cite{Angle:2011th,Essig:2017kqs} ($4\lesssim m_{\rm dm}\lesssim 30\,{\rm MeV}$);
XENON100~\cite{Aprile:2016wwo,Essig:2017kqs} and DarkSide-50~\cite{Agnes:2018oej} ($30\,{\rm MeV}\lesssim m_{\rm dm}$).
We rescale the constraints on $\sigma_{\chi_1 e}$ with the factor $r_1^{-1}$.
In Figure~\ref{fig:paramnoSH_s}, XENON10, XENON100, and DarkSide-50 are relevant for $m_{\chi_1}$ that is not constrained by the ${\rm MeV}$-scale freeze-out, i.e., $m_{\chi_1}\gtrsim4.6\,{\rm MeV}$.
However, the direct-detection constraints can be weakened for large elastic scattering cross section and thus there is also an upper bound on $\sigma_{\chi_1e}$ that the experiments can probe~\cite{Emken:2019tni};
strong DM-nucleus/electron interaction significantly attenuate the DM flux reaching the detector.
We translate the upper boundary of the range by the factor of $r_1$ and present it in Figure~\ref{fig:paramnoSH_s};
the upper boundary does not appear in the presented parameter range.

\end{itemize}
In the assisted regime, the annihilation cross section is enhanced for small $r_1$ as $(\sigma_1v_{\rm rel})_s\propto1/r_1^2$.
Since the photo-dissociation, CMB, and diffuse Galactic photon background constraints are basically given in terms of the volumetric rate $n_{\chi_1}^2(\sigma_1 v_{\rm rel})_s$, the constraints are virtually independent of $r_1$ in the assisted regime;
for example, see the diffuse Galactic photon background constraints (deep-blue) in Figure~\ref{fig:paramnoSH_s}.
The required $(\sigma_1v_{\rm rel})_s$ in the assisted regime also increases for lighter $m_{\chi_0}$~[Eq.~\eqref{eq:assistedxsections}];
compare the left and the right panel.
This is because the $\chi_1$-production from $\chi_0$-annihilation is more significant for lighter $\chi_0$ due to the enhanced number density of $\chi_0$, and thus larger $(\sigma_1v_{\rm rel})_s$ is required to achieve the desired $r_1$.
We see that for $s$-wave annihilating $\chi_1$, the strongest constraint on $\chi_1$-annihilation comes from the CMB bound which disfavor the whole parameter space for the sub-${\rm GeV}$ two-component DM scenario.

\subsubsection{$p$-wave annihilation of $\chi_1$}

In the previous section, we have seen that the cosmological/astrophysical constraints disfavor $s$-wave annihilating $\chi_1$ in the sub-${\rm GeV}$ mass range, even for $r_1\ll1$.
If DM annihilation is $p$-wave suppressed, the annihilation cross section may be small enough at the cosmological epochs of interest and therefore sub-${\rm GeV}$ DM can be consistent with the existing bounds.
The difference from the $s$-wave annihilation case is that in the assisted regime, the required annihilation cross section increases even more sharply towards smaller $r_1$, $(\sigma_1 v_{\rm rel})_p \propto1/r_1^3$~[Eq.~\eqref{eq:assistedxsection}].
The highly enhanced $\chi_1$-annihilation cross section for $r_1\ll1$ could render several caveats to be kept in mind on the $\chi_1$-${\rm SM}$ interaction, as will be discussed below.
In the rest of this section, we will describe the thermal history of $p$-wave annihilating $\chi_1$, and discuss the various constraints on $\chi_1$-annihilation described in the previous section.

\begin{figure}[t!]
\centering
\includegraphics[scale=0.61]{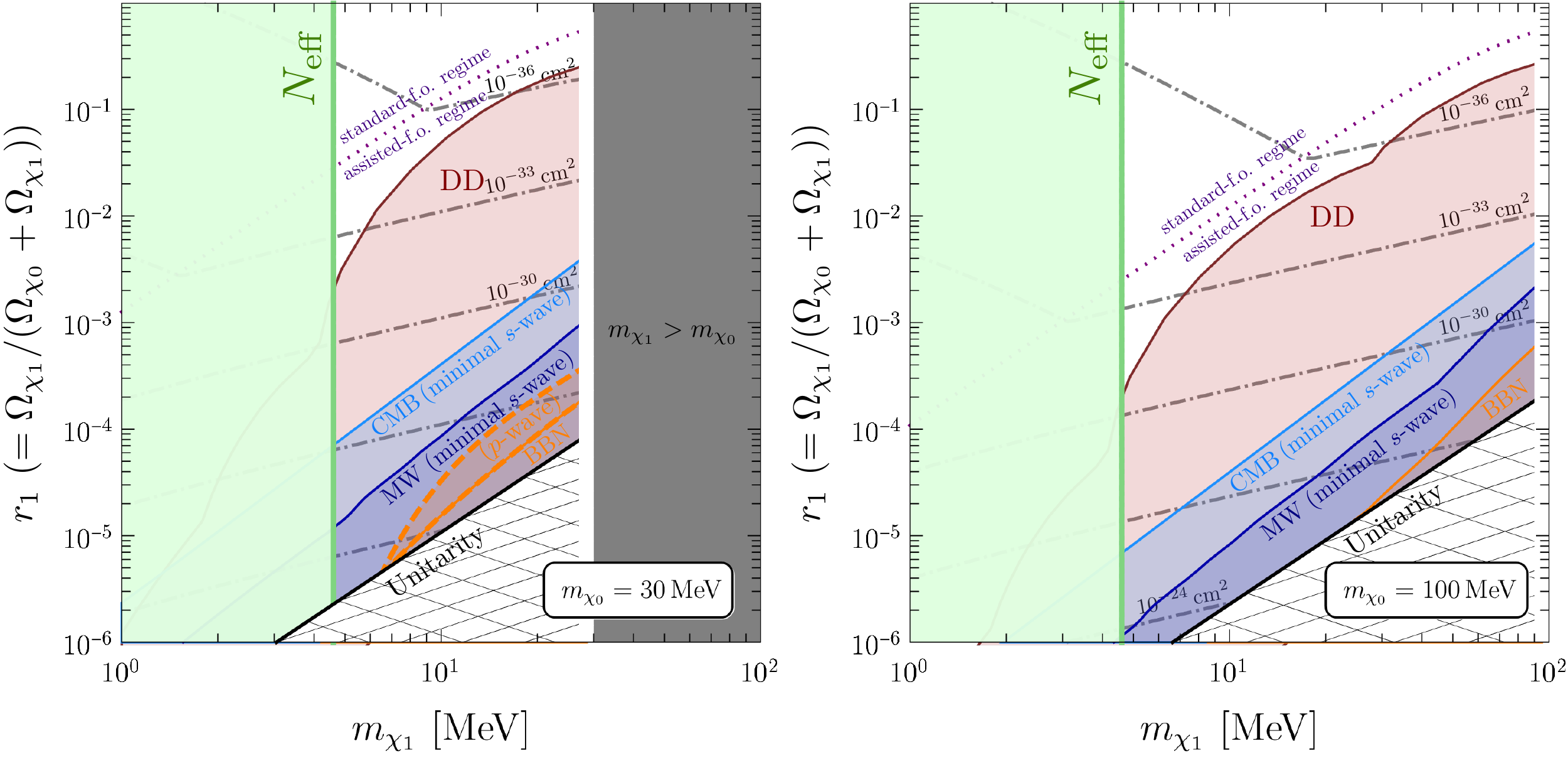}
\caption{
Same as Figure~\ref{fig:paramnoSH_s} but for $p$-wave annihilating $\chi_1$ while the hatched region at the right-bottom corner is the unitarity bound on the DM annihilation cross section.
In the assisted regime (below the dotted curve), the annihilation cross section is sharply enhanced towards small $r_1$, i.e., $(\sigma_1 v_{\rm rel})_p\propto1/r_1^3$.
The only robust constraint in the presented parameters is the $N_{\rm eff}$ constraint from ${\rm MeV}$-scale freeze-out (green).
Constraints on $(\sigma_1 v_{\rm rel})_p$ from photo-dissociation, CMB, and DM annihilations in the MW do not appear in the presented range.
For $r_1<0.01$, the minimal $s$-wave contribution can become relevant.
As an example, we present the bounds coming from the unsuppressed $s$-wave contribution in the heavy mediator limit (the transparent regions in sky-blue, deep-blue, and orange).
}
\label{fig:paramnoSH}
\end{figure}

\begin{figure}[t!]
\centering
\includegraphics[scale=0.61]{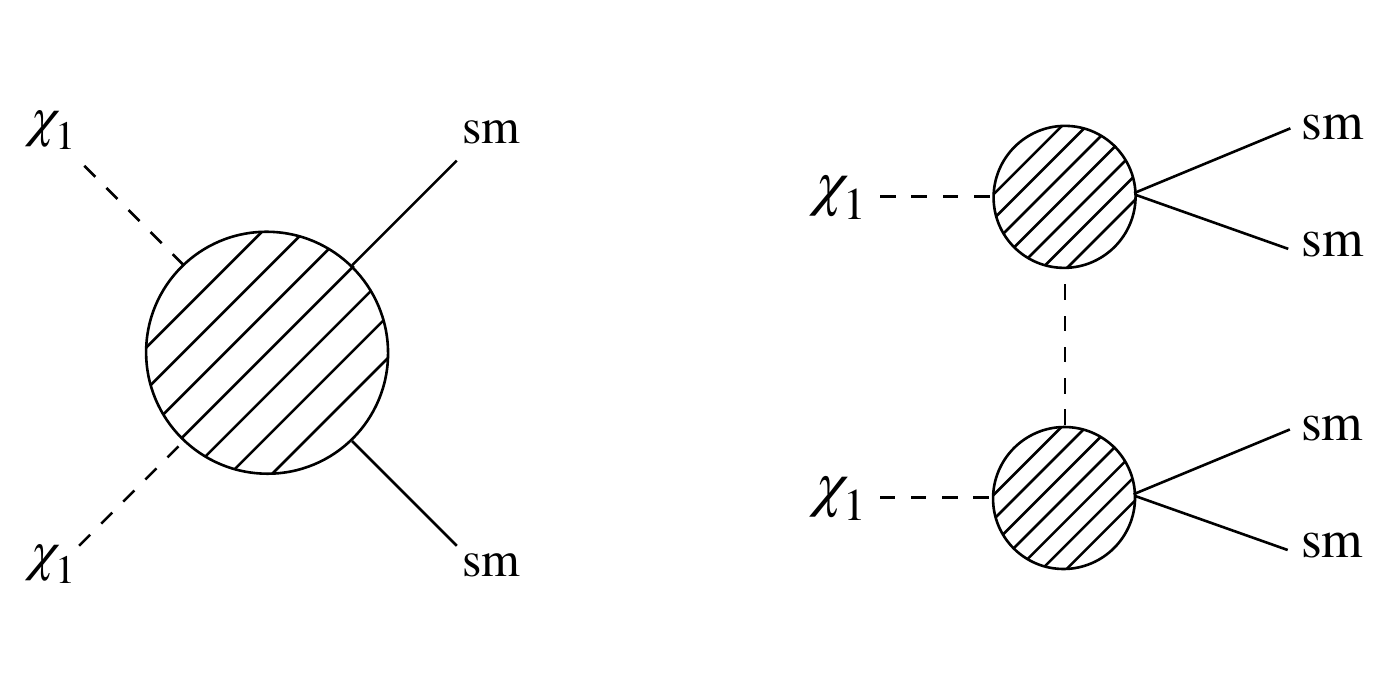}
\caption{
2-body (left) and 4-body (right) annihilation channels of $\chi_1$.
While the $p$-wave 2-body annihilation channel of $\chi_1$ dominates the annihilation of $\chi_1$ around the freeze-out of $\chi_1$, the unsuppressed $s$-wave 4-body annihilation channel may become relevant afterwards, e.g., during the photo-dissociation epoch, recombination epoch, and inside the MW halo.
}
\label{fig:dmann}
\end{figure}

For $p$-wave annihilating DM, the bounds on DM annihilation from the observations on light element abundances and CMB depend on the DM temperature evolution during the relevant cosmological epochs.
We expand the annihilation cross section of $\chi_1$ in the non-relativistic limit as~\footnote{Note that $\langle\sigma_1 v_{\rm rel}\rangle$ represents the total annihilation cross section and does not specify a final state. Dominant annihilation processes for $(\sigma_1 v_{\rm rel})_s$ and $(\sigma_1v_{\rm rel})_p$ may have different final states. For $p$-wave annihilating $\chi_1$, the dominant annihilation channels for the $s$ and $p$-wave contributions may be given as in Figure~\ref{fig:dmann}.}
\begin{equation}
\langle\sigma_1 v_{\rm rel}\rangle\simeq (\sigma_1 v_{\rm rel})_s+(\sigma_1v_{\rm rel})_p\,\langle v_{\rm rel}^2\rangle\,.
\label{eq:xsectionexpand}
\end{equation}
Note that what we mean by $p$-wave annihilation is that the term proportional to $(\sigma_1v_{\rm rel})_p$ is dominant around $T = T_{\rm fo,1} \sim m_{\chi_1}/20$.
Even in the $p$-wave annihilation case, the unsuppressed $s$-wave annihilation contribution $(\sigma_1 v_{\rm rel})_s$ may be dominant over the other around the cosmological epoch of interest for $r_1\ll1$, as will be discussed shortly.
In the absence of the DM self-heating epoch, the temperature evolution of $\chi_1$ is
\begin{equation}
T_{\chi_{1}}=\begin{cases}
T & {\rm for}\,T>T_{{\rm kd}}\,,\\
T_{{\rm kd}}\left[a(T_{\rm kd})/a(T)\right]^{2} & {\rm for}\,T<T_{{\rm kd}}\,,
\end{cases}
\label{eq:Tchi1noSH}
\end{equation}
where $T_{\rm kd}$ is the SM temperature at the kinetic decoupling of $\chi_1$ and $a$ is the scale factor.
Hereafter, we assume that the elastic scattering process that keeps $\chi_1$ in kinetic equilibrium, i.e., $\chi_1 {\rm sm}\rightarrow\chi_1 {\rm sm}$, is related to the $p$-wave annihilating process of $\chi_1$ by the crossing symmetry.
In the heavy mediator limit, the elastic scattering cross section has the minimal contribution given by Eq.~\eqref{eq:sigmachi1e}.

If the kinetic decoupling of $\chi_1$ takes place before the electron-position annihilation, $T\gtrsim m_e/20$, the decoupling point is in turn virtually determined by the $\chi_1 e\rightarrow \chi_1 e$ process.
For $r_1\ll1$, due to the enhanced annihilation cross section (and thus the enhanced $\sigma_{\chi_1 {\rm sm}}$), the kinetic decoupling could happen after the electron-position annihilation.
In such a case, the elastic scattering of $\chi_1$ with proton also has to be taken into account.
We determine the kinetic-decoupling temperature $T_{\rm kd}$ is determined by the condition $\gamma_{\chi_1 {\rm sm}}\simeq H$ where $\gamma_{\chi_1 {\rm sm}}$ is the momentum transfer rate given by~\cite{Dvorkin:2013cea,Binder:2016pnr,Boddy:2018wzy}
\begin{equation}
\gamma_{\chi_{1}{\rm sm}}\simeq\left(\frac{\delta E}{T}\right)n_{{\rm sm}}\sigma_{\chi_{1}{\rm sm}}\left\langle v_{{\rm rel,\chi_1{\rm sm}}}\right\rangle \,,
\end{equation}
where $\delta E$ is the change in $\chi_1$ kinetic energy per elastic scattering and $\left\langle v_{{\rm rel,\chi_1{\rm sm}}}\right\rangle$ is the averaged relative scattering velocity between $\chi_1$ and an SM particle.
For elastic scattering with electrons, we may estimate $\delta E/T$ as $\simeq T/m_{\chi_1}$ ($\simeq m_e/m_{\chi_1}$) for relativistic (non-relativistic) electrons. 
For the scattering with non-relativistic protons, $\delta E / T \simeq m_{\chi_1}/m_p$.
For general $T_{\chi_1}$, the relative scattering velocity is given as
\begin{equation}
\left\langle v_{{\rm rel},\chi_1 {\rm sm}}\right\rangle ^{2}=\frac{8}{\pi}\left(\frac{T_{\chi_{1}}}{m_{\chi_{1}}}+\frac{T}{m_{{\rm sm}}}\right)\,,
\end{equation}
where we may put $T_{\chi_1}=T$ when estimating $T_{\rm kd}$ in the absence of DM self-heating;
if the $\chi_1$ exhibits the self-heating epoch, the kinetic-decoupling point can be determined by a different condition, as will be discussed in the next section.

Since the photo-dissociation constraints are sensitive to the DM annihilation rate $n_{\chi_1}^2\langle\sigma_1 v_{\rm rel}\rangle$, the constraints depends on the DM temperature evolution in the temperature range relevant to photo-dissociation of light nuclei, $100\,{\rm eV} \lesssim T\lesssim 10\,{\rm keV}$.
Therefore, in order to put the photo-dissociation constraints on $\chi_1$-annihilation, one needs to estimate $T_{\rm kd}$.
For $T_{\rm kd}\gtrsim 10\,{\rm keV}$, the redshift behavior is $T_{\chi_1}\propto 1/a^2$ during the relevant epoch and we simply rescale the photo-dissociation constraints (as an upper bound) on $(\sigma_1v_{\rm rel})_p$ for $T_{\rm kd} = 10\,{\rm keV}$~\cite{Depta:2019lbe} with the factor $\sim (T_{\rm kd}/10\,{\rm keV})$ (aside from the rescaling with $r_1$ discussed above).
For $T_{\rm kd}\lesssim 100\,{\rm eV}$, since the redshift behavior is $T_{\chi_1}=T$ during the relevant epoch, we may simply take the upper bound on $(\sigma_1v_{\rm rel})_p$ for $T_{\rm kd} = 100\,{\rm eV}$.
If $T_{\rm kd}$ lies within the range $100\,{\rm eV} \lesssim T\lesssim 10\,{\rm keV}$, we aggressively underestimate the upper bound by applying the same upper bound with the $T_{\rm kd} = 100\,{\rm eV}$ case (orange region with dashed boundary in Figure~\ref{fig:paramnoSH});
this is to display the potentially constrained parameter region, while a robust bound requires dedicated analyses.

The CMB bounds on DM annihilations are also sensitive to the DM annihilation rate at the last scattering and one needs to evaluate the DM temperature around the recombination epoch $T\sim 0.235\,{\rm eV}$.
For the CMB bounds on $\chi_1$-annihilation, we estimate $T_{\chi_1}$ at the CMB epoch, $T=0.235\,{\rm eV}$ using Eq.~\eqref{eq:Tchi1noSH}.
The Galactic $\chi_1$-annihilations could provide stronger upper bounds on $(\sigma_1v_{\rm rel})_p$ than the CMB bound since the annihilation rates can be larger in the Galactic halo compared to the one in the recombination epoch;
this is because the velocities of DM particles in the Galactic halo can be larger than the DM velocities around the recombination.
We take $\langle v_{\rm rel}\rangle\sim220\,{\rm km/s}$ to estimate the annihilation cross section on the Galactic scales~\cite{Essig:2013goa}.

As we have done in the case of $s$-wave annihilating $\chi_1$, we summarize the aforementioned indirect-detection constraints in Figure~\ref{fig:paramnoSH}.
As a reference, we plot the contours for the minimal contribution to $\sigma_{\chi_1 e}$ (dot-dashed) according to Eq.~\eqref{eq:sigmachi1e} [but with $(\sigma_1v_{\rm rel})_p$ instead of $(\sigma_1v_{\rm rel})_s$] and present the corresponding direct-detection constraints based on this minimal contribution (brown).
In the assisted regime, the annihilation cross section is enhanced for small $r_1$ as $(\sigma_1v_{\rm rel})_p\propto1/r_1^3$.
Since the volumetric annihilation rate scales as $n_{\chi_1}^2\langle\sigma_1 v_{\rm rel}\rangle\propto1/r_1$ for $p$-wave annihilation, the constraints are more relevant towards the small $r_1$.

We find that for $p$-wave annihilating $\chi_1$, the only robust constraint appearing in Figure~\ref{fig:paramnoSH} is the $N_{\rm eff}$ bound from the ${\rm MeV}$-scale freeze-out of DM (in green).
However, for $r_1< 0.01$, the unsuppressed $s$-wave component $(\sigma_1v_{\rm rel})_s$ can be dominant over the $p$-wave part during the cosmological epoch of interest.
In the heavy mediator limit, we may have the following minimal contribution to $(\sigma_1v_{\rm rel})_s$ given by
\begin{equation}
(\sigma_1v_{\rm rel})_s\sim\frac{m_{\chi_{1}}^{2}}{(4\pi)^{3}}(\sigma_1v_{\rm rel})_p^{2}\,,
\end{equation}
where we have in mind the unsuppressed 4-body annihilation channel contributing to $(\sigma_1v_{\rm rel})_s$ (see Figure~\ref{fig:dmann}).
We plot the possible constraints from the minimal $s$-wave contribution as well (labeled by `minimal $s$-wave').
The photo-dissociation constraint with the dashed boundary is the region where the $s$-wave contribution starts to dominate during the relevant photo-dissociation epoch, $100\,{\rm eV} \lesssim T\lesssim 10\,{\rm keV}$;
in such a case, we take the more constraining bound among the pure $s$-wave case and the pure $p$-wave case.

\section{Self-heating from boosted DM particles}
\label{section:SHDM}

After the chemical freeze-out of $\chi_0$, residual annihilations of $\chi_0$ continuously produce boosted $\chi_1$ particles.
Before the kinetic decoupling of $\chi_1$, the boosted $\chi_1$ particles have no effect on the evolution of $T_{\chi_1}$.
As we have discussed in the previous section, the mere effect of the produced $\chi_1$ is to contribute to the relic abundance of $\chi_1$.
However, if $\chi_1$ exhibits sizable self-scattering so that the self-scattering is efficient even after the kinetic decoupling of $\chi_1$, the temperature evolution of $\chi_1$ after the kinetic decoupling exhibits interesting dynamics.
In the presence of efficient self-scattering, the excess kinetic energy of energetic $\chi_1$ particles produced from residual $\chi_0$-annihilations are shared with the majority of the $\chi_1$ particles and heat the $\chi_1$ particles as a whole.
Such processes, which we dub as the DM self-heating, could enhance the temperature of $\chi_1$ compared to the SM one.
For example, if $\chi_1$ elastically scatter with electrons, the kinetic decoupling typically occurs around the electron-position annihilation due to dwindling electron number density.~\footnote{A notable exception is when the final DM abundance is set by the DM annihilation through a resonant mediator~\cite{Binder:2017rgn}; while the annihilation cross section is resonantly enhanced, the DM-SM elastic scattering is relatively suppressed and the kinetic decoupling may take place very close to the freeze-out.}
Assuming $T_{\chi_1}=T$, the decoupling of self-scattering takes place when the SM temperature is
\begin{equation}
T_{\rm dec,self} \simeq \frac{m_e}{20}\,\left(\frac{m_{\chi_1}}{100\,{\rm MeV}}\right)^{1/3}\left(\frac{0.1}{r_1}\right)^{2/3}\left(\frac{10^{-6}\,{\rm cm^2/g}}{\sigma_{\rm self}/m}\right)^{2/3}\,,
\label{eq:Tselfpreliminary}
\end{equation}
where $\sigma_{\rm self}/m$ is the self-scattering cross section per $\chi_1$ mass and $m_e/20$ is the SM temperature around the electron-positron annihilation.
Thus, if the self-scattering cross section is large enough to delay the decoupling of self-scattering beyond the kinetic-decoupling point, DM undergoes self-heating until the decoupling of self-scattering.
After then, $\chi_1$ particles adiabatically cool as $T_{\chi_1}\propto1/a^2$.

In this section, we demonstrate the cosmological evolution of $\chi_1$ in the case of $s$-wave annihilation of $\chi_0$ and $p$-wave annihilation of $\chi_1$;
as we have discussed in the last section, the case of $s$-wave annihilation of $\chi_1$ is strongly disfavored by the CMB bounds on DM annihilation.
With the $s$-wave annihilation of $\chi_0$, the temperature of $\chi_1$ could redshift like radiation $T_{\chi_1}\propto1/a$ even after the kinetic decoupling.
The modified evolution of $T_{\chi_1}$ from the self-heating adds several interesting aspects to the cosmological constraints on $\chi_1$.
For self-scattering cross section of $\chi_1$ as large as $\sigma_{\rm self}/m\sim0.1\,{\rm cm^2/g}$, the self-heating epoch could persist until the matter-radiation equality.
Such elongated self-heating epoch may suppress the structure formation of $\chi_1$ and could be subject to the warm dark matter (WDM) constraints, e.g., from the Lyman-$\alpha$ forest observations.
The warmness of $\chi_1$ could also suppress the clustering of $\chi_1$ in the Galactic scale and may relax the direct/indirect-detection constraints.
Enhanced $T_{\chi_1}$ also affects the constraints that directly depend on the annihilation rate of $\chi_1$.
For example, if the self-heating epoch overlaps with the epoch relevant for photo-dissociation of light nuclei, the constraints on the annihilation cross section would become severer.

We describe the self-heating of $\chi_1$ in Section~\ref{section:SHhistory};
details of the Boltzmann equations and the analytic arguments are collected in the Appendix~\ref{appendix:Boltzmanndetail}.
We discuss the implications of the DM self-heating epoch on the cosmological constraints in Section~\ref{section:SHDMconstraint}.

\subsection{Thermal history of $\chi_1$ with self-heating}
\label{section:SHhistory}

Efficient self-scattering of $\chi_1$, i.e., $\Gamma_{\rm self}\gtrsim H$, allows $\chi_1$ particles to efficiently exchange their energy and momentum among themselves.
Regardless of the energy exchanges with external systems, efficient self-scattering forces $\chi_1$ particles to follow the thermal energy distribution $f_{\chi_1}(E)\propto e^{-E/T_{\chi_1}}$;
this is the case even in the presence of the $\chi_0\chi_0\rightarrow\chi_1\chi_1$ process.
After the chemical decoupling of $\chi_0$, residual $\chi_0$-annihilations produce a minority of boosted $\chi_1$ particles.
Efficient self-scattering quickly redistribute the excess kinetic energy to the majority of $\chi_1$ particles, heating the $\chi_1$ particles as a whole.
In such a case, the evolution of $\chi_1$ temperature is described by the following equation:
\begin{equation}
\dot{T}_{\chi_{1}}+2HT_{\chi_{1}}\simeq\gamma_{\rm heat}T-2\gamma_{\chi_1 {\rm sm}}\left(T_{\chi_{1}}-T\right)\,,
\label{eq:Teqnonrel}
\end{equation}
where $\gamma_{\rm heat}$ is defined as
\begin{equation}
\begin{aligned}
\gamma_{{\rm heat}}&=\frac{2n_{\chi_0}^2 \left(\sigma_0 v_{{\rm rel}}\right)\delta m}{3n_{\chi_{1}}T}\,.
\end{aligned}
\label{eq:gammaheat}
\end{equation}
After the chemical freeze-out, the abundances of $\chi_0$ and $\chi_1$ are virtually conserved and thus Eq.~\eqref{eq:Teqnonrel} alone determines the evolution of $T_{\chi_1}$.
Note that we have assumed that both $\chi_0$ and $\chi_1$ are non-relativistic in Eq.~\eqref{eq:Teqnonrel} (see Appendix~\ref{appendix:Boltzmanndetail} for details).
The inverse of the heating rate, $\gamma_{\rm heat}^{-1}$, represents the timescale during which a $\chi_1$ particle obtains kinetic energy comparable to $\sim T$.
The two terms in the RHS of Eq.~\eqref{eq:Teqnonrel} represents the two paths for the energy exchange of $\chi_1$ with external systems.
The term proportional to $\gamma_{\chi_1 {\rm sm}}$ represents the energy exchange with the SM plasma through the $\chi_1{\rm sm}\rightarrow\chi_1{\rm sm}$ process.

Initially, $\gamma_{\chi_1 {\rm sm}}$ is dominant over both $H$ and $\gamma_{\rm heat}$, and the kinetic equilibrium is achieved.
As the Universe cools, $\gamma_{\chi_1{\rm sm}}$ drops and the term proportional to $\gamma_{\chi_1{\rm sm}}$ could become negligible from Eq.~\eqref{eq:Teqnonrel}.
The self-heating epoch starts from then, and the heat injection from $\chi_0$-annihilation can modify the evolution of $T_{\chi_1}$ from what we expect for free-streaming non-relativistic particles, i.e., $T_{\chi_1}\propto1/a^2$.
In the case of $s$-wave annihilation of $\chi_0$, the temperature ratio $T_{\chi_1}/T$ asymptotes to the following:
\begin{equation}
\begin{aligned}
\left(\frac{T_{\chi_{1}}}{T}\right)_{{\rm asy.}}&\sim\frac{\gamma_{{\rm heat}}}{H}\,,\\
&\simeq\begin{dcases}
\frac{2\left(1-r_{1}\right)}{3r_{1}}\frac{m_{\chi_{1}}\delta m}{m_{\chi_0} T_{{\rm fo,0}}}\left(\frac{g_{\star}\left(T_{{\rm fo,0}}\right)}{g_{\star}\left(T_{{\rm asy.}}\right)}\right)^{1/2}\frac{g_{\star S}\left(T_{{\rm asy.}}\right)}{g_{\star S}\left(T_{{\rm fo,0}}\right)} & {\rm for}\,\,T>T_{{\rm eq}}\,,\\
\frac{4\left(1-r_{1}\right)}{3r_{1}}\frac{m_{\chi_{1}}\delta m}{m_{\chi_0} T_{{\rm fo,0}}}\left(\frac{g_{\star}\left(T_{{\rm fo,0}}\right)}{g_{\star}\left(T_{{\rm eq}}\right)}\right)^{1/2}\frac{g_{\star S}\left(T\right)}{g_{\star S}\left(T_{{\rm fo,0}}\right)}\left(\frac{T}{T_{{\rm eq}}}\right)^{1/2} & {\rm for}\,\,T<T_{{\rm eq}}\,,
\end{dcases}
\end{aligned}
\label{eq:Tratioasy}
\end{equation}
where we have used Eq.~\eqref{eq:chi2relic} and $T_{\rm eq}\sim0.75\,{\rm eV}$ is the SM temperature at the matter-radiation equality.
In the case of $p$-wave annihilation of $\chi_0$, $T_{\chi_1}$ scales as the $T_{\chi_1}\propto1/a^2$, while there is an enhancement compared to the case of no self-heating (see Appendix~\ref{section:pwaveann} for the discussion on the case of $p$-wave annihilation of $\chi_0$).
Hereafter, we focus on the case of $s$-wave annihilation since it exhibits the maximal impact of self-heating.

Due to practical reasons, we do not attempt to follow the full evolution of $T_{\chi_1}$ from Eq.~\eqref{eq:Teqnonrel}.
Instead, we specify an interval in $T$ where we can reliably estimate $T_{\chi_1}$:
\begin{equation}
T_{\chi_{1}}=\begin{dcases}
\left(\frac{T_{\chi_{1}}}{T}\right)_{{\rm asy.}}T & {\rm for}\,\,T_{{\rm dec,self}}<T<T_{{\rm min}}\,,\\
\left(\frac{T_{\chi_{1}}}{T}\right)_{{\rm asy.}}T_{{\rm dec,self}}\left[a(T_{{\rm dec,self}})/a(T)\right]^{2} & {\rm for}\,\,T<T_{{\rm dec,self}}\,,
\end{dcases}
\label{eq:naiveTchi1}
\end{equation}
where $T_{\rm min}$ is the SM temperature below which the $T_{\chi_1}$ follows the asymptotic solution given in Eq.~\eqref{eq:Tratioasy}, and $T_{\rm dec, self}$ is the SM temperature at the decoupling of self-scattering.
To define $T_{\rm min}$, we rewrite Eq.~\eqref{eq:Teqnonrel} as
\begin{equation}
\dot{T}_{\chi_{1}}\simeq-2\left(H+\gamma_{\chi_{1}{\rm sm}}\right)T_{\chi_{1}}+\left\{ \gamma_{{\rm heat}}\Theta\left(\Gamma_{{\rm self}, v_{\rm rel}=c}-\Gamma_{\chi_{1}{\rm sm},v_{\rm rel}=c}\right)+2\gamma_{\chi_{1}{\rm sm}}\right\} T\,.
\label{eq:Tchi1evolre}
\end{equation}
The first term in the RHS is the friction term for $T_{\chi_1}$ and the second term is the source term.
$\gamma_{\rm heat}$ is multiplied by the unit-step function by hand to incorporate the stopping of boosted $\chi_1$ by the SM plasma;
if the boosted $\chi_1$ particles dominantly scatter with SM particles, self-heating is ineffective.
We remark that such SM-stopping of $\chi_1$ is only relevant for small $r_1\ll0.1$, where large annihilation cross section is required deplete $\chi_1$ to the desired $r_1$.
Since we consider $\sigma_{\rm self}/m$ as strong as $\sim 0.1\,{\rm cm^2/g}$, boosted $\chi_1$ dominantly scatters with $\chi_1$ for $r_1\gtrsim0.1$.
When evaluating the rates for boosted $\chi_1$ in the step-function, we set $v_{\rm rel}=c$ for simplicity.

\begin{figure}
\centering
\includegraphics[scale=0.62]{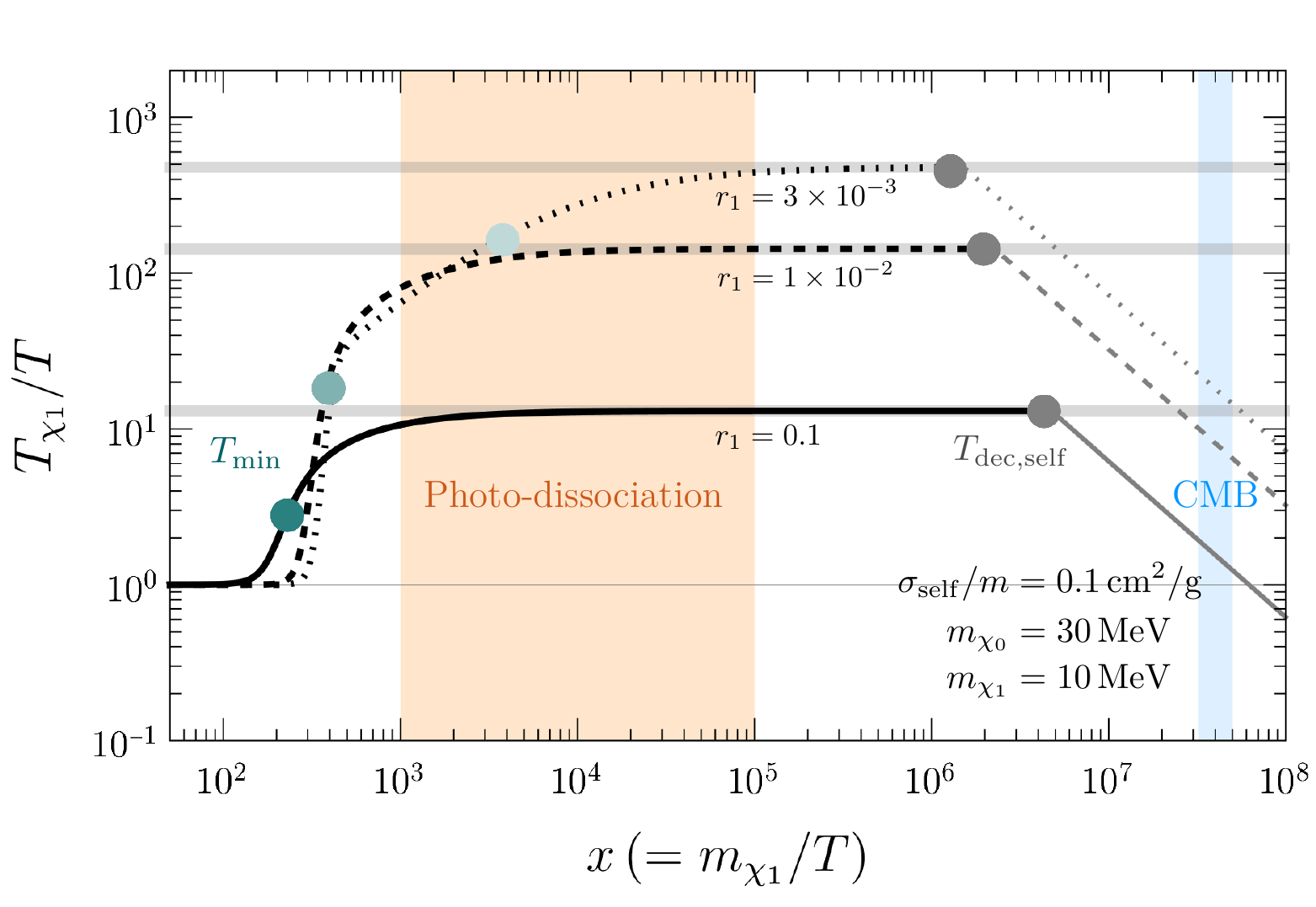}
\caption{Evolution of $T_{\chi_1}$ for various abundance fractions (black); $r_1=0.1$ (solid), $r_1=1.5\times10^{-2}$ (dashed), and $r_1=3\times10^{-3}$ (dotted);
the selected parameters corresponds to the depicted parameters (as stars) in the upper-left panel of Figure~\ref{fig:paramSH}.
Note that there is no qualitative difference in the evolutions of $T_{\chi_1}$ among the depicted parameters.
The shaded regions corresponds to cosmological epochs that constrain DM annihilation, i.e., the photo-dissociation epoch relevant for DM annihilation (orange), and the recombination epoch (blue).
The green circles represent the underestimated $T_{\rm min}$~[Eq.~\eqref{eq:Tmin}], after which we may apply the naive estimation for $T_{\chi_1}$ given in Eq.~\eqref{eq:naiveTchi1};
the horizontal gray lines are the asymptotic temperature ratio given in Eq.~\eqref{eq:Tratioasy}.
The gray circles represent the decoupling point of self-scattering~[Eq.~\eqref{eq:Tself}] assuming Eq.~\eqref{eq:naiveTchi1}.
The presented parameters corresponds to the depicted parameters in the upper-left panel of Figure~\ref{fig:paramSH}.
}
\label{fig:Tdmevol}
\end{figure}

The asymptotic solution given by Eq.~\eqref{eq:Tratioasy} is defined when $\gamma_{\chi_1{\rm sm}}$ is negligible from Eq.~\eqref{eq:Tchi1evolre}.
We define $T_{\rm dec,el}$ as the SM temperature below which $\gamma_{\chi_1{\rm sm}}$ is negligible as a friction term, i.e, $\gamma_{\chi_1{\rm sm}}\lesssim H$.
We define $T_{\rm sh}$ as the SM temperature below which $\gamma_{\chi_1{\rm sm}}$ becomes negligible as a source term;
according to the definition, we have
\begin{equation}
T_{{\rm sh}}=\min\left[T_{{\rm stop}},T_{{\rm heat}}\right]\,,
\label{eq:Tsh}
\end{equation}
where $T_{\rm stop}$ and $T_{\rm heat}$ are determined by the condition $\Gamma_{{\rm self}, v_{\rm rel}=c}=\Gamma_{\chi_{1}{\rm sm},v_{\rm rel}=c}$ and $\gamma_{\rm heat}=\gamma_{\chi_1{\rm sm}}$, respectively.
Therefore, the asymptotic solution for $T_{\chi_1}$ is defined for $T\lesssim T_{\rm min}$, where $T_{\rm min}$ is given by
\begin{equation}
T_{{\rm min}}=\min\left[T_{\rm dec, el},T_{{\rm stop}},T_{{\rm heat}}\right]\,.
\label{eq:Tmin}
\end{equation}
Note that to determine the true value of $T_{\rm min}$, one needs a priori knowledge on the exact evolution of $T_{\chi_1}$, since the rate $\gamma_{\chi_1{\rm sm}}$ generally depends on $T_{\chi_1}$.
Nevertheless, we underestimate $T_{\rm min}$ as follows;
we overestimate $\gamma_{\chi_1{\rm sm}}$ by assuming the maximal temperature of $T_{\chi_1}=(T_{\chi_1}/T)_{\rm asy.} T$ to underestimate $T_{\rm dec,el}$ and $T_{\rm heat}$.
At temperatures lower than the underestimated $T_{\rm min}$, one can reliably estimate $T_{\chi_1}$ with Eq.~\eqref{eq:naiveTchi1}, regardless of the exact evolution of $T_{\chi_1}$ for $T\gtrsim T_{\rm min}$.
By taking into account the estimation given in Eq.~\eqref{eq:naiveTchi1}, we modify the estimation of $T_{\rm dec, self}$ from Eq.~\eqref{eq:Tselfpreliminary} as
\begin{equation}
T_{{\rm dec,self}}\simeq1\,{\rm eV}\,\left(\frac{T_{\chi_{1}}}{T}\right)_{{\rm asy.}}^{-n}\left(\frac{0.3}{r_{1}}\right)^{2n}\left(\frac{m_{\chi_{1}}}{100\,{\rm MeV}}\right)^{n}\left(\frac{1\,{\rm cm^{2}/g}}{\sigma_{{\rm self}}/m_{\chi_{1}}}\right)^{2n}\,,
\label{eq:Tself}
\end{equation}
where $n=1/3$ for when $T_{\rm dec, self}>T_{\rm eq}$, and $n=2/9$ for when $T_{\rm dec,self}<T_{\rm eq}$.
If $T_{\rm dec,self}$ evaluated in this way is larger than $T_{\rm min}$, our estimation of Eq.~\eqref{eq:naiveTchi1} is not self-consistent and thus not reliable.
Hereafter, we use Eq.~\eqref{eq:naiveTchi1} to estimate $T_{\chi_1}$ when the consistency condition $T_{\rm min}>T_{\rm dec,self}$ is satisfied at the most conservative level;
we underestimate $T_{\rm min}$ by taking the highest possible value for $T_{\chi_1}$, i.e., the asymptotic solution Eq.~\eqref{eq:Tratioasy}.
At the same time, we overestimate $T_{\rm dec,self}$ by taking the lowest possible value, i.e., the evolution in the absence of DM self-heating Eq.~\eqref{eq:Tchi1noSH} (see Appendix~\ref{section:Tdmevolapp} for more discussion).
In Figure~\ref{fig:Tdmevol}, we present the numerical solutions to Eq.~\eqref{eq:Tchi1evolre};
for temperatures lower than the underestimated $T_{\rm min}$ (green circles), we find that Eq.~\eqref{eq:naiveTchi1} approximates well the evolution of $T_{\chi_1}$.

On the other hand, there may be cases where $\gamma_{\rm heat}$ never becomes dominant in the source term before the decoupling of self-scattering, i.e., $T_{\rm sh}<T_{\rm dec,self}$.
In such a case, $T_{\chi_1}$ is reliably estimated with Eq.~\eqref{eq:Tchi1noSH}.
Again, since we cannot a priori determine $T_{\rm sh}$ and $T_{\rm dec,self}$ before knowing the evolution of $T_{\chi_1}$, we conservatively overestimate (underestimate) $T_{\rm sh}$ ($T_{\rm dec,self}$).
More specifically, we determine $T_{\rm sh}$ ($T_{\rm dec,self}$) by taking the lowest (highest) possible values for $T_{\chi_1}$, which is estimated by Eq.~\eqref{eq:Tchi1noSH}~[$T_{\chi_1}=(T_{\chi_1}/T)_{\rm asy.} T$].

\subsection{Cosmological constraints on DM self-heating}
\label{section:SHDMconstraint}

\subsubsection{Warm dark matter constraints}

\noindent
Before the matter-radiation equality, $T_{\chi_1}$ redshifts like radiation during DM self-heating epoch.
For $\sigma_{\rm self}/m$ as large as $\sim1\,{\rm cm^2/g}$, DM self-heating epoch could persist until the vicinity of the matter-radiation equality, i.e., $T_{\rm dec,self}\sim T_{\rm eq}$~[Eq.~\eqref{eq:Tself}].
The resultant $T_{\chi_1}$ around $T_{\rm eq}$ is much larger than that without DM self-heating, and may be sizable to make $\chi_1$ behave as warm dark matter (WDM).
Therefore, in the presence of DM self-heating, the total relic dark matter is composed of two components with distinct temperatures: warm $\chi_1$ and cold $\chi_0$.
One way to represent the warmness of DM is the cutoff in the resultant matter power spectrum, which can be estimated by the (co-moving) Jeans scale $k_{\rm J}$ at the matter-radiation equality.
$k_{\rm J}$ is the wave number that appears in the evolution equation of $\chi_1$'s density perturbation, and corresponds to the length scale $\lambda_{\rm J}=2\pi/k_{\rm J}$ below which the pressure gradient of DM wins over gravity.
For density perturbations of wave numbers $k>k_{\rm J}$, $\chi_1$ cannot experience gravitational collapse due to its own velocity dispersion.
The reason that $k_{\rm J}$ is evaluated at the matter-radiation equality is that DM density perturbations start to rapidly grow only after the matter-radiation equality, and $k_{\rm J}$ of $\chi_1$ takes the minimum value (the largest length scale) then since $k_{\rm J}\propto a^{1/2}$ during the matter-dominated era.
Assuming the temperature evolution of $\chi_1$ follows the estimation given in Eq.~\eqref{eq:naiveTchi1}~\footnote{We note that the WDM constraints we will consider will only be relevant for $r_1\gtrsim0.07$;
for such $r_1$, estimation of $T_{\chi_1}$ given in Eq.~\eqref{eq:naiveTchi1} is a good approximation and the WDM constraints with respect to Eq.~\eqref{eq:kJ} will be robust.}, the Jeans wave number of $\chi_1$ is given as
\begin{equation}
\begin{aligned}
k_{{\rm J}}&=a\sqrt{\frac{4\pi G\bar{\rho}_{m}}{\left\langle \overrightarrow{v}^{2}\right\rangle_1 }}\bigg|_{{\rm eq}}\,,\\
&\simeq76\,{\rm Mpc}^{-1}\,\left(\frac{r_{1}}{1-r_{1}}\right)^{1/2}\left(\frac{m_{\chi_0}}{\delta m}\right)^{1/2}\left(\frac{m_{\chi_0}}{100\,{\rm MeV}}\right)^{1/2}\max\left[1,\sqrt{\frac{T_{{\rm dec,self}}}{T_{{\rm eq}}}}\right]\,,
\end{aligned}
\label{eq:kJ}
\end{equation}
where $\bar{\rho}_{m}$ is the average matter density, and $\langle \overrightarrow{v}^{2}\rangle_1$ is the variance of $\chi_1$ velocity.
We remark that although the $k_{\rm J}$ of $\chi_1$ implicitly depends on $m_{\chi_1}$, the value of $m_{\chi_1}$ itself is not important;
this is because $\langle \overrightarrow{v}^{2}\rangle_1\propto T_{\chi_1}/m_{\chi_1}=(T_{\chi_1}/T)_{\rm asy.}(T/m_{\chi_1})$ and the additional factor of $m_{\chi_1}$ from the $(T_{\chi_1}/T)_{\rm asy.}$ through DM self-heating~[Eq.~\eqref{eq:Tratioasy}] cancels the explicit $m_{\chi_1}$ dependence.
For a fixed $\delta m$, $k_{\rm J}$ of $\chi_1$ increases towards heavier $\chi_0$;
this is because heavier $\chi_0$ corresponds to smaller $\chi_0$ number density and hence smaller heating rate~[Eq.~\eqref{eq:gammaheat}].

On the other hand, $k_{\rm J}$ alone cannot entirely represent the overall effect of DM self-heating on structure formation.
This is because we have an additional parameter, $r_1$;
no matter how $k_{\rm J}$ is small, warmness of $\chi_1$ would have negligible effect on the overall matter power spectrum for $r_1\ll1$.
Therefore the two parameters, $k_{\rm J}$ and $r_1$, are needed to characterize the resultant matter power spectrum.~\footnote{We remark that one may use a different definition of $k_{\rm J}$ to estimate the suppression scale in mixed warm$+$cold DM scenarios. Instead of taking the velocity dispersion of $\chi_1$ in Eq.~\eqref{eq:kJ}, we may use the velocity dispersion of total DM, i.e., $\langle \overrightarrow{v}^{2}\rangle_{\rm tot.}\simeq r_1\langle \overrightarrow{v}^{2}\rangle_1$~\cite{Harada:2014lma}. The Jeans wave number defined with the $r_1$-weighted velocity dispersion represents the scale at which the matter power spectrum exhibits {\it sizable} suppression from the CDM one, while the one defined in Eq.~\eqref{eq:kJ} represents the exact suppression scale from the CDM one. Nevertheless, once we specify $r_1$, the WDM constraints are the same as long as we consistently use one definition of $k_{\rm J}$ for mixed DM scenarios. See also Ref.~\cite{Dienes:2020bmn} for discussion on the Jeans scale of DM in the case of multiple species with distinctive distribution function behaving as WDM.}
Given the abundance ratio $r_1$, we investigate how $k_{\rm J}$ is constrained by observations.

As shown in Eq.~\eqref{eq:kJ}, the cutoff scale defined by the linear matter power spectrum could be at the galactic scales, i.e., $k_{\rm J}={\cal O}\left(1\right)\,{\rm Mpc}^{-1}$, and thus $\chi_1$ could behave as WDM.
To differentiate WDM from cold dark matter (CDM), it is better to look into the matter distribution at high redshifts or the abundance of sub-galactic scale non-linear objects.
This is because the formation of large-size halos enhance the small-size correlation in the non-linear matter power spectrum to compensate the original discrepancy of WDM from CDM (in the linear matter power spectrum).
Also, the abundance of small-size gravitationally bound objects is sensitive to the linear matter power spectrum before the non-linear growth of structures~\cite{Press:1973iz}.
We summarize the considered WDM constraints below;
we choose these observations since the constraints are explicitly given in terms of the mixed DM scenarios.
\begin{itemize}

\item {\bf Lyman-$\alpha$ forest observations}~\cite{Baur:2017stq} :
One of the most stringent constraint on the warmness of DM comes from the observations on rather high redshifts, i.e., $z \sim 3$.
As discussed above, it is more advantageous to look into the structure of the Universe at higher redshifts to discriminate WDM and CDM.
One of the promising methods is the Lyman-$\alpha$ forest method.
Lyman-$\alpha$ absorption lines in the spectrum of distant quasars can be used as a tracer of cosmological fluctuations on scales $k\sim 0.1$--$10\,h \,{\rm Mpc}^{-1}$, at redshifts $z=2$--$4$.
We translate the constraints for warm$+$cold DM (or mixed DM) into our scenario.
For example, in Fig.~6 of Ref.~\cite{Baur:2017stq}, the constraints on mixed DM is given in the $m_{\rm wdm}$ versus $r_{\rm warm}$ plane, where $m_{\rm wdm}$ is the mass of conventional thermal WDM and $r_{\rm warm}$ is the fraction of them in mass density.
What we mean by conventional thermal warm DM is that the DM particles of mass $m_{\rm wdm,th}$ follow the Fermi-Dirac distribution with temperature $T_{\rm wdm,th}$ (motivated by, e.g., light gravitino DM from gauge-mediated supersymmetry breaking models~\cite{Pagels:1981ke,Bond:1982uy,Kamada:2013sya,Osato:2016ixc}):
\begin{equation}
f_{{\rm wdm,th}}\left(p\right)=\frac{1}{1+\exp\left[p/T_{{\rm wdm,th}}\right]}\,,
\label{eq:thwdmdis}
\end{equation}
and the relic density of the warm DM is given as
\begin{equation}
\Omega_{{\rm wdm,th}}=r_{{\rm warm}}\Omega_{{\rm DM}}=\left(\frac{T_{{\rm wdm,th}}}{T_{\nu}}\right)^3\left(\frac{m_{{\rm wdm,th}}}{94\,{\rm eV}}\right)\,,
\label{eq:thwdmden}
\end{equation}
where $T_\nu$ is the temperature of SM neutrinos.
We convert the $m_{\rm wdm}$ axis into the $k_{\rm J,wdm}/4$ axis, following the definition given in the first equality of Eq.~\eqref{eq:kJ}:
\begin{equation}
k_{{\rm J,wdm}}\simeq20\,{\rm Mpc}^{-1}\,\left(\frac{m_{{\rm wdm,th}}}{1\,{\rm keV}}\right)^{4/3}\left(\frac{1}{r_{{\rm warm}}}\right)^{1/3}\,.
\end{equation}
We then correspond $k_{\rm J,wdm}/4$ with $k_{\rm J}$ for $\chi_1$, and $r_{\rm warm}$ with $r_1$ in our scenario.~\footnote{The reason we divide factor $4$ for WDM is that the actual cutoff scale in matter power spectrum occurs at $k<k_{\rm J}$ due to their free-streaming during the radiation dominated era with relativistic distribution function.
This amounts to a cutoff wave number that is smaller than $k_{\rm J}$, which is the free-streaming horizon scale $k_{\rm FSH}\simeq k_{\rm J}/4$~\cite{Boyarsky:2008xj};
it is the present value of the particle horizon of WDM.}
The data provides constraints for $r_1\gtrsim0.07$, and we assume vanishing WDM constraint for smaller $r_1$.

\item {\bf Number of satellite galaxies in the Milky Way}~\cite{Diamanti:2017xfo} : 
While the Lyman-$\alpha$ forest observations look into the diffuse distribution of DM, the number of compact DM halos also keeps information of the linear matter power spectrum before the non-linear growth of structures.
By comparing the (expected to be) observed number of satellites in the MW with the predicted number in the mixed DM scenarios, one may constrain the warmness of DM;
if the predicted number of satellites is smaller than $N_{\rm sat}\simeq 63$, such WDM is excluded.
Ref.~\cite{Diamanti:2017xfo} combines the analyses of the predictions on the number of satellites with the {\it Planck} CMB data on temperature, polarization, and lensing measurements and the baryon acoustic oscillation data to constrain the warmness of non-conventional WDM motivated by non-resonantly produced sterile neutrinos~\cite{Dodelson:1993je};
the WDM constraint is given in Figure~5 of Ref.~\cite{Diamanti:2017xfo} in the $m_{\rm wdm,nth}$ versus $r_{\rm warm}$ plane, where $m_{\rm wdm,nth}$ is the mass of the non-conventional WDM;
the subscript `${\rm nth}$' stands for `non-thermal'.
The non-conventional WDM follow the distribution function given as
\begin{equation}
f_{{\rm wdm,nth}}\left(p\right)=\frac{N}{1+\exp\left[p/T_{\nu}\right]}\,,
\end{equation}
where $N\ll1$ is the normalization factor that reproduces the correct abundance for a given $m_{\rm wdm,nth}$.
The corresponding Jeans scale is given as
\begin{equation}
k_{{\rm J, wdm}}\simeq4.6\,{\rm Mpc}^{-1}\,\left(\frac{m_{{\rm wdm,nth}}}{1\,{\rm keV}}\right)\,,
\end{equation}
which is independent of $r_{\rm warm}$.

\end{itemize}

\begin{figure}[t!]
\centering
\includegraphics[scale=0.6]{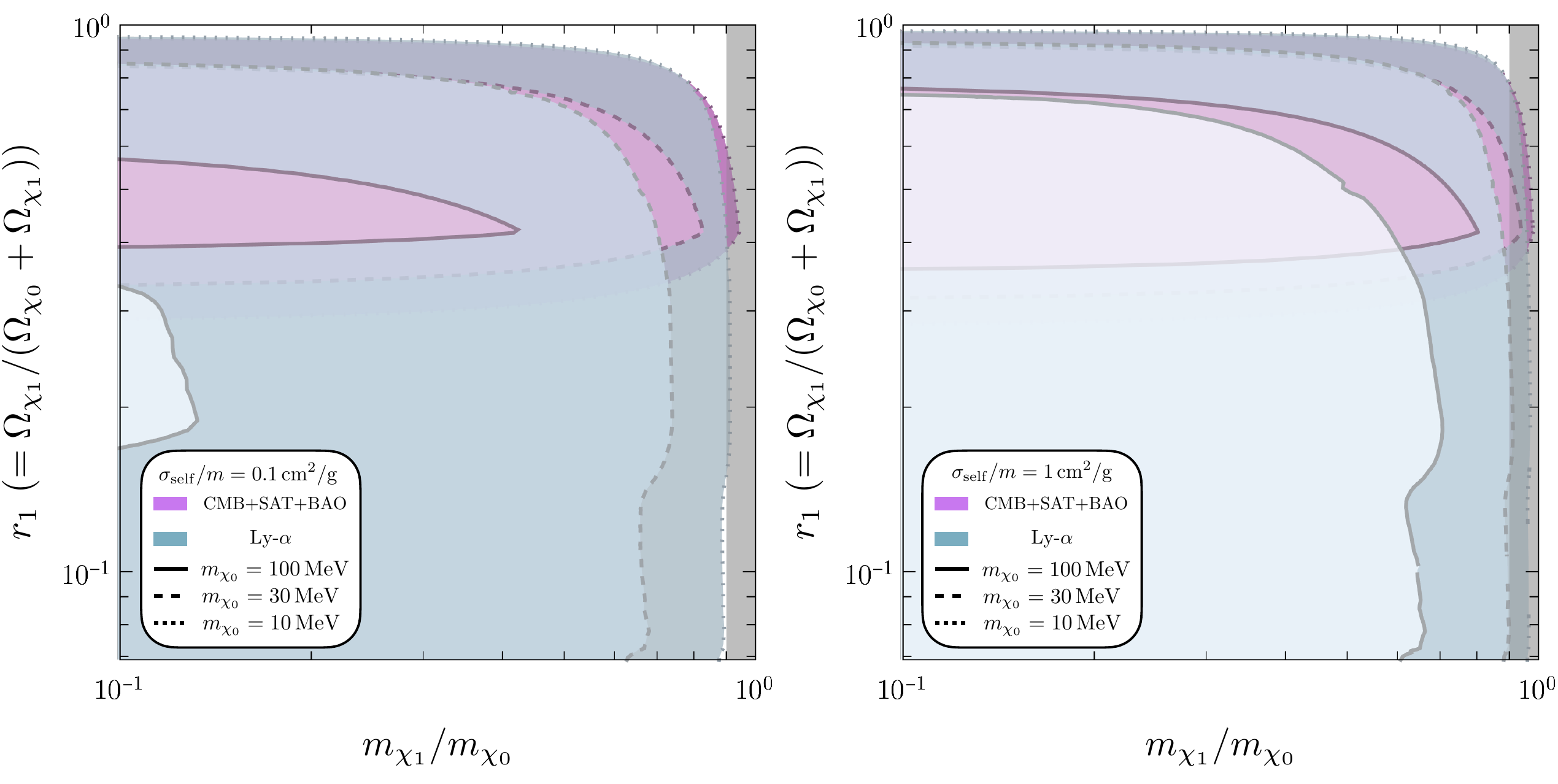}
\caption{
({\it Left}) - WDM constraints~\cite{Baur:2017stq,Diamanti:2017xfo} on $\chi_1$ for $\sigma_{\rm self}/m=0.1\,{\rm cm^2/g}$.
Shaded regions with different boundary styles correspond WDM constraints with different values of $m_{\chi_0}$ (see the lengends).
According to Eq.~\eqref{eq:kJ}, the WDM constraints become weaker for heavier $\chi_0$;
we find that the WDM constraints vanish for $m_{\chi_0}\gtrsim 110\,{\rm MeV}$.
The region in gray is where $\delta m\lesssim m_{\chi_1}/10$;
the WDM constraints in this region may be significantly modified since the of chemical freeze-outs of $\chi_0$ and $\chi_1$ interfere and the estimation of Eq.~\eqref{eq:chi2relic}~[and hence Eq.~\eqref{eq:Tratioasy} and Eq.~\eqref{eq:kJ}] would be modified.
({\it Right}) - Same as the left panel but for $\sigma_{\rm self}/m=1\,{\rm cm^2/g}$.
The WDM constraints are stronger than the left panel due to longer self-heating epoch.
We find that the WDM constraints vanish for $m_{\chi_0}\gtrsim 300\,{\rm MeV}$.
For larger values of $\sigma_{\rm self}/m$, the WDM constraints virtually do not change.
}
\label{fig:lya}
\end{figure}

In Figure~\ref{fig:lya}, we present the WDM constraints;
in each panels, we fix $\sigma_{\rm self}/m_{\chi_1}$ and display the constraints for different values of $m_{\chi_0}$ in the $m_{\chi_1}/m_{\chi_0}$ versus $r_1$ plane.
The WDM constraints vanish as we consider $r_1$ close to unity because there is no $\chi_0$ to annihilate and heat $\chi_1$.
On the other hand, the constraints vanish for $r_1\lesssim 0.07$ since the impact of the warmness of $\chi_1$ on the overall structure formation is negligible.
For a given $r_1$, the WDM constraints become weaker for larger $m_{\chi_0}$~[see Eq.~\eqref{eq:kJ}] since the heating rate is suppressed.
For $\sigma_{\rm self}/m\gtrsim1\,{\rm cm^2/g}$, the WDM constraints virtually do not change from the right panel of Figrue~\ref{fig:lya}.
This is because the self-heating epoch persists beyond the matter-radiation equality for $\sigma_{\rm self}/m\gtrsim1\,{\rm cm^2/g}$ and the cutoff scale do not change as we consider larger values of $\sigma_{\rm self}/m$.

While DM self-heating enhances $T_{\chi_1}$ and hence the $\chi_1$ annihilation rate during the photo-dissociation epoch and at the last scattering,
DM self-heating does not affect the velocity dispersion of $\chi_1$ inside our Galaxy.
However, DM self-heating may suppress the structure formation of $\chi_1$ on the Galactic scales and hence suppress the abundance fraction of $\chi_1$ inside our Galaxy compared to the cosmological one.
Such suppression of $\chi_1$ structure formation may affect the direct-detection constraints and the DM annihilation constraints from observations on diffuse X-ray and $\gamma$-ray background;
while the latter is already significantly weaker than the other constraints on DM annihilation and hence not shown in Figure~\ref{fig:paramSH}, we display the possible change in the former constraint.
We define the Jeans mass $M_{\rm J}$ of $\chi_1$ given by~\footnote{The definition of $M_{\rm J}$ given in Eq.~\eqref{eq:MJ} is smaller than the one given in Ref.~\cite{Harada:2014lma} by a factor of 8.}
\begin{equation}
M_{\rm J}=\frac{4\pi}{3} \bar{\rho}_{m,0} \left(\lambda_{\rm J}/2\right)^3 \simeq4\times10^{10}\,{\rm M}_{\odot}\,\left(\frac{\lambda_{{\rm J}}}{1\,{\rm Mpc}}\right)^{3} \,, 
\label{eq:MJ}
\end{equation}
where $\bar{\rho}_{m,0}$ is the average matter density at present.
$M_{\rm J}$ is the total mass contained within a sphere of diameter $\lambda_{\rm J}=2\pi/k_{\rm J}$ before the non-linear gravitational collapse.
The gravitational collapse of $\chi_1$ along a DM clump of mass smaller than $M_{\rm J}$ would be suppressed, and thus the abundance fraction of $\chi_1$ inside such a clump would be smaller than the cosmological one.
In Figure~\ref{fig:paramSH}, we display the possible modification to the direct-detection constraints by aggressively estimating the abundance fraction of $\chi_1$ inside our Galaxy to be vanishing when the Jeans mass of $\chi_1$ is larger than the mass of MW.
The brown shaded region enclosed by solid curve is the same direct-detection constraint in Figure~\ref{fig:paramnoSH}, but with the requirement of $M_{\rm J}\lesssim 1\times 10^{12}\,{\rm M}_\odot$.
For $M_{\rm J}\gtrsim 1\times 10^{12}\,{\rm M}_\odot$ (region bounded by dashed brown curve), we aggressively estimate $\chi_1$ abundance inside the MW to be vanishing and thus the direct-detection constraints are vanishing.

Although we took an aggressive estimation for the abundance fraction in Figure~\ref{fig:paramSH}, we expect that the abundance fraction of $\chi_1$ inside our Galaxy to be rather gradually suppressed towards increasing $M_{\rm J}$.
This is because for baryons (prior to its decoupling from photons), it is known that their abundance fraction (among total matter) along a clump of mass $M$, $r_{b,M}$, is suppressed as $r_{b,M} \sim r_b/[1+(M_{{\rm J},b}/M)^{2/3}]$ at the linear perturbation level~\cite{Weinberg:2008zzc};
$r_b=\Omega_b/\Omega_m \simeq 0.16$ is the cosmological abundance fraction of baryons (among total matter), and $M_{{\rm J},b}\sim6\times10^5\,{\rm M}_\odot$ is the Jeans mass of baryons.
Therefore, in order to reflect the gradual suppression towards increasing $M_{\rm J}$, one may take a more conservative requirement for vanishing Galactic $\chi_1$ abundance fraction.
For example, when we take an order of magnitude smaller Jeans mass for the aggressive requirement, i.e., $M_{\rm J}/10\gtrsim1\times 10^{12}\,{\rm M}_\odot$, the lower boundaries of the region constrained by direct-detection experiments (enclosed by solid curve) extend to smaller $r_1$ by a factor of $\sim 10^{-1}$ ($10^{-2/3}$) when $T_{\rm dec,self}>T_{\rm eq}$ ($T_{\rm dec,self}<T_{\rm eq}$)~[Eq.~\eqref{eq:kJ}].
Nonetheless, it would be interesting to investigate the resultant abundance fraction of $\chi_1$ inside halos in mixed DM scenarios at the non-linear level, since it directly affects the interpretation of direct/indirect-detection constraints.

\subsubsection{Constraints on DM annihilation in the presence of DM self-heating}

The enhancement of $T_{\chi_1}$ from DM self-heating could modify the constraints on DM annihilation.
As shown in Figure~\ref{fig:Tdmevol}, DM self-heating may significantly enhance $T_{\chi_1}$ during cosmological epochs that are sensitive to DM annihilations, i.e., the photo-dissociation epoch and recombination epoch. 
Thus, DM self-heating could significantly enhance the $p$-wave annihilation rate of $\chi_1$ during the epochs and allows us to probe the parameter space that is not constrained in the case of no self-heating.

\begin{figure}[h!]
\centering
\includegraphics[scale=0.62]{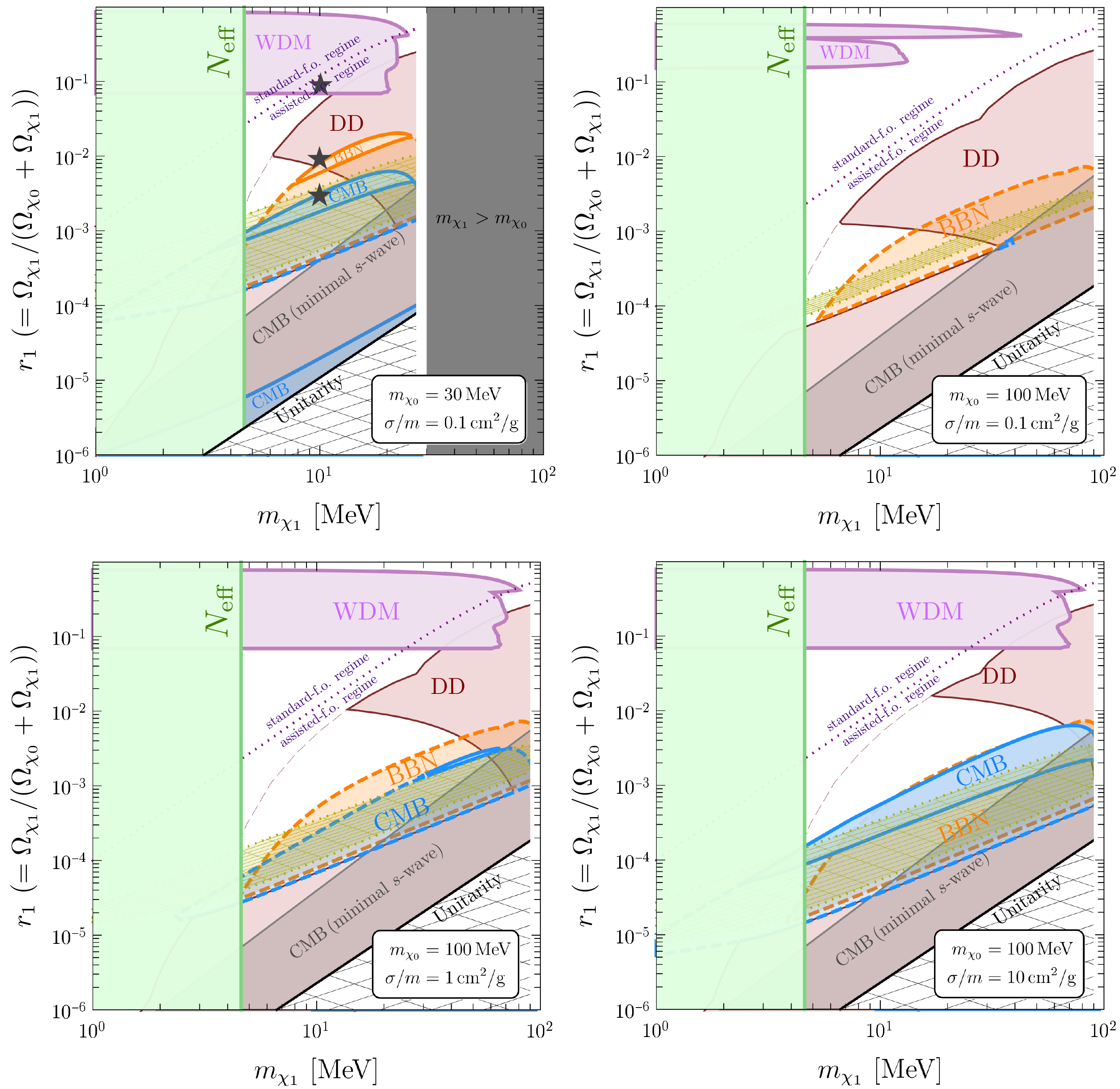}
\caption{Same as Figure~\ref{fig:paramnoSH} but in the presence of self-heating of $\chi_1$.
Due to the self-heating of $\chi_1$, the WDM constraints on $\chi_1$ emerges~[see Figure~\ref{fig:lya}] (pink).
Furthermore, the constraints on $\chi_1$-annihilation from photo-dissociation, and CMB are modified from Figure~\ref{fig:paramnoSH};
we employ Eq.~\eqref{eq:naiveTchi1} and Eq.~\eqref{eq:Tchi1noSH} to estimate $T_{\chi_1}$ during the relevant cosmological epochs.
The constrained region with solid boundaries are the ones where we can reliably estimate $T_{\chi_1}$ in the relevant cosmological epochs through Eq.~\eqref{eq:naiveTchi1} and Eq.~\eqref{eq:Tchi1noSH}.
The constrained region with dashed boundaries are the ones where Eq.~\eqref{eq:naiveTchi1} is not a robust estimation for $T_{\chi_1}$.
On the other hand, for $r_1<0.01$, the minimal $s$-wave contribution can become relevant.
We over plot the CMB bound coming from the unsuppressed $s$-wave contribution in the heavy mediator limit~[see Figure~\ref{fig:paramnoSH}] (gray).
We also plot the direct-detection constraints on the minimal contribution to $\sigma_{\chi_1e}$ in the heavy mediator limit (brown);
the region enclosed by the dashed brown curve is also the region constrained by direct-detection experiments, but $\chi_1$ may not cluster on the Galactic scales.
The evolution of $T_{\chi_1}$ for the parameters depicted by stars are presented in Figure~\ref{fig:Tdmevol}.
}
\label{fig:paramSH}
\end{figure}

DM self-heating could start before the photo-dissociation epoch relevant to DM annihilations.
Although we do not follow the exact evolution of $T_{\chi_1}$, e.g., according to Eq.~\eqref{eq:Tchi1evolre}, we can robustly constrain $(\sigma_1 v_{\rm rel})_p$ in the presence self-heating as follows.
When $T_{\chi_1}$ follows the asymptotic solution, i.e., Eq.~\eqref{eq:Tratioasy}, before the onset of the photo-dissociation epoch ($T\gtrsim 10\,{\rm keV}$), we may straight-forwardly translate the constraints discussed in Section~\ref{section:noSHcosmology};
in such a case, $T_{\chi_1}/T$ remains a constant throughout the photo-dissociation epoch~[see Eq.~\eqref{eq:naiveTchi1}] and we simply rescale the upper bound on $(\sigma_1 v_{\rm rel})_p$ in the case of $T_{\rm kd}=100\,{\rm eV}$~\cite{Depta:2019lbe} with the factor of $(T_{\chi_1}/T)_{\rm asy}^{-1}$.
The solid and dashed curves in Figure~\ref{fig:Tdmevol} corresponds to this case.
We present the constrained parameter space as the orange region enclosed by solid curves in Figure~\ref{fig:paramSH}.
On the other hand, there are cases where DM self-heating starts before $T\gtrsim 10\,{\rm keV}$ but $(T_{\chi_1}/T)$ varies throughout the photo-dissociation epoch.
One example of such cases is the dotted curve in Figure~\ref{fig:Tdmevol};
while the constraints on DM annihilation will still be more stringent compared to the case of no self-heating, dedicated analysis would be needed for a robust constraint.
Instead, we aggressively display the constraints (as orange regions enclosed by the dashed curves) while assuming $T_{\chi_1}$ follows the asymptotic solution during the photo-dissociation epoch to show the potentially constrained parameter space.
In addition, we also require $T_{\rm sh}\gtrsim 100\,{\rm eV}$ so that the self-heating may start at least before the photo-dissociation epoch where $T_{\rm sh}$ is determined by assuming $T_{\chi_1}=T$~[see Appendix~\ref{section:Tdmevolapp} for more discussion].
We note that in the constrained regions, the minimal $s$-wave contribution to the DM annihilation cross section is negligible compared to the $p$-wave part during the photo-dissociation epoch.

DM self-heating enhances $T_{\chi_1}$ around the recombination epoch.
As long as $T_{\rm min}>T_{\rm dec,self}$ is satisfied at the most conservative level~[see the discussion below Eq.~\eqref{eq:Tself}], we may use Eq.~\eqref{eq:Tchi1evolre} to estimate $T_{\chi_1}$ at the last scattering and apply Eq.~\eqref{eq:CMB} to constrain $(\sigma_1 v_{\rm rel})_p$;
in Figure~\ref{fig:paramSH}, we present the constrained region as the blue shaded regions enclosed by solid curves.
If $T_{\rm min}<T_{\rm dec,self}$, there may still be enhancement on $T_{\chi_1}/T$ compared to the case of no self-heating.
However, $T_{\chi_1}/T$ may not reach the asymptotic solution of Eq.~\eqref{eq:Tratioasy} and thus Eq.~\eqref{eq:naiveTchi1} may overestimate $T_{\chi_1}$ at the last scattering.
Nevertheless, we use Eq.~\eqref{eq:naiveTchi1} even for $T_{\rm min}<T_{\rm dec,self}$ and aggressively display the potentially constrained parameter space as blue regions enclosed by the dashed curves in Figure~\ref{fig:paramSH}.
On the other hand, for $r_1\lesssim10^{-3}$, the $\chi_1$-SM interaction is highly enhanced so that the heating rate $\gamma_{\rm heat}$ may never become important for the evolution of $T_{\chi_1}$ throughout the cosmological history, i.e., $T_{\rm sh}<T_{\rm dec,self}$~[see the discussion below Eq.~\eqref{eq:Tself}];
in Figure~\ref{fig:paramSH}, the lower boundary of the region enclosed by the dashed blue curve corresponds to the boundary where $T_{\rm sh}=T_{\rm dec,self}$.
In such a case, we may estimate $T_{\chi_1}$ at the last scattering as in the case of no self-heating~[Eq.~\eqref{eq:Tchi1noSH}];
the constrained parameter region is also shown as blue region with solid boundary.
As a reference, we overplot the possible CMB constraint on DM annihilation from the minimal $s$-wave contribution as well (gray region).

The parameter space (potentiallly) constrained from photo-dissociation and CMB bounds generally extends as we consider larger $\sigma_{\rm self}/m$;
compare the upper-right and lower panels of Figure~\ref{fig:paramSH}.
Larger values of $\sigma_{\rm self}/m$ lead to the extension of the self-heating epoch, i.e., by delaying the decoupling point $T_{\rm dec,self}$, and thus enhance $T_{\chi_1}$ around the last scattering;
this is why the upper (lower) boundary for the CMB bound on DM annihilation extends to larger (smaller) $r_1$ as we consider larger $\sigma_{\rm self}/m$.
The upper boundary (both solid and dashed) for the photo-dissociation constraint does not depend on $\sigma_{\rm self}/m$ since the asymptotic temperature ratio remains unchanged as we vary $\sigma_{\rm self}/m$.
However, the dashed lower boundary extends to smaller $r_1$ because larger $\sigma_{\rm self}/m$ leads to larger $T_{\rm sh}$ ($\sim T_{\rm stop}$).

The chemical freeze-out of $\chi_1$ can interfere with the self-heating epoch. 
In such a case, we should not assume that the DM relic densities are fixed during the self-heating epoch.
Such a case is realized when the freeze-out of $\chi_1$ is very delayed in the assisted regime, so that the freeze-out of $\chi_1$ occurs during the self-heating epoch,
i.e., $T_{\rm fo}^\prime<T_{\rm sh}$~[see Eq.~\eqref{eq:Xfoprime} and Eq.~\eqref{eq:Tsh}].
We display such a parameter space by the yellow hatched region in Figure~\ref{fig:paramSH};
we conservatively over-estimate $T_{\rm sh}$ by taking the lowest possible values for $T_{\chi_1}$, which is estimated by Eq.~\eqref{eq:Tchi1noSH}.
A robust analysis for this region would require one to follow the co-evolution of $\chi_1$ yield and temperature and may be done elsewhere.

\section{Impact on dark photon searches}
\label{section:dphdemon}

The light DM component $\chi_1$ can be directly probed in high-intensity accelerator experiments~\cite{Batell:2009di,Battaglieri:2017aum,Dutta:2019nbn,DUNE:2020fgq}, which provides a complementary approach in identifying the multi-component dark matter scenarios.
In order to show such a complementarity, we fix a reference model where both $\chi_0$ and $\chi_1$ are the SM gauge singlet complex scalars and a dark photon $A'$ mediates the interaction between $\chi_1$ and the SM sector.
The relevant terms in the effective Lagrangian are:
\begin{equation}
{\cal L}\supset\epsilon A_{\mu}^{\prime}J_{{\rm em}}^{\mu}-ig_{D}A_{\mu}^{\prime}\left(\chi_{1}^{\ast}\partial^{\mu}\chi_{1}-\chi_{1}\partial^{\mu}\chi_{1}^{\ast}\right)-\frac{\lambda_{\rm ast.}}{4}\left|\chi_{1}\right|^{2}\left|\chi_{0}\right|^{2}\,,
\label{eq:scalarDMint}
\end{equation}
where $m_{A^\prime}$ is the dark photon mass and $g_D$ is the dark gauge coupling.
The dark photon $A^\prime$ kinetically mixes with the SM photon and induces the coupling with the SM electromagnetic current $J_{{\rm em}}^{\mu}$, which is set by the kinetic mixing parameter $\epsilon$.
As will be discussed in more detail, we found the parameter region of $m_{A'} - \epsilon$ which is expected to be reached by the current and future experiments can be sensitive to our WDM constraints for $0 \lesssim r_1 \lesssim 1$.

There are several tree-level annihilation channels for $\chi_1$ that may determine $\chi_1$ relic abundance, e.g., $\chi_1 \chi_1^\ast \rightarrow f \bar{f}$ and $\chi_1\chi_1^\ast\rightarrow A^\prime A^\prime$.
The former annihilation channel is generated through the kinetic mixing of $A^\prime$;
the $\chi_1$ pair annihilates through an off-shell $A^\prime$ and it is $p$-wave suppressed for a complex scalar $\chi_1$.
This is the dominant number changing process for the case of $m_{A^\prime}/m_{\chi_1}>2$, while $\chi_1\chi_1^\ast\rightarrow A^\prime A^\prime$ is kinematically forbidden.
We focus on such a case throughout this section.
For $1<m_{A^\prime}/m_{\chi_1}<2$, if $\chi_1 \chi_1^\ast \rightarrow f \bar{f}$ determines the relic density of $\chi_1$, while the unsuppressed $s$-wave annihilation $\chi_1 \chi_1^\ast \rightarrow A^\prime f \bar{f}$ around the last scattering strongly disfavor this case.~\footnote{The dark sector processes like the $\chi_1\chi_1\chi_1^\ast\rightarrow \chi_1 A^\prime$ and $\chi_1\chi_1^\ast \rightarrow A^\prime A^\prime$ could also determine the relic density of $\chi_1$~\cite{Fitzpatrick:2020vba}, in which case the stringent CMB bounds may be evaded.}
The relic abundance of $\chi_0$ particles is determined through the $\lambda_{\rm ast.}$ coupling, i.e., the $\chi_0\chi_0^\ast \rightarrow \chi_1 \chi_1^\ast$ process which is $s$-wave.
The non-relativistic annihilation cross sections of $\chi_1$ and $\chi_0$ are given as
\begin{align}
\left(\sigma_{\chi_{0}\chi_{0}^{\ast}\rightarrow\chi_{1}\chi_{1}^{\ast}}v_{{\rm rel}}\right)&\simeq\frac{\lambda_{\rm ast.}^{2}}{32\pi m_{\chi_0}^2}\sqrt{1-R^{-2}_{\chi_0}}\,,\label{eq:scalarchi2ann}\\
\left(\sigma_{\chi_{1}\chi_{1}^{\ast}\rightarrow f\bar{f}}v_{{\rm rel}}\right)&\simeq\frac{4\pi\alpha\alpha_{D}\epsilon^{2}}{3}\frac{\underset{f}{\sum}q_{f}^{2}\left(2+R^2_{f}\right)\sqrt{1-R_{f}^2}}{m_{\chi_{1}}^{2}\left(R_{A^\prime}^2-4\right)^{2}}\times\left(s/m_{\chi_{1}}^{2}-4\right)\,,\label{eq:scalarchi1ann}
\end{align}
where $q_f$ is the EM charge of the SM fermions, $\alpha_D = g_D^2 / 4\pi$, and $R_i=m_i/m_{\chi_1}$.
The factor of $(s/m_{\chi_1}^2-4)$ in Eq.~\eqref{eq:scalarchi1ann} may be replaced with $\langle v_{\rm rel}^2\rangle \simeq 6T_{\chi_1}/m_{\chi_1}$ when we take the thermal average.

\begin{figure}[h!]
\centering
\includegraphics[scale=0.55]{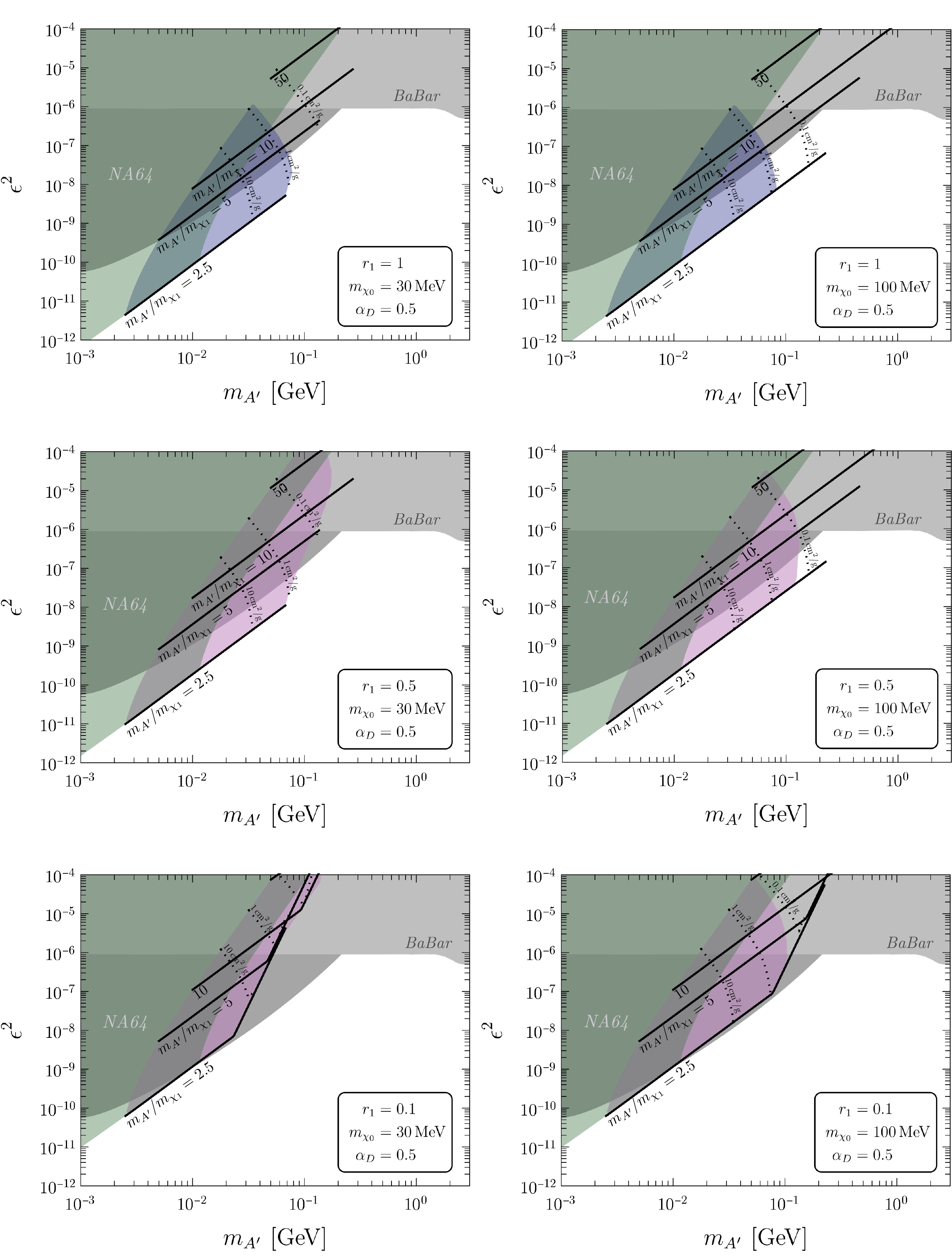}
\caption{Collection of various constraints on the toy model with dark photon portal~[Eq.~\eqref{eq:scalarDMint}] for various $r_1$.
Along the black curves, desired thermal relic abundance of $\chi_1$ is achieved;
in the top and the center panel, the presented curves are in the standard freeze-out regime;
in the bottom panel, the break of the thermal relic curves represent the transition to the assisted freeze-out regime.
The gray regions are constrained by the missing energy/momentum experiments~\cite{Lees:2017lec,NA64:2019imj}.
The green region represents the $N_{\rm eff}$ constraint from the ${\rm MeV}$-scale freeze-out of $\chi_1$~\cite{Sabti:2019mhn}.
({\it Top}): $\chi_1$ is the dominant component of DM and thus the WDM constraint is vanishing. Meanwhile, the large self-scattering among $\chi_1$ provides a constraint~\cite{Randall:2007ph} (blue) complementary to the missing energy/momentum experiments.
({\it Center}): As we consider smaller $r_1$, the self-scattering constraint becomes relaxed, while the effect of DM self-heating become more relevant.
The WDM constraint on $\chi_1$ emerges~\cite{Baur:2017stq,Diamanti:2017xfo} (pink).
({\it Bottom}): Larger annihilation cross section is required as we consider smaller $r_1$ and eventually conflicts with the constraints from {\it NA64} for $r_1\lesssim 0.1$.
}
\label{fig:dphdemon}
\end{figure}

We present the thermal relic curves for given values of the mass ratio $m_{A^\prime}/m_{\chi_1}$ (black) in Figure~\ref{fig:dphdemon} for various values of $r_1$ in the $m_{A^\prime}$ versus $\epsilon^2$ plane.
We also over plot the constraints on $\epsilon$ from the low-energy experiments (gray), which search for missing-energy/momentum events via the production of dark photons.
One of the relevant constraint comes from the {\it NA64} collaboration~\cite{NA64:2019imj}, which is the missing-energy experiment.
It is based on the detection of the missing energy carried away by the soft production of $A^\prime$ by scattering high-energy electrons to the active beam dump target (via bremsstrahlung emission of $A^\prime$ in the process $e^- Z\rightarrow e^- Z A^\prime$);
as we focus on $m_{A^\prime}/m_{\chi_1}>2$, dark photon decays invisibly, i.e., $A^\prime \rightarrow \chi_1 \chi_1^\ast$, with the branching ratio close to unity.
The difference of this type of experiment from the conventional beam-dump experiments~\footnote{There, $A^\prime$ is produced by a high-intensity beam in a dump and generate a flux of DM particles through the $A^\prime\rightarrow \chi\chi$ decay. The produced DM through the decay could be detected through the scattering off electrons in the far target.} (see, e.g., Ref.~\cite{Battaglieri:2016ggd,Battaglieri:2020lds}), is that there is no need for additional DM scattering at a far target.
Therefore, the sensitivity is proportional to the production cross section of $A^\prime$, which scales as $\propto\epsilon^2/m_{A^{\prime}}^2$ for a given mass ratio $m_{A^\prime}/m_{\chi_1}$;
this is why the thermal relic curves in the standard freeze-out regime (top and center panels of Figure~\ref{fig:dphdemon}) are nearly parallel to the lower boundary of the constraint from {\it NA64}.
In the bottom panel, the break of the thermal relic curves represents the transition to the assisted freeze-out regime, since the required annihilation cross section also depends on $m_{\chi_1}$ in the assisted regime~[Eq.~\eqref{eq:assistedxsection}].
We also display the constraint from the {\it BaBar} collaboration~\cite{Lees:2017lec}, which searches for events with a single high-energy photon and a large missing momentum and energy that is consistent with hard production of $A^\prime$ through the process $e^-e^+\rightarrow \gamma A^\prime$ followed by $A^\prime \rightarrow \chi_1 \chi_1^\ast$.
The production cross section of $A^\prime$ is proportional to $\propto \epsilon^2/s$ and thus the sensitivity is virtually independent of $m_{A^\prime}$.

We remark that, as can be seen in the bottom panel of Figure~\ref{fig:dphdemon}, the constraint from {\it NA64} disfavors the abundance ratio smaller than $r_1\lesssim0.1$ unless the annihilation process $\chi_1 \chi_1^\ast \rightarrow f \bar{f}$ is near the resonance to push the required $\epsilon$ to smaller values, i.e., $m_{A^\prime}/m_{\chi_1} \rightarrow 2$;
investigating the robust thermal relic curve near the resonance may require dedicated analyses~\cite{Binder:2021bmg}.
Therefore, we focus on $r_1\gtrsim 0.1$ where the only relevant cosmological constraint from DM self-heating is the WDM constraint on $\chi_1$.

The $\chi_1$ particles exhibit self-scattering via the $A^\prime$-exchange.
The self-scattering cross section of $\chi_1$ is given by
\begin{equation}
\sigma_{{\rm self}}/m_{\chi_{1}}=\frac{6\pi\alpha_{D}^{2}m_{\chi_{1}}}{m_{A^\prime}^{4}}+\frac{3\lambda_{\chi_{1}}^{2}}{16\pi m_{\chi_{1}}^{3}}\,,
\end{equation}
where the second term in the RHS is the possible contribution to $\chi_1$ self-scattering from the $\lambda_{\chi_1}|\chi_1|^4$ coupling.
Hereafter, we set $\lambda_{\chi_1}=0$ for the simplicity of the discussion.~\footnote{For $\lambda_{\chi_1}={\cal O}(1)$, $\sigma_{\rm self}/m_{\chi_1}$ will considerably increase and hence lead to stronger WDM constraints (pink) in Figure~\ref{fig:dphdemon}. Nevertheless, the qualitative discussions do not change.}
After the freeze-out of DM, the residual annihilation of $\chi_0$ produces boosted $\chi_1$ particles and induces DM self-heating in collaboration with the $\chi_1$ self-scattering.
For $r_1$ close to unity (the top panel of Figure~\ref{fig:dphdemon}), the $\chi_0$-annihilation rate is suppressed.
Thus the effect of DM self-heating is not significant~[Eq.~\eqref{eq:Tratioasy}] and the WDM constraints vanish (also see Figure~\ref{fig:lya}).
Meanwhile, the self-scattering among $\chi_1$ particles can be as large as $\sigma_{\rm self}/m\sim1\,{\rm cm^2/g}$;
see the contours for $\sigma_{\rm self}/m$ in Figure~\ref{fig:dphdemon} (dotted).
Since $\chi_1$ is the dominant component of DM, the large self-scattering among $\chi_1$ may conflict with the observations on galaxy clusters~\cite{Randall:2007ph,Harvey:2018uwf,Sagunski:2020spe}.
As a reference, we display the constraint on $\sigma_{\rm self}/m$ from the Bullet cluster based on mass loss (blue)~\cite{Randall:2007ph} which covers the region not yet constrained by {\it NA64}.

As we consider smaller $r_1$, only a sub-dominant component of DM exhibits self-scattering and the constraint from the Bullet cluster may get relaxed.
At the same time, the effect of DM self-heating becomes more relevant.
In the center and bottom panel of Figure~\ref{fig:dphdemon}, the WDM constraints of $\chi_1$ emerge (pink), redeeming the relaxed constraint on the self-scattering cross section.
The WDM constraints also depend on $m_{\chi_0}$ since the annihilation rate decreases as we consider larger $m_{\chi_0}$;
the WDM constraints virtually vanish for $m_{\chi_0}\gtrsim 300\,{\rm MeV}$.
As we further decrease $r_1\lesssim 0.1$, the WDM constraints are virtually vanishing while the constraint from {\it NA64} disfavors smaller values of $r_1$.

\clearpage

\clearpage

\section{Conclusions}
\label{section:conclusion}

We have studied the cosmology of the two-component DM scenario, which serves as an illustrating example of a non-minimal dark sector.
In this scenario, two stable components, i.e., $\chi_0$ and $\chi_1$, consist DM and only the lighter state $\chi_1$ interacts with SM.
If $\chi_1$ interacts sufficiently strong with SM, the dark sector may be in thermal equilibrium in the early Universe, and the DM relic density would be determined through the thermal freeze-out of DM, i.e., through the $\chi_0\chi_0\rightarrow\chi_1\chi_1$ and the $\chi_1\chi_1\rightarrow {\rm sm}\,{\rm sm}$ processes.
We have carefully studied the dynamics of the two-component DM scenario, especially focusing on the detectability of the sub-dominant component of DM, $\chi_1$, in cosmological/astrophysical observations.

We have shown that as we consider a smaller $\chi_1$ abundance fraction, i.e., $r_1\lesssim 0.1$, the freeze-out of $\chi_1$ transits to the assisted-regime where the required annihilation cross section of $\chi_1$ is sharply enhanced towards smaller values of $r_1$.
Contrary to the usual case of the standard thermal freeze-out of DM where the annihilation cross section scales as $\propto1/r_1$, the annihilation cross section of $\chi_1$ in the assisted-regime scales as $(\sigma_1 v_{\rm rel})_s\propto1/r_1^2$ in the case of $s$-wave annihilation, and as $(\sigma_1 v_{\rm rel})_p\propto1/r_1^3$ in the case of $p$-wave annihilation.
The sharp scaling of the annihilation cross section implies better detectability of $\chi_1$ for smaller abundance fractions, e.g., in direct/indirect-detection experiments.
Having in mind the sharp scaling, we have reviewed the cosmological/astrophysical constraints on $\chi_1$-annihilation.
It is worthwhile to note that considering smaller values of $r_1$ in the two-component DM scenario is sometimes considered to be a minimal remedy to evade the stringent constraints on sub-${\rm GeV}$ DM annihilations;
however, considering smaller values of $r_1$ does not relax the constraints.

We have demonstrated that self-scattering among $\chi_1$ could considerably affect the detectability of $\chi_1$.
The collaboration of the residual $\chi_0$-annihilation and the self-scattering leads to DM self-heating, which may enhance the temperature of $\chi_1$ compared to the SM one.
The self-scattering cross section as large as $\sigma_{\rm self}/m\sim0.1\,{\rm cm^2/g}$ can be naturally realized for a sub-{\rm GeV} mass scale, and we have shown that WDM constraints from the Lyman-$\alpha$ forest data and the number of satellite galaxies in the MW are significant for $m_{\chi_0}\lesssim200\,{\rm MeV}$ and $r_1\gtrsim 0.1$.
For abundance fraction smaller than $r_1\lesssim 0.1$, although the warmness (or Jeans mass) of $\chi_1$ increases towards smaller $r_1$, such a sub-dominant fraction of $\chi_1$ have a negligible effect on the overall matter power spectrum, and hence the warmness is not constrained by the cosmological observations on the structure formation of our Universe.
Nevertheless, the warmness of sub-dominant fraction of $\chi_1$ has interesting implications on the interpretation of direct-detection experiments.
For $r_1\ll 0.1$, the elastic scattering rate of an SM particle with $\chi_1$ increases towards smaller $r_1$ and thus direct-detection constraints on $\chi_1$ is expected to be severer at the first sight.
However, we have shown that the resultant warmness of $\chi_1$ could suppress $\chi_1$'s gravitational clustering in our Galaxy and thus relax the direct-detection constraints.
How much the direct-detection constraints are relaxed depends on the suppression of the Galactic abundance fraction of $\chi_1$ compared to the cosmological one.
We have aggressively estimated the Galactic $\chi_1$ abundance fraction to vanish when the Jeans mass of $\chi_1$ exceeds the mass of the MW.
However, we expect the suppression of the Galactic abundance with respect to an increasing Jeans mass to be more gradual.
It would be interesting to investigate the gravitational clustering of $\chi_1$ at the non-linear level and put more robust direct-detection constraints.

Moreover, DM self-heating may enhance $\chi_1$ temperature during cosmological epochs sensitive to DM annihilations, e.g., during the photo-dissociation epoch of light nuclei ($100\,{\rm eV} \lesssim T\lesssim 10\,{\rm keV}$) and at the last scattering.
In the case of $p$-wave annihilation of $\chi_1$, the enhanced $\chi_1$ temperature from DM self-heating increases the annihilation rate, and we have demonstrated that the cosmological constraints on $\chi_1$-annihilation could become relevant for $r_1\ll 0.1$.
We have aggressively displayed the parameter regions that can be potentially constrained by the BBN and CMB observations.
Interestingly, such parameter regions redeem the relaxed direct-detection constraints for $r_1\ll 0.1$.
Therefore, it would be interesting to do more robust analyses of the cosmological constraints of DM annihilations.

Despite the interesting cosmology for $r_1\ll 0.1$, we remark that for such small values of $r_1$, the $\chi_1$-SM interaction is highly enhanced and it is usually incompatible with accelerator-based experiments.
We have demonstrated this by taking a case where $\chi_1$ interacts with SM through the dark photon portal.
We have focused on the case of $m_{A^\prime}/m_{\chi_1}>2$, where the missing-energy/momentum experiments provide relevant constraints on the kinetic mixing parameter.
In particular, the missing-energy experiment at {\it NA64} disfavors $r_1\lesssim 0.1$, unless the $\chi_1$ annihilation via an off-shell dark photon is close to resonance.
On the other hand, we have found that WDM constraints on $\chi_1$ provide complementary constraints on the kinetic mixing parameter for $r_1\gtrsim 0.1$.
We emphasize that the WDM constraints redeem the relaxed constraints on the self-scattering cross section of $\chi_1$ for $r_1<1$.
This motivates the further study of structure-formation constraints on the mixed DM scenarios.
Actually, some observations, e.g., the flux anomaly of quadrupole lens systems~\cite{Inoue:2014jka,Kamada:2016vsc,Kamada:2017icv,Birrer:2017rpp,Gilman:2017voy,Vegetti:2018dly,Rivero:2018bcd,Gilman:2019nap}, and the redshifted $21\,{\rm cm}$ signal~\cite{Sitwell:2013fpa,Sekiguchi:2014wfa,Safarzadeh:2018hhg,Schneider:2018xba,Lidz:2018fqo,Lopez-Honorez:2018ipk,Nebrin:2018vqt,Chatterjee:2019jts}, provides severer constraints in the case of pure WDM compared to the constraints we took in this paper.
If one reanalyzes data from such probes in the case of mixed DM, we may get stronger constraints on the warmness of $\chi_1$ and hence increase the synergy between the warmness constraints from structure formation and terrestrial experiments.

\subsection*{Acknowledgments}
The authors would like to thank Doojin Kim for fruitful discussions and comments.
The work of A.K. and H.K. is supported by IBS under the project code, IBS-R018-D1.
A.K. also acknowledges partial support from Grant-in-Aid for Scientific Research from the Ministry of Education, Culture, Sports, Science, and Technology (MEXT), Japan, 18K13535 and 19H04609; from World Premier International Research Center Initiative (WPI), MEXT, Japan; from Norwegian Financial Mechanism for years 2014-2021, grant nr 2019/34/H/ST2/00707; and from National Science Centre, Poland, grant DEC-2018/31/B/ST2/02283.
J.C.P. acknowledges support from the National Research Foundation of Korea (NRF-2019R1C1C1005073 and NRF-2021R1A4A2001897). 
S.S. acknowledges support from the National Research Foundation of Korea (NRF-2020R1I1A3072747).

\appendix

\section{Boltzmann equations for the boosted DM (BDM)}
\label{appendix:Boltzmanndetail}

In this appendix, we derive the evolution equations for boosted DM (BDM).
One can find the number density evolution equations in Eq.~\eqref{eq:chi2numden} and Eq.~\eqref{eq:chi1numden},
and the temperature evolution equation in Eq.~\eqref{eq:BoltzmannT}.

\subsection{Thermal averaged quantities \label{section:thermalidentities}}

We lay-out the identities that we will utilize in this appendix.
The thermal averaged quantities for particle with mass $m$ with the Boltzmann distribution $f^{\rm eq}=\exp\left[-E/T\right]$ are given as
\begin{equation}
\begin{aligned}
\int_{{\bf p}}f^{{\rm eq}}&=\frac{m^{3}}{2\pi^{2}}\left(\frac{T}{m}\right)K_{2}\left(m/T\right)\equiv n^{{\rm eq}}\,,\\
\int_{{\bf p}}E\,f^{{\rm eq}}&=\frac{m^{4}}{2\pi^{2}}\left(\frac{T}{m}\right)\left[K_{1}\left(m/T\right)+3\left(\frac{T}{m}\right)K_{2}\left(m/T\right)\right]\equiv\rho^{{\rm eq}}\,,\\
\int_{{\bf p}}\frac{\left|{\bf p}\right|^{2}}{2m}\,f^{{\rm eq}}&=\frac{3}{2}\frac{m^{4}}{2\pi^{2}}\left(\frac{T}{m}\right)^{2}K_{3}\left(m/T\right)\equiv K^{{\rm eq}}\,,\\
\int_{{\bf p}}\frac{\left|{\bf p}\right|^{2}}{3E}\,f^{{\rm eq}}&=\frac{m^{4}}{2\pi^{2}}\left(\frac{T}{m}\right)^{2}K_{2}\left(m/T\right)\equiv P^{{\rm eq}}\,,
\end{aligned}
\end{equation}
where $\int_{\bf p} \equiv \int d^3p / (2\pi)^3$  and $K_n$ are the modified Bessel function of the second kind.
From the top, each quantities represent number density ($n^{\rm eq}$), energy density ($\rho^{\rm eq}$), kinetic-energy density in the non-relativistic limit ($K^{\rm eq}$), and pressure ($P^{\rm eq}$).
We will often perform derivatives of the thermal averaged quantities:
\begin{equation}
K_{n}^{\prime}\left(x\right)=-\frac{1}{2}\left[K_{n-1}\left(x\right)+K_{n+1}\left(x\right)\right]\,.
\end{equation}
It is also useful to note the following recurrence relation for integer $n$:
\begin{equation}
K_{n-1}\left(x\right)-K_{n+1}\left(x\right)=-\frac{2n}{x}K_{n}\left(x\right)\,.
\end{equation}
We are interested in an epoch where dark matter is non-relativistic.
Therefore, it is useful to note the asymptotic behavior of $K_n(x)$:
\begin{equation}
\underset{x\rightarrow\infty}{\lim}K_{n}\left(x\right)=e^{-x}\left[\sqrt{\frac{\pi}{2x}}+{\cal O}\left(x^{-3/2}\right)\right]\,.
\end{equation}

\subsection{Interaction of DM and the corresponding collision terms}

We consider a case where two DM particles, $\chi_0$ and $\chi_1$, which were initially in thermal equilibrium with a thermal plasma, e.g., SM plasma.
The heavy state $\chi_0$ has no direct couplings to SM, and annihilates into the light state $\chi_1$.
On the other hand, $\chi_1$ interacts with SM, e.g., through a dark photon portal, and its relic abundance is determined by the annihilation into SM particles.
The yield of the DM particles are determined by the chemical freeze-out of the following processes:
\begin{equation}
\begin{aligned}
\chi_{0}\left(1\right)+\chi_{0}\left(2\right)&\leftrightarrow\chi_{1}\left(3\right)+\chi_{1}\left(4\right)\,,
\\
\chi_{1}\left(1\right)+\chi_{1}\left(2\right)&\leftrightarrow\phi\left(3\right)+\phi\left(4\right)\,,
\label{eq:chiann}
\end{aligned}
\end{equation}
where $\phi$ is some SM state, and the number indices will be used when defining the collisional integrals, as will be shown shortly.
For simplicity, we assume $\chi_0$, $\chi_1$, and $\phi$ to be real scalars.
Processes above accompany elastic scatterings from the crossing symmetry:
\begin{equation}
\begin{aligned}
\chi_{0}\left(1\right)+\chi_{1}\left(2\right)&\leftrightarrow\chi_{0}\left(3\right)+\chi_{1}\left(4\right)\,,
\\
\chi_{1}\left(1\right)+\phi\left(2\right)&\leftrightarrow\chi_1\left(3\right)+\phi\left(4\right)\,.
\label{eq:chiel}
\end{aligned}
\end{equation}
Efficient processes of Eq.~\eqref{eq:chiel} keep DM in kinetic equilibrium with the SM plasma ($T_{\chi_1}=T$) during (and after) their chemical freeze-out.

We are interested in a case where $\chi_1$ exhibits sizable self-scattering $\sigma_{\rm self}/m\sim1\,{\rm cm^2/g}$:
\begin{equation}
\chi_{1}\left(1\right)+\chi_{1}\left(2\right)\rightarrow\chi_{1}\left(3\right)+\chi_{1}\left(4\right)\,.
\end{equation}
Efficient self-scattering keeps the distribution function of $\chi_1$ in the equilibrium form, i.e., $f_{\chi_1}=\exp\left[-(E-\mu)/T_{\chi_1}\right]$.
In the presence of the efficient self-scattering, the excess kinetic energy of the boosted $\chi_1$'s produced from the first process of Eq.~\eqref{eq:chiann} would be efficiently re-distributed to the other $\chi_1$ particles;
this results in the self-heating epoch after the kinetic decoupling of $\chi_1$.

Now that we have introduced the interactions of DM, we present their corresponding Boltzmann equations.
The Boltzmann equation for $\chi_0$ is given as
\begin{equation}
\left[\frac{\partial}{\partial t}+Hp_{1}\frac{\partial}{\partial p_{1}}\right]f_{\chi_{0}}=\frac{1}{2E_{1}}\left(C_{\rm ann, \chi_0}+C_{\rm el,\chi_0}\right)\,,
\label{eq:chi2Boltzmann}
\end{equation}
where $C_{\rm ann, \chi_0}$ and $C_{\rm el,\chi_0}$ are the collisional integral for annihilation~[the first process of Eq.~\eqref{eq:chiann}] and elastic scattering~[the first process of Eq.~\eqref{eq:chiel}] of $\chi_0$, respectively.
$C_{\rm ann, \chi_0}$ is given as
\begin{equation}
\begin{aligned}
C_{\rm ann, \chi_0}\left[f_{\chi_{0}}\left(p_{1}\right)\right]&=2\int d\Pi_{2}d\Pi_{3}d\Pi_{4}\left(2\pi\right)^{4}\delta^{4}\left(p_{1}+p_{2}-p_{3}-p_{4}\right)\\
&\times\left|{\cal M}_{\chi_{0}\chi_{0}\rightarrow\chi_{1}\chi_{1}}\right|^{2}\left[f_{\chi_{1}}\left(p_{3}\right)f_{\chi_{1}}\left(p_{4}\right)-f_{\chi_{0}}\left(p_{1}\right)f_{\chi_{0}}\left(p_{2}\right)\right]\,.
\end{aligned}
\label{eq:chi2annC}
\end{equation}
Note that if two identical particles participate in initial/final state, additional factor of $1/2$'s are present compared to the case of non-identical particles in the initial/final state;
this is to correct the over-counting of the equivalent phase-space configurations.
In Eq.~\eqref{eq:chi2annC}, we implicitly assume that the phase space integrals $d\Pi_i$'s are done over inequivalent configurations so that the integrations implicitly take into account the $1/2$ factor(s);
hereafter, we assume this convention.
$C_{\rm el,\chi_0}$ is the collision term for kinetic interactions that keeps $T_{\chi_0}=T_{\chi_1}$, e.g., the first process of Eq.~\eqref{eq:chiel}.
For our purpose, instead of explicitly writing down $C_{\rm el, \chi_0}$, we will assume that kinetic interactions are efficient so that $T_{\chi_0}=T_{\chi_1}$ during the freeze-out of $\chi_0$, and decouples afterwards so that $T_{\chi_0}\propto1/a^2$.

Similarly, the Boltzmann equation for $\chi_1$ is given as
\begin{equation}
\left[\frac{\partial}{\partial t}+Hp_{1}\frac{\partial}{\partial p_{1}}\right]f_{\chi_{1}}=\frac{1}{2E_{1}}\left(C_{\rm inv, \chi_0}+C_{\rm ann, \chi_1}+C_{\rm el, \chi_1}+C_{{\rm self}}\right)\,,
\label{eq:chi1Boltzmann}
\end{equation}
where the collisional integrals are given as
\begin{equation}
\begin{aligned}
C_{\rm inv, \chi_0}\left[f_{\chi_{1}}\left(p_{1}\right)\right]=&2\int d\Pi_{2}d\Pi_{3}d\Pi_{4}\left(2\pi\right)^{4}\delta^{4}\left(p_{1}+p_{2}-p_{3}-p_{4}\right)\\
&\times\left|{\cal M}_{\chi_{1}\chi_{1}\rightarrow\chi_{0}\chi_{0}}\right|^{2}\left[f_{\chi_{0}}\left(p_{3}\right)f_{\chi_{0}}\left(p_{4}\right)-f_{\chi_{1}}\left(p_{1}\right)f_{\chi_{1}}\left(p_{2}\right)\right]\,,
\end{aligned}
\end{equation}
\begin{equation}
\begin{aligned}
C_{\rm ann, \chi_1}\left[f_{\chi_{1}}\left(p_{1}\right)\right]=&2\int d\Pi_{2}d\Pi_{3}d\Pi_{4}\left(2\pi\right)^{4}\delta^{4}\left(p_{1}+p_{2}-p_{3}-p_{4}\right)\\
&\times\left|{\cal M}_{\chi_{1}\chi_{1}\rightarrow\phi\phi}\right|^{2}\left[f_{\phi}\left(p_{3}\right)f_{\phi}\left(p_{4}\right)-f_{\chi_{1}}\left(p_{1}\right)f_{\chi_{1}}\left(p_{2}\right)\right]\,,
\end{aligned}
\end{equation}
\begin{equation}
\begin{aligned}
C_{\rm el, \chi_1}\left[f_{\chi_{1}}\left(p_{1}\right)\right]=&\int d\Pi_{2}d\Pi_{3}d\Pi_{4}\left(2\pi\right)^{4}\delta^{4}\left(p_{1}+p_{2}-p_{3}-p_{4}\right)\\
&\times\left|{\cal M}_{\chi_{1}\phi\rightarrow\chi_{1}\phi}\right|^{2}\left[f_{\chi_{1}}\left(p_{3}\right)f_{\phi}\left(p_{4}\right)-f_{\chi_{1}}\left(p_{1}\right)f_{\phi}\left(p_{2}\right)\right]\,,
\end{aligned}
\end{equation}
\begin{equation}
\begin{aligned}
C_{{\rm self}}\left[f_{\chi_{1}}\left(p_{1}\right)\right]&=2\int d\Pi_{2}d\Pi_{3}d\Pi_{4}\left(2\pi\right)^{4}\delta^{4}\left(p_{1}+p_{2}-p_{3}-p_{4}\right)\\
&\times\left|{\cal M}_{{\rm self}}\right|^{2}\left[f_{\chi_{1}}\left(p_{3}\right)f_{\chi_{1}}\left(p_{4}\right)-f_{\chi_{1}}\left(p_{1}\right)f_{\chi_{1}}\left(p_{2}\right)\right]\,.
\end{aligned}
\end{equation}

Integrating Eq.~\eqref{eq:chi2Boltzmann} (Eq.~\eqref{eq:chi1Boltzmann}) with respect to $d^3 p_1/(2\pi)^3$ would give the evolution equation for number density of $\chi_0$ ($\chi_1$), and integrating with $E_1$-weighting will give the evolution equation for temperatures of $\chi_0$ ($\chi_1$).
We will present the derivations in the following subsections.

\subsection{Equations for number density \label{section:numberden}}

We now derive the number density equations for $\chi_0$ and $\chi_1$.
Specifically, we are interested in the chemical freeze-out of the DM particles.
For $\chi_0$, we assume that $\chi_0$ follows the temperature of $\chi_1$, i.e., we assume $f_{\chi_0}=\exp\left[-(E-\mu)/T_{\chi_1}\right]$, during its chemical freeze-out.
this could be due to the efficient kinetic interactions of $\chi_0$ with $\chi_1$ represented by $C_{\rm el, \chi_0}$ in Eq.~\eqref{eq:chi2Boltzmann}.
While we do not specify the interactions for simplicity, we instead assume that the kinetic interactions are efficient during the chemical freeze-out of $\chi_0$.
Integrating Eq.~\eqref{eq:chi2Boltzmann} over $d^3p_1 / (2\pi)^3$, we find the evolution equation for number density of $\chi_0$:
\begin{equation}
\begin{aligned}
\dot{n}_{\chi_{0}}+3Hn_{\chi_{0}}&=\int_{\bf{p}_{1}}\frac{1}{2E_{1}}C_{\rm ann, \chi_0}\left[f_{\chi_{0}}\left(p_{1}\right)\right]\\
&=2\int\underset{i=1}{\overset{4}{\prod}}\,d\Pi_{i}\left(2\pi\right)^{4}\delta^{4}\left(p_{1}+p_{2}-p_{3}-p_{4}\right)\\
&\;\;\;\;\times\left|{\cal M}_{\chi_{0}\chi_{0}\rightarrow\chi_{1}\chi_{1}}\right|^{2}\left[f_{\chi_{1}}\left(p_{3}\right)f_{\chi_{1}}\left(p_{4}\right)-f_{\chi_{0}}\left(p_{1}\right)f_{\chi_{0}}\left(p_{2}\right)\right]\\
&=-\left\langle \sigma_0 v_{{\rm rel}}\right\rangle _{T_{\chi_{0}}}\left[n_{\chi_{0}}^{2}-\frac{\left\langle \sigma_0 v_{{\rm rel}}\right\rangle _{T_{\chi_{1}}}}{\left\langle \sigma_0 v_{{\rm rel}}\right\rangle _{T_{\chi_{0}}}}\left(\frac{n_{\chi_{0}}^{{\rm eq}}\left(T_{\chi_{1}}\right)}{n_{\chi_{1}}^{{\rm eq}}\left(T_{\chi_{1}}\right)}\right)^{2}n_{\chi_{1}}^{2}\right]\\
&=-\left\langle \sigma_0 v_{{\rm rel}}\right\rangle _{T_{\chi_{1}}}\left[n_{\chi_{0}}^{2}-\left(\frac{n_{\chi_{0}}^{{\rm eq}}\left(T_{\chi_{1}}\right)}{n_{\chi_{1}}^{{\rm eq}}\left(T_{\chi_{1}}\right)}\right)^{2}n_{\chi_{1}}^{2}\right]\,,
\end{aligned}
\label{eq:chi2numden}
\end{equation}
where $(\sigma_2 v_{\rm rel})$ annihilation cross section of $\chi_0$, and we have set $T_{\chi_0}=T_{\chi_1}$ in the last equality.
In the first equality, integration over $C_{\rm el, \chi_0}$ vanishes under the assumption that kinetic interactions conserve number of $\chi_0$.
$\left\langle\sigma v_{\rm rel}\right\rangle$ denotes the thermal average of the cross section;
we multiply the distribution function of initial particle states to the cross section, and integrate over {\it all possible} phase space configuration (this is in contrast to our convention of integration over $d\Pi_i$'s in $C$'s, where we only integrate over inequivalent phase space configuration):
\begin{equation}
\left\langle \sigma v_{{\rm rel}}\right\rangle _{T}=\frac{1}{n_{1}^{{\rm eq}}\left(T\right)n_{2}^{{\rm eq}}\left(T\right)}\int d^{3}p_{1}d^{3}p_{2}\left(\sigma v_{{\rm rel}}\right)f_{1}^{{\rm eq}}\left(p_{1};T\right)f_{2}^{{\rm eq}}\left(p_{2};T\right)\,.
\end{equation}

For $\chi_1$, efficient self-scattering of $\chi_1$ keeps its distribution proportional to Boltzmann distribution, i.e., $f_{\chi_1}=\exp\left[-(E-\mu)/T_{\chi_1}\right]$.
Efficient elastic scattering with SM states, e.g.,  $\chi_1 \phi\rightarrow\chi_1\phi$, keep $\chi_1$ in kinetic equilibrium ($T_{\chi_1}=T$) with the SM plasma during its chemical freeze-out.
From Eq.~\eqref{eq:chi1Boltzmann}, we find the evolution equation for number density of $\chi_1$:
\begin{equation}
\begin{aligned}
\dot{n}_{\chi_{1}}+3Hn_{\chi_{1}}&=\int_{\boldsymbol{p}_{1}}\frac{1}{2E_{1}}\left(C_{\rm inv, \chi_0}\left[f_{\chi_{1}}\left(p_{1}\right)\right]+C_{\rm ann, \chi_1}\left[f_{\chi_{1}}\left(p_{1}\right)\right]\right)\\
&=\left\langle \sigma_0 v_{\rm rel}\right\rangle _{T_{\chi_{1}}}\left[n_{\chi_{0}}^{2}-\left(\frac{n_{\chi_{0}}^{{\rm eq}}\left(T_{\chi_{1}}\right)}{n_{\chi_{1}}^{{\rm eq}}\left(T_{\chi_{1}}\right)}\right)^{2}n_{\chi_{1}}^{2}\right]\\
&\qquad-\left\langle \sigma_1 v_{\rm rel}\right\rangle _{T_{\chi_{1}}}\left[n_{\chi_{1}}^{2}-\frac{\left\langle \sigma_1 v_{{\rm rel}}\right\rangle _{T}}{\left\langle \sigma_1 v_{\rm rel}\right\rangle _{T_{\chi_{1}}}}n_{\chi_{1}}^{{\rm eq}2}\left(T\right)\right]\\
&=\left\langle \sigma_0 v_{\rm rel}\right\rangle _{T}\left[n_{\chi_{0}}^{2}-\left(\frac{n_{\chi_{0}}^{{\rm eq}}\left(T\right)}{n_{\chi_{1}}^{{\rm eq}}\left(T\right)}\right)^{2}n_{\chi_{1}}^{2}\right]-\left\langle \sigma_1 v_{\rm rel}\right\rangle _{T}\left[n_{\chi_{1}}^{2}-n_{\chi_{1}}^{{\rm eq}2}\left(T\right)\right]\,,
\end{aligned}
\label{eq:chi1numden}
\end{equation}
where we have set $T_{\chi_1}=T$ in the last equality.
In the first equality, $C_{\rm el, \chi_1}$ and $C_{\rm self}$ do not contribute to the number density equation since they conserve number of $\chi_1$.

\subsection{Equations for temperature}

In Appendix~\ref{section:numberden}, we derived the number density equations under the assumption that $\chi_0$ and $\chi_1$ share the same temperature with the SM plasma ($T_{\chi_0}=T_{\chi_1}=T$);
this may be realized around the chemical freeze-outs of $\chi_0$ and $\chi_1$, due to their efficient kinetic interactions; $\chi_0\chi_1\rightarrow \chi_0\chi_1$, and $\chi_1\phi\rightarrow\chi_1\phi$.
However, the kinetic interactions will eventually decouple as the Universe cools down, and we would need to follow the temperature evolution of $\chi_0$ ($T_{\chi_0}$) and $\chi_1$ ($T_{\chi_1}$), independent from the SM plasma temperature ($T$).

In the Boltzmann equation for $\chi_0$~[Eq.~\eqref{eq:chi2Boltzmann}], the kinetic interactions of $\chi_0$ is represented by $C_{\rm el, \chi_0}$.
While we do not specify the interactions for simplicity, one inevitable contribution is the $\chi_0 \chi_1\rightarrow\chi_0 \chi_1$ process.
This process may be efficient around the freeze-out of $\chi_0$ and keep $T_{\chi_0}=T_{\chi_1}$, but likely to decouple at the similar time of the freeze-out of $\chi_0$.~\footnote{Kinetic equilibrium during $\chi_0$'s freeze-out may be realized for large mass difference between $\chi_0$ and $\chi_1$. Meanwhile, for almost degenerate masses, kinetic equilibrium of $\chi_0$ may not be achieved, since $\chi_0\chi_1\rightarrow\chi_0\chi_1$ may not be efficient.}
Again, for simplicity of our analysis, we assume that kinetic equilibrium is achieved between $\chi_0$ and $\chi_1$ during the freeze-out of $\chi_0$, and decouples afterwards so that $T_{\chi_0}\propto1/a^2$.

For $\chi_1$, we have a kinetic interaction between $\chi_1$ and the SM plasma, $\chi_1\phi\rightarrow\chi_1\phi$.
Due to un-suppressed number density of the light SM particle $\phi$, the kinetic equilibrium is likely to be maintained until long after the freeze-out of $\chi_1$.
Until the kinetic decoupling, the temperature redshifts as $T_{\chi_1}\propto1/a$.
Unlike $\chi_0$, the temperature of $\chi_1$ would not redshift like non-relativistic free-streaming particles ($\propto 1/a^2$) because boosted $\chi_1$ are constantly produced from $\chi_0$ annihilation into $\chi_1$.
The excess kinetic energy of the boosted $\chi_1$'s will be redistributed to the other $\chi_1$'s through efficient self-scattering, heating the $\chi_1$ particles as a whole.
To investigate the evolution of $\chi_1$ around the kinetic decoupling, we derive the evolution equation for $T_{\chi_1}$.
Starting from Eq.~\eqref{eq:chi1Boltzmann}, we integrate it with $E_1$-weighting.
Let us perform the integration for the LHS of Eq.~\eqref{eq:chi1Boltzmann}:
\begin{equation}
\begin{aligned}
\int_{\bf{p}_{1}}E_{1}\left[\frac{\partial}{\partial t}+Hp_{1}\frac{\partial}{\partial p_{1}}\right]f_{\chi_{1}}&=\dot{\rho}_{\chi_{1}}+3H\left(\rho_{\chi_{1}}+P_{\chi_{1}}\right)\\
&=\left(\dot{n}_{\chi_{1}}+3Hn_{\chi_{1}}\right)\left\langle E_{\chi_{1}}\right\rangle _{T_{\chi_{1}}}+n_{\chi_{1}}\left(\frac{\dot{T}_{\chi_{1}}}{T_{\chi_{1}}^{2}}\sigma_{E,T_{\chi_{1}}}^{2}+3HT_{\chi_{1}}\right)\,,
\end{aligned}
\end{equation}
where in the second equality, we assumed that the distribution function of $\chi_1$ is $f_{\chi_1}=[n_{\chi_1}/n^{\rm eq}_{\chi_1}(T_{\chi_1})] f^{\rm eq}(T_{\chi_1})$;
this is a reasonable assumption if the self-scattering of $\chi_1$ is efficient.
Using the identities presented in Appendix~\ref{section:thermalidentities}, it is straightforward to achieve the second equality where $\sigma_{E,T_{\chi_1}}^2=\left\langle E_{\chi_1}^2\right\rangle _{T_{\chi_1}}-\left\langle E_{\chi_1}\right\rangle _{T_{\chi_1}}^2$, and
\begin{align}
\left\langle E_{\chi_{1}}\right\rangle _{T_{\chi_{1}}}/m_{\chi_{1}}&=\frac{\rho_{\chi_{1}}^{{\rm eq}}\left(T_{\chi_{1}}\right)}{m_{\chi_{1}}n_{\chi_{1}}^{{\rm eq}}\left(T_{\chi_{1}}\right)}=\frac{3}{x_{\chi_{1}}}+\frac{K_{1}\left(x_{\chi_{1}}\right)}{K_{2}\left(x_{\chi_{1}}\right)}\rightarrow1+\frac{3}{2x_{\chi_{1}}}\,,\label{eq:Eidentity}\\
\left\langle E_{\chi_{1}}^{2}\right\rangle _{T_{\chi_{1}}}/m_{\chi_{1}}^{2}&=1+\frac{2K_{\chi_{1}}^{{\rm eq}}}{m_{\chi_{1}}n_{\chi_{1}}^{{\rm eq}}}=1+\frac{3}{x_{\chi_{1}}}\frac{K_{3}\left(x_{\chi_{1}}\right)}{K_{2}\left(x_{\chi_{1}}\right)}\rightarrow1+\frac{3}{x_{\chi_{1}}}\,,\label{eq:Esqidentity}\\
\sigma_{E,T_{\chi_{1}}}^{2}&\rightarrow\frac{3}{2}T_{\chi_{1}}^{2}\,,
\label{eq:sigmaidentity}
\end{align}
where $x_{\chi_1}=m_{\chi_1}/T_{\chi_1}$, and the RHS of the arrows denote the non-relativistic limits.
Putting together with the RHS of the $E_1$-weighted integral of Eq.~\eqref{eq:chi1Boltzmann}, the overall Boltzmann equation for $T_{\chi_1}$ is given as
\begin{equation}
\begin{aligned}
\frac{\dot{T}_{\chi_{1}}}{T_{\chi_{1}}^{2}}+\frac{3HT_{\chi_{1}}}{\sigma_{E,T_{\chi_{1}}}^{2}}=&\frac{1}{n_{\chi_{1}}\sigma_{E,T_{\chi_{1}}}^{2}} \int_{\bf{p}_{1}}\frac{1}{2E_1}\\
&\times\left[\left(E_{1}-\left\langle E_{\chi_{1}}\right\rangle _{T_{\chi_{1}}}\right)\left\{ C_{\rm inv, \chi_0}+C_{\rm ann, \chi_1}\right\} +E_{1}C_{\rm el, \chi_1}\right]\,.
\label{eq:BoltzmannTpre}
\end{aligned}
\end{equation}
Let us have a look at the collisional integrals one by one.
The $E_1$-weighted integral of $C_{\rm inv, \chi_0}$ can be manipulated as
\begin{align}
\int_{\bf{p}_{1}}\left(E_{1}-\left\langle E_{\chi_{1}}\right\rangle _{T_{\chi_{1}}}\right)\frac{C_{\rm inv, \chi_0}}{2E_1}&=\int_{\bf{p}_{1}}\left(\frac{E_{1}+E_{2}}{2}-\left\langle E_{\chi_{1}}\right\rangle _{T_{\chi_{1}}}\right)\frac{C_{\rm inv, \chi_0}}{2E_1}\,, \nonumber\\
&=\int_{\bf{p}_{1}}\left(\frac{E_{3}+E_{4}}{2}-\left\langle E_{\chi_{1}}\right\rangle _{T_{\chi_{1}}}\right)\frac{C_{\rm inv, \chi_0}}{2E_1}\,, \label{eq:EweightedC}\\
&=\left\langle \Delta E_{0}\sigma_0 v_{{\rm rel}}\right\rangle _{T_{\chi_{0}}}\left[n_{\chi_{0}}^{2}-\left(\frac{n_{\chi_{0}}^{{\rm eq}}\left(T_{\chi_{1}}\right)}{n_{\chi_{1}}^{{\rm eq}}\left(T_{\chi_{1}}\right)}\right)^{2}\frac{\left\langle \Delta E_{0}\sigma_0 v_{{\rm rel}}\right\rangle _{T_{\chi_{1}}}}{\left\langle \Delta E_{0}\sigma_0 v_{{\rm rel}}\right\rangle _{T_{\chi_{0}}}}n_{\chi_{1}}^{2}\right]\,, \nonumber
\end{align}
where we have defined $\Delta E_{0}=E_{\chi_{0}}-\left\langle E_{\chi_{1}}\right\rangle _{T_{\chi_{1}}}$.
In the second equality, we have used the property of the $4$-momentum conserving $\delta$-function.
A similar expression holds for the $E_1$-weighted integral of $C_{\rm ann, \chi_1}$.

The remaining is the $E_1$-weighted integral of $C_{\rm el, \chi_1}$.
For numerical convenience, we adopt the analytic approximation for $C_{\rm el, \chi_1}\left[f_{\chi_1}\right]$ from Ref.~\cite{Binder:2016pnr}:
\begin{equation}
C_{\rm el, \chi_1}\left[f_{\chi_{1}}\right]\simeq2 m_{\chi_1}\frac{\partial}{\partial{\bf p}_{1,i}}\left[\gamma_{\chi_{1}{\rm sm}}\left(m_{\chi_1}T\frac{\partial f_{\chi_{1}}}{\partial{\bf p}_{1,i}}+{\bf p}_{1,i}f_{\chi_{1}}\right)\right]\,,
\label{eq:Celapprox}
\end{equation}
where we took the non-relativistic limit of $\chi_1$, $T$ is the temperature of $\phi$, and $\gamma_{\chi_1 {\rm sm}}$ is the momentum transfer rate given as
\begin{equation}
\gamma_{\rm \chi_1 sm}\simeq\frac{1}{6 m_{\chi_1} T}\underset{s_{2}}{\sum}\int\frac{d^{3}p_{2}}{\left(2\pi\right)^{3}}f_{2}^{{\rm eq}}\int_{-4|{\bf p}_2|^{2}}^{0}dt\left(-t\right)\frac{d\sigma_{\chi_{1}\phi\rightarrow\chi_{1}\phi}}{dt}v_{{\rm rel}}\,.
\end{equation}
The sum denotes the spin degrees of freedom of $\phi$.
Note that the approximation of Eq.~\eqref{eq:Celapprox} is done in the limit where the momentum transfer ${\bf p}_{\chi_1,1}-{\bf p}_{\chi_1,3}$ is smaller than the typical DM momentum;
this is a reasonable approximation for non-relativistic $\chi_1$ scattering with a much lighter relativistic SM particle $\phi$.
The $E_1$-weighted integral of $C_{\rm el, \chi_1}$ is then approximated as
\begin{equation}
\int_{{\bf p}_{1}}E_{1}\frac{C_{\rm el, \chi_1}[f_{\chi_1}(p_1)]}{2E_{1}}\simeq-3n_{\chi_{1}}\gamma_{\rm \chi_1 {\rm sm}} \left(T_{\chi_{1}}-T\right)\,.
\label{eq:EweightedCel}
\end{equation}
Putting altogether the Eq.~\eqref{eq:BoltzmannTpre}, Eq.~\eqref{eq:EweightedC}, and Eq.~\eqref{eq:EweightedCel}, the temperature evolution equation is given as
\begin{align}
\frac{\dot{T}_{\chi_{1}}}{T_{\chi_{1}}^{2}}+\frac{3HT_{\chi_{1}}}{\sigma_{E,T_{\chi_{1}}}^{2}}&\simeq\frac{1}{n_{\chi_{1}}\sigma_{E,T_{\chi_{1}}}^{2}}\bigg\{\left\langle \Delta E_{0}\sigma_0 v_{{\rm rel}}\right\rangle _{T_{\chi_{0}}}\left[n_{\chi_{0}}^{2}-\left(\frac{n_{\chi_{0}}^{{\rm eq}}\left(T_{\chi_{1}}\right)}{n_{\chi_{1}}^{{\rm eq}}\left(T_{\chi_{1}}\right)}\right)^{2}\frac{\left\langle \Delta E_{0}\sigma_0 v_{{\rm rel}}\right\rangle _{T_{\chi_{1}}}}{\left\langle \Delta E_{0}\sigma_0 v_{{\rm rel}}\right\rangle _{T_{\chi_{0}}}}n_{\chi_{1}}^{2}\right]\nonumber\\
&\quad-\left\langle \Delta E_{1}\sigma_1 v_{{\rm rel}}\right\rangle _{T_{\chi_{1}}}\left[n_{\chi_{1}}^{2}-\frac{\left\langle \Delta E_{1}\sigma_1 v_{{\rm rel}}\right\rangle _{T}}{\left\langle \Delta E_{1}\sigma_1 v_{{\rm rel}}\right\rangle _{T_{\chi_{1}}}}n_{\chi_1}^{{\rm eq}2}\left(T\right)\right] \label{eq:BoltzmannT}\\
&\quad-3n_{\chi_{1}}\gamma_{\rm \chi_1 sm}\left(T_{\chi_{1}}-T\right)\bigg\}\,,\nonumber
\end{align}
where $\Delta E_1=E_{\chi_1}-\left\langle E_{\chi_1}\right\rangle _{T_{\chi_1}}$.
In general, Eq.~\eqref{eq:BoltzmannT} (together with the evolution equation for $T_{\chi_0}$ that we have not specified) and the number density evolution equations~[Eq.~\eqref{eq:chi2numden} and Eq.~\eqref{eq:chi1numden}] form a coupled system of Boltzmann equations.
However, if the kinetic decoupling of $\chi_1$ takes place well after the freeze-out of DM, Eq.~\eqref{eq:BoltzmannT} can be effectively considered to be decoupled from the number density equations, while taking the freeze-out number density for $n_{\chi_i}$.
In the case where the kinetic decoupling of $\chi_1$ interferes with the chemical freeze-out, one has to follow the co-evolution of temperature and number density;
such a case is shown as yellow hatched region in Figure~\ref{fig:paramSH}.

For most of the parameter space we consider in the main text, the kinetic decoupling of $\chi_1$ is well separated from the freeze-out of $\chi_1$.
Furthermore, the DM particles remain non-relativistic throughout their evolution, i.e., the DM temperatures are negligible compared to their masses and the mass difference $\delta m=m_{\chi_0}-m_{\chi_1}$.
In such case, the RHS of Eq.~\eqref{eq:BoltzmannT} is further simplified, leading to Eq.~\eqref{eq:Teqnonrel}:
inside the squared parentheses of the first line, the second term which represents the cooling of $\chi_1$ through the inverse process $\chi_1\chi_1\rightarrow \chi_0 \chi_0$, is negligible to the first term since it is a kinematically forbidden process;
the second line, which corresponds to the heating and cooling through the $\chi_1\chi_1\leftrightarrow \phi\phi$ process, is negligible compared to the first line.

\section{Freeze-out of DM}
\label{appendix:Boltzmannsolutions}

In this appendix, we give semi-analytic estimations for final yield of DM.
We find that our semi-analytic estimations agrees reasonably well with numerical solutions.
Semi-analytic understanding to the numerical solutions will be useful when scanning the viable parameter space for the two-component DM scenario.
One can find the relic density estimations in Eq.~\eqref{eq:WIMPyield}, Eq.~\eqref{eq:chi1yieldestimationY12const}, and Eq.~\eqref{eq:chi1assistedY12nc}.

We rewrite the number density equations for $\chi_0$ and $\chi_1$~[Eqs.~\eqref{eq:chi2numden} and~\eqref{eq:chi1numden}] in terms of the yield $Y=n/s$:
\begin{align}
\frac{dY_{\chi_{0}}}{dx}&=-\frac{\lambda_{\chi_0}(x)}{x}\left[Y_{\chi_{0}}^{2}-\left(\frac{Y_{\chi_{0}}^{{\rm eq}}\left(x\right)}{Y_{\chi_{1}}^{{\rm eq}}\left(x\right)}\right)^{2}Y_{\chi_{1}}^{2}\right]\,,\label{eq:chi2numeq}\\
\frac{dY_{\chi_{1}}}{dx}&=\frac{\lambda_{\chi_0}(x)}{x}\left[Y_{\chi_{0}}^{2}-\left(\frac{Y_{\chi_{0}}^{{\rm eq}}\left(x\right)}{Y_{\chi_{1}}^{{\rm eq}}\left(x\right)}\right)^{2}Y_{\chi_{1}}^{2}\right]-\frac{\lambda_{\chi_{1}}(x)}{x}\left[Y_{\chi_{1}}^{2}-\left(Y_{\chi_{1}}^{{\rm eq}}\left(x\right)\right)^{2}\right]\,, \label{eq:chi1numeq}
\end{align}
where $x=m_{\chi_1}/T$, and we assumed that ($\chi_0$, $\chi_1$) and ($\chi_1$, $\phi$) are in kinetic equilibrium, $T_{\chi_0}=T_{\chi_1}=T$.
The $\lambda(x)$'s are given as
\begin{align}
\lambda_{\chi_{0}}\left(x\right)&=\frac{s\left\langle \sigma_0 v_{\rm rel}\right\rangle_{T}}{H}\left[1-\frac{1}{3}\frac{d\ln g_{\star S}\left(x\right)}{d\ln x}\right]\,,\\
\lambda_{\chi_{1}}\left(x\right)&=\frac{s\left\langle \sigma_1 v_{\rm rel}\right\rangle _{T}}{H}\left[1-\frac{1}{3}\frac{d\ln g_{\star S}\left(x\right)}{d\ln x}\right]\,.
\end{align}
Hereafter, we will consider $g_{\star}$ and $g_{\star S}$ as constants for simplicity.
\\

{\bf -- Abundance of $\chi_0$.}
\medskip

Before numerically solving the number density equations, let us take a semi-analytical approach.
First, let us discuss the chemical freeze-out of $\chi_0$.
We focus on the case where the freeze-out of $\chi_0$ is well separated from (and prior to) that of $\chi_1$.
In such case, $Y_{\chi_1}$ would follow $Y^{\rm eq}_{\chi_1}(x)$ around the freeze-out of $\chi_0$, and the final yield of $\chi_0$ will be virtually the same as the standard case of WIMP.
Let us recall the estimation of final yield ($Y_\infty$) for WIMP, as we will do a similar analysis when estimating the yield for $\chi_1$.
Around the freeze-out of $\chi_0$, $Y_{\chi_1}\simeq Y^{\rm eq}_{\chi_1}(x)$ and Eq.~\eqref{eq:chi2numeq} is approximated as
\begin{equation}
\frac{dY_{\chi_0}}{dx}=-\frac{\lambda_{\chi_0}\left(x\right)}{x}\left[Y_{\chi_0}^{2}-\left(Y_{\chi_0}^{\rm eq}(x)\right)^2\right]\,.
\label{eq:chi2WIMP}
\end{equation}
In the region where $\chi_0$ is near the chemical equilibrium, $Y_{\chi_0}$ can be written as a small deviation from $Y^{\rm eq}_{\chi_0}$:
\begin{equation}
Y_{\chi_0}(x)=Y_{\chi_0}^{\rm eq}(x)+\Delta Y_{\chi_0}(x)\,.
\end{equation}
In the lowest order in $\Delta Y_{\chi_0}$, Eq.~\eqref{eq:chi2WIMP} is written as
\begin{equation}
\frac{dY_{\chi_0}^{{\rm eq}}}{dx}\simeq-\frac{\lambda_{\chi_0}(x)}{x}2\Delta Y_{\chi_0}Y_{\chi_0}^{{\rm eq}}\,.
\label{eq:WIMPtight}
\end{equation}
Since $dY_{\chi_0}^{{\rm eq}}/dx\sim-(m_{\chi_0}/m_{\chi_1})Y_{\chi_0}^{{\rm eq}}$ (note again that $x=m_{\chi_1}/T$) when $\chi_0$ is non-relativistic, we find
\begin{equation}
\frac{\Delta Y_{\chi_0}}{Y_{\chi_0}^{{\rm eq}}}\left(x\right)\simeq\frac{(m_{\chi_0}/m_{\chi_1})x}{2\lambda_{\chi_0}\left(x\right)Y_{\chi_0}^{{\rm eq}}\left(x\right)}=\frac{(m_{\chi_0}/m_{\chi_1})x}{2}\left(\frac{\Gamma\,(=n_{\chi_0}^{\rm eq}\left\langle\sigma_0 v_{\rm rel}\right\rangle)}{H}\bigg|_{x}\right)^{-1}\,,
\label{eq:WIMPdeviation}
\end{equation}
which implies that the relative deviation grows exponentially with $x$, since $Y_{\chi_0}^{\rm eq}(x)\propto x^{3/2} \exp[-x]$.
We {\it define} the freeze-out point, $x_{\rm fo,0}$, as the point when the relative deviation of $Y_{\chi_0}$ from $Y^{\rm eq}_{\chi_0}$ starts to exceed unity:
\begin{equation}
\frac{\Delta Y_{\chi_0}}{Y_{\chi_0}^{{\rm eq}}}\left(x_{{\rm fo},0}\right)\simeq \frac{(m_{\chi_0}/m_{\chi_1})x_{\rm fo,0}}{2}\left(\frac{\Gamma}{H}\bigg|_{x_{\rm fo,0}}\right)^{-1}=1\,,
\label{eq:WIMPfo}
\end{equation}
where we see the familiar Gamow's criterion with an extra factor of $x^{-1}$ multiplied to the reaction rate $\Gamma$.
Note that for $\left\langle\sigma_0 v_{\rm rel}\right\rangle_T$ that achieves the abundance of $\chi_0$ that is similar to the observed DM abundance, $x_{\rm fo,0}\sim20\,(m_{\chi_1}/m_{\chi_0})$.
Now, let us examine the region well after the freeze-out, $x\gtrsim x_{\rm fo,0}$.
Since the growth of the relative deviation with respect to $x$ is exponential, $Y^{\rm eq}_{\chi_0}$ would be ignorable compared to $Y_{\chi_0}$:
\begin{equation}
\frac{dY_{\chi_0}}{dx}\simeq-\frac{\lambda_{\chi_0}(x)}{x}Y_{\chi_0}^{2}\,.
\end{equation}
Given the boundary condition at $x_{\rm fo,0}$, this is a separable equation that we can solve:
\begin{equation}
\begin{aligned}
\frac{1}{Y_{\chi_0}\left(x\right)}&=\frac{1}{Y_{\chi_0}\left(x_{{\rm fo,0}}\right)}+\int_{x_{{\rm fo,0}}}^{x}dx^{\prime}\frac{\lambda_{\chi_0}\left(x^{\prime}\right)}{x^{\prime}}\,,\\
&=\frac{1}{Y_{\chi_0}\left(x_{{\rm fo,0}}\right)}+\underset{=1/Y_{\rm ann,\chi_0}(x;x_{\rm fo,0})}{\underbrace{\frac{\lambda_{\chi_0}\left(x_{{\rm fo,0}}\right)}{n_{0}+1}\left[1-\left(\frac{x_{{\rm fo,0}}}{x}\right)^{n_{0}+1}\right]}}\,,
\label{eq:fosolve}
\end{aligned}
\end{equation}
where $\left\langle\sigma_0 v_{\rm rel}\right\rangle\propto x^{-n_0}$.
Let us define $Y_{{\rm WIMP},\chi_i}(x)$ for a notational convenience:
\begin{equation}
Y_{{\rm WIMP},\chi_i}(x)=\frac{n_i+1}{\lambda_{\chi_i}(x)}\,.
\end{equation}
We get the final yield of $\chi_0$ by taking $x\rightarrow \infty$ in Eq.~\eqref{eq:fosolve};
note that in this limit, the second term in the RHS, $1/Y_{\rm ann,\chi_0}(x;x_{\rm fo,0})\simeq1/Y_{{\rm WIMP},\chi_0}(x_{\rm fo,0})$, dominates over the first term, $1/Y_{\chi_0}(x_{\rm fo,2})$, since
\begin{equation}
Y_{\chi_0}\left(x_{{\rm fo}}\right)\simeq\frac{\left(m_{\chi_0}/m_{\chi_1}\right)x_{{\rm fo},0}}{\lambda_{\chi_0}\left(x_{{\rm fo},0}\right)}=\left(m_{\chi_0}/m_{\chi_1}\right)x_{{\rm fo},0}\times\frac{Y_{\rm WIMP,\chi_0}(x_{\rm fo,0})}{n_0+1}\,,
\end{equation}
which can be deduced from the definition of freeze-out point in Eq.~\eqref{eq:WIMPfo}.
Then the final abundance of $\chi_0$, $Y_{\chi_0}(\infty)$, is given as
\begin{equation}
Y_{\chi_0}(\infty)\simeq Y_{\rm WIMP,\chi_0}(x_{\rm fo,0})=\frac{n_0+1}{\lambda_{\chi_0}\left(x_{{\rm fo},0}\right)}\,.
\label{eq:WIMPyield}
\end{equation}
\\

{\bf -- Abundance of $\chi_1$; the case of constant $Y_{\rm ast.}$ with $Y_{\rm ast.}<Y_{\rm WIMP,\chi_1}(x_{\rm fo,1})$.}
\medskip

The chemical freeze-out of $\chi_1$ could be very different from $\chi_0$, since annihilation of $\chi_0$ continuously produce $
\chi_1$.
Around the freeze-out of $\chi_1$, $\chi_0$ already froze-out and we may approximate Eq.~\eqref{eq:chi1numeq} as
\begin{equation}
\frac{dY_{\chi_{1}}}{dx}\simeq-\frac{\lambda_{\chi_{1}}(x)}{x}\left[Y_{\chi_{1}}^{2}-\left(Y_{\chi_{1}}^{{\rm eq}}\left(x\right)\right)^{2}-Y_{\rm ast.}^{2}\left(x\right)\right]\,,
\label{eq:chi1numeqfo}
\end{equation}
where $Y_{\rm ast.}$ is defined as
\begin{equation}
Y_{\rm ast.}\left(x\right)=\sqrt{\frac{\left\langle \sigma_0 v_{\rm rel}\right\rangle _{T}}{\left\langle \sigma_1 v_{\rm rel}\right\rangle _{T}}}Y_{\chi_0}(\infty)\,.
\end{equation}

First, let us first focus on a case where $Y_{\rm ast.}$ is constant, e.g., annihilation of $\chi_0$ and $\chi_1$ are both $s$-wave.
We define the standard freeze-out point of $\chi_1$ as $x_{\rm fo,1}=m_{\chi_1}/T_{\rm fo,1}$ as in $T_{\rm fo,1}\sim m_{\chi_1}/20$, as in Eq.~\eqref{eq:WIMPfo}.
If $Y_{\rm ast.}$ is smaller than $Y_{\rm WIMP,\chi_1}(x_{\rm fo,1})\simeq1/\lambda_{\chi_1}(x_{\rm fo,1})$, then $Y_{\rm ast.}$ is smaller than $Y_{\chi_1}(x)$ throughout the freeze-out of $\chi_1$ and the final yield of $\chi_1$ would be just the standard estimation given in Eq.~\eqref{eq:WIMPyield}.
The reasoning is the following:
if $Y_{\rm ast.}<Y_{\rm WIMP,\chi_1}(x_{\rm fo,1})$, then $Y_{\rm ast.}<Y^{\rm eq}_{\chi_1}(x_{\rm fo,1})$ where $x_{\rm fo,1}$ is defined by the WIMP freeze-out condition Eq.~\eqref{eq:WIMPfo} for $\chi_1$.
Therefore, around $x=x_{\rm fo,1}$, we can drop the $Y_{\rm ast.}^2$ term in the RHS of Eq.~\eqref{eq:chi1numeqfo}.
Then, the number density equation for $\chi_0$ is the same with Eq.~\eqref{eq:chi2WIMP}.
\\

{\bf -- Abundance of $\chi_1$; the case of constant $Y_{\rm ast.}$ with $Y_{\rm ast.}>Y_{\rm WIMP,\chi_1}(x_{\rm fo,1})$.}
\medskip

If $Y_{\rm ast.}>Y_{\rm WIMP,\chi_1}$, there would be a point in $x$ where $Y_{\chi_1}(x)$ decreases to become $Y_{\chi_1}(x)=Y_{\rm ast.}$.
After such moment, $Y_{\chi_1}$ will freeze at the value of $Y_{\rm ast.}$.
We can justify this statement by using the similar procedure that we have done for the freeze-out of $\chi_0$.
If the moment when $Y_{\chi_1}(x)=Y_{\rm ast.}$ takes places when $\chi_1$ is non-relativistic, soon after then $Y^{\rm eq}_{\chi_1}(x)$ would become negligible in Eq.~\eqref{eq:chi1numeqfo} due to the exponential suppression.
Let us take of a small deviation of $Y_{\chi_1}(x)$ from $Y_{\rm ast.}$:
\begin{equation}
Y_{\chi_{1}}(x)=Y_{\rm ast.}+\Delta Y_{\chi_{1}}(x)\,.
\end{equation}
Then Eq.~\eqref{eq:chi1numeqfo} leads to
\begin{equation}
\frac{dY_{\rm ast.}}{dx}\simeq-\frac{\lambda_{\chi_{1}}\left(x\right)}{x}2\Delta Y_{\chi_{1}}Y_{\rm ast.}\,.
\label{eq:deltaY12eq}
\end{equation}
Since we are thinking of a case where $Y_{\rm ast.}$ is constant, Eq.~\eqref{eq:deltaY12eq} implies that $\Delta Y_{\chi_1}(x)\simeq0$ afterwards.
Thus, $Y_{\rm ast.}$ is a final yield of $\chi_1$ in the case of $Y_{\rm ast.}>Y_{\rm WIMP,\chi_1}(x_{\rm fo,1})$.
Putting our results together for the cases of $Y_{\rm ast.}<Y_{\rm WIMP,\chi_1}(x_{\rm fo,1})$ and $Y_{\rm ast.}>Y_{\rm WIMP,\chi_1}(x_{\rm fo,1})$, we may write
\begin{equation}
Y_{\chi_{1},\infty}\simeq\max\left[Y_{\rm ast.},Y_{{\rm WIMP},\chi_{1}}(x_{\rm fo,1})\right]\,,
\label{eq:chi1yieldestimationY12const}
\end{equation}
where $Y_{\rm ast.}$ is constant.
The discrepancy between Eq.~\eqref{eq:chi1yieldestimationY12const} and the numerical solutions to Eq.~\eqref{eq:chi2numeq} and Eq.~\eqref{eq:chi1numeq} is within ${\cal O}(10)\%$;
see Figure~\ref{fig:yields}.
\\

{\bf -- Abundance of $\chi_1$; the case of decreasing $Y_{\rm ast.}(x)$.}
\medskip

The remaining is the case where $Y_{\rm ast.}(x)$ is not constant.
Again, we estimate the final yield of $\chi_1$ while assuming that $Y_{\chi_1}(x)$ is initially following $Y_{\rm ast.}(x)$;
we will compare this yield with $Y_{{\rm WIMP},\chi_1}(x_{\rm fo,1})$ and take a larger one as the final yield of $\chi_1$.

\begin{figure}[t!]
\centering
\includegraphics[scale=0.6]{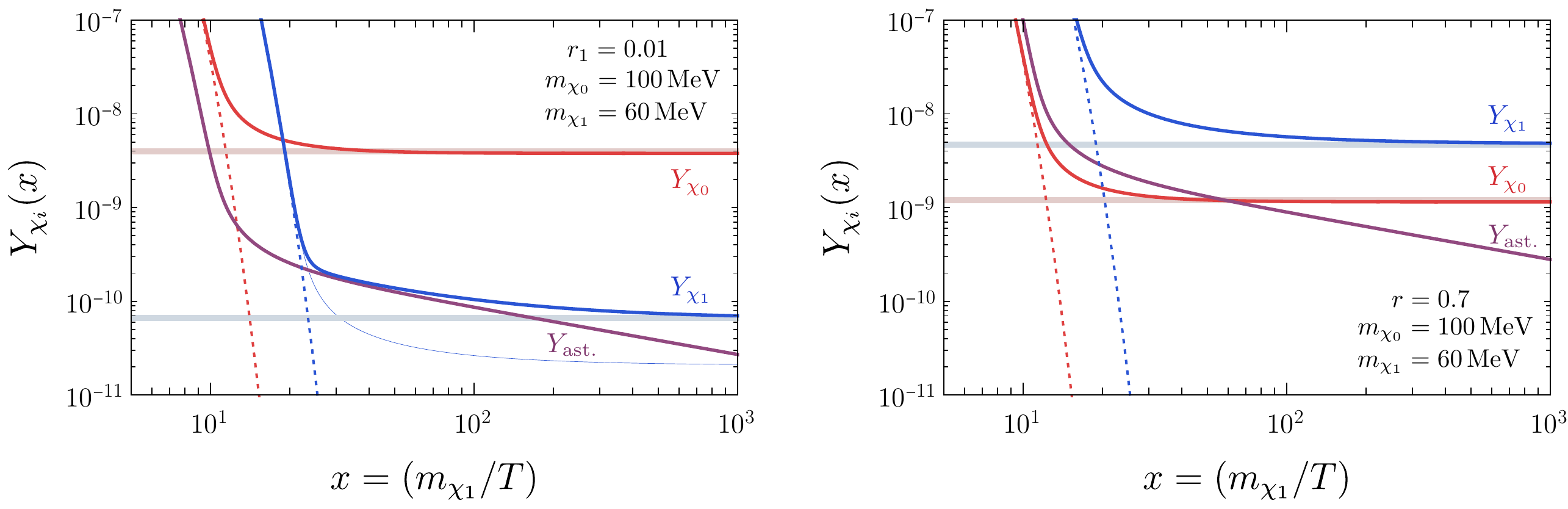}
\caption{Numerical solutions to Eq.~\eqref{eq:chi2numeq} and Eq.~\eqref{eq:chi1numeq} for decreasing $Y_{12}(x)$, e.g., $p$-wave annihilation for $\chi_0$ and $s$-wave annihilation for $\chi_1$. Solid curves represent numerical solutions for $Y_{\chi_0}(x)$ (in red), $Y_{\chi_1}(x)$ (in blue), and $Y_{12}(x)$ (in purple).
The thin solid blue curve is the solution for the standard WIMP freeze-out of $\chi_1$.
The dotted curves are $Y^{\rm eq}_{\chi_i}(x)$'s.
The horizontal lines are our estimations for final yield of $\chi_0$ (in red, Eq.~\eqref{eq:WIMPyield}), and $\chi_1$ (in blue, Eq.~\eqref{eq:chi1assistedde}).
({\it Left}) - The case where the freeze-out of $\chi_1$ is in the assisted regime, i.e., Eq.~\eqref{eq:chi1assistedde}. We have set $c^\prime=0.63$ to match Eq.~\eqref{eq:chi1assistedde} with the numerical results; the departure point of $Y_{\chi_1}$ from $Y_{\rm ast.}$ is given by $x_{\rm fo}^\prime\simeq 72$.
({\it Right}) - The case where the freeze-out of $\chi_1$ is in the standard regime.
}
\label{fig:yieldD}
\end{figure}

Let us consider a case where $Y_{\rm ast.}(x)\propto x^{n_{\rm ast.}}$ is decreasing, i.e., $n_{\rm ast.}<0$.
If $Y_{\chi_1}(x)$ is initially following $Y_{\rm ast.}(x)$, $Y_{\chi_1}(x)$ would be decreasing as with $Y_{\rm ast.}(x)$.
But $Y_{\chi_1}(x)$ would not be following $Y_{\rm ast.}(x)$ indefinitely, but eventually depart from $Y_{\rm ast.}(x)$.
From Eq.~\eqref{eq:deltaY12eq}, the relative deviation of $Y_{\chi_1}(x)$ from $Y_{\rm ast.}(x)$ becomes order unity when $x=x_{\rm fo}^\prime$, i.e., $\Delta Y_{\chi_1}(x)/Y_{\rm ast.}(x)=c^\prime$ where $c^\prime$ is a ${\cal O}(1)$ constant for fitting our analytic estimations on final relic abundance with numerical results. 
For general values of $n_{\rm ast.}$, the deviation point $x_{\rm fo}^\prime$ is determined by the condition given as
\begin{equation}
\frac{\left|n_{\rm ast.}\right|}{2\lambda_{\chi_{1}}\left(x_{{\rm fo}}^{\prime}\right)Y_{\rm ast.}\left(x_{{\rm fo}}^{\prime}\right)}=c^\prime\,,
\label{eq:xfoprime}
\end{equation}
where $n_{\rm ast.}=(n_1-n_2)/2$.
Afterwards, Eq.~\eqref{eq:chi1numeqfo} becomes
\begin{equation}
\frac{dY_{\chi_{1}}}{dx}\simeq-\frac{\lambda_{\chi_{1}}\left(x\right)}{x}Y_{\chi_{1}}^{2}\,,
\end{equation}
and the solution for $Y_{\chi_1}$ is given as
\begin{equation}
\frac{1}{Y_{\chi_{1}}\left(x\right)}=\frac{1}{Y_{\chi_{1}}\left(x_{{\rm fo}}^{\prime}\right)}+\frac{\lambda_{\chi_{1}}\left(x_{{\rm fo}}^{\prime}\right)}{n_{1}+1}\left[1-\left(\frac{x_{{\rm fo}}^{\prime}}{x}\right)^{n_{1}+1}\right]\,.
\label{eq:chi1assistedde}
\end{equation}
with $\left\langle \sigma_1 v_{\rm rel}\right\rangle_{T}\propto x^{-n_1}$.
Taking $x\rightarrow \infty$, the second term in the RHS is similar to the standard case for freeze-out~[Eq.~\eqref{eq:WIMPyield}] while $\lambda_{\chi_1}$ is evaluated at $x_{\rm fo}^\prime$, i.e., $1/Y_{{\rm WIMP},\chi_1}(x_{\rm fo}^{\prime})$.
Meanwhile, we can see that the first term and the second term in the RHS of Eq.~\eqref{eq:chi1assistedde} gives similar contribution to the final yield of $\chi_1$;
they are both of order $\sim1/Y_{{\rm WIMP},\chi_1}(x_{\rm fo}^{\prime})$.
This is different from the standard freeze-out of WIMP;
in Eq.~\eqref{eq:fosolve}, the first term in the RHS is negligible compared to the second term.
The difference stems from the fact that $Y_{\chi_0}^{\rm eq}(x)$ decreases exponentially, while $Y_{\rm ast.}(x)$ decreases by a power-law.
In the LHS of Eq.~\eqref{eq:WIMPtight}, since $Y_{\chi_0}^{\rm eq}(x)$ decreases exponentially, we had $dY_{\chi_0}^{{\rm eq}}/dx\simeq-Y_{\chi_0}^{{\rm eq}}$.
On the contrary, in the LHS of Eq.~\eqref{eq:deltaY12eq}, since $Y_{\rm ast.}(x)$ decreases by a power-law, we had $dY_{\rm ast.}/dx=(n_{\rm ast.}/x)Y_{\rm ast.}$.
The additional factor of $1/x$ in the LHS of Eq.~\eqref{eq:deltaY12eq} makes the difference from the case of WIMP.
From this observation, we will simply estimate the final yield of $\chi_1$ as $Y_{\rm WIMP,\chi_1}(x_{\rm fo}^\prime)$.
The left panel of Figure~\ref{fig:yieldD} shows the comparison between~[Eq.~\eqref{eq:chi1assistedde}] (horizontal blue) and the numerical solution of $Y_{\chi_1}$ (solid blue);
we have set $c^\prime \simeq 0.63$ to match our estimation $Y_{\rm WIMP,\chi_1}(x_{\rm fo}^\prime)$ with the numerical solutions.
With the calibration of $c^\prime$ at hand, we estimate the required annihilation cross section of $\chi_1$ by requiring $Y_{\rm WIMP,\chi_1}(x_{\rm fo}^\prime)$ to be equal to the desired final yield of $\chi_1$.
Note that we may also use Eq.~\eqref{eq:chi1assistedde} to estimate the final yield of $\chi_1$, while we would arrive at a different value of $c^\prime$;
nevertheless, we would end up with the same required annihilation cross section of $\chi_1$.
\\

{\bf -- Abundance of $\chi_1$; the case of increasing $Y_{\rm ast.}(x)$.}
\medskip

Similarly, in the case where $Y_{\rm ast.}(x)$ is increasing ($n_{\rm ast.}>0$), the departure point $x_{\rm fo}^{\prime}$ of $Y_{\chi_1}(x)$ from $Y_{\rm ast.}(x)$ is determined by Eq.~\eqref{eq:xfoprime}.
Afterward the departure point, Eq.~\eqref{eq:chi1numeqfo} becomes
\begin{equation}
\frac{dY_{\chi_{1}}}{dx}\simeq\frac{\lambda_{\chi_{1}}\left(x\right)}{x}Y_{\rm ast.}^{2}(x)\,,
\end{equation}
and the solution is 
\begin{equation}
Y_{\chi_{1}}\left(x\right)=Y_{\chi_{1}}\left(x_{{\rm fo}}^{\prime}\right)+\frac{\lambda_{\chi_{1}}\left(x_{{\rm fo}}^{\prime}\right)Y_{\rm ast.}^{2}\left(x_{{\rm fo}}^{\prime}\right)}{n_0+1}\left[1-\left(\frac{x_{{\rm fo}}^{\prime}}{x}\right)^{n_0+1}\right]\,.
\label{eq:chi1assistedin}
\end{equation}
Taking $x\rightarrow \infty$, the second term in the RHS is similar to the case of WIMP~[Eq.~\eqref{eq:WIMPyield}] while $\lambda_{\chi_1}$ is evaluated at $x_{\rm fo}^\prime$.
We see again that the first term and the second term in the RHS of Eq.~\eqref{eq:chi1assistedin} have comparable contribution to the final yield of $\chi_1$, i.e., $Y_{\rm ast.}(x_{\rm fo}^\prime)\sim Y_{\chi_1}(x_{\rm fo}^\prime)\sim1/\lambda_{\chi_1}(x_{\rm fo}^\prime)$:
we again simply estimate the final yield of $\chi_1$ as $Y_{\rm WIMP,\chi_1}(x_{\rm fo}^\prime)$.
The left panel of Figure~\ref{fig:yield} shows the comparison between our estimation $Y_{\rm WIMP,\chi_1}(x_{\rm fo}^\prime)$ (horizontal blue) and the numerical solution of $Y_{\chi_1}$ (solid blue);
we have set $c^\prime \simeq 0.35$ to match our estimation $Y_{\rm WIMP,\chi_1}(x_{\rm fo}^\prime)$ with the numerical solutions.
\\

In both of the cases when $Y_{\rm ast.}(x)$ is decreasing/increasing, we see that if $Y_{\chi_1}(x)$ was initially following $Y_{\rm ast.}(x)$, $Y_{{\rm WIMP},\chi_1}(x_{\rm fo}^{\prime})\simeq1/\lambda_{\chi_1}(x_{\rm fo}^\prime)$ is a reasonable estimate for the final yield of $\chi_1$.
It is natural to ask what is the precise condition for $Y_{\chi_1}(x)$ to initially follow $Y_{12}(x)$ to happen, and how to determine the final yield of $\chi_1$ when $Y_{\chi_1}(x)$ is not initially following $Y_{\rm ast.}(x)$.
This is a question that is difficult to study analytically, and we would need to depend on numerical analyses.
Nevertheless, we give the following estimation for final yield of $\chi_1$ when $Y_{\rm ast.}(x)$ is time-dependent:
\begin{equation}
Y_{\chi_{1},\infty}\simeq\max\left[Y_{{\rm WIMP},\chi_{1}}\left(x_{{\rm fo}}^{\prime}\right),Y_{{\rm WIMP},\chi_{1}}\left(x_{{\rm fo,1}}\right)\right]\,,
\label{eq:chi1assistedY12nc}
\end{equation}
where we note again that $x_{\rm fo}^\prime$ ($x_{\rm fo,1}$) is defined by Eq.~\eqref{eq:xfoprime} (Eq.~\eqref{eq:WIMPfo}).
Although somewhat crude, Eq.~\eqref{eq:chi1assistedY12nc} physically makes sense.
For example, if we think of a limit where $Y_{\rm ast.}(x)\rightarrow 0$, $x_{\rm fo}^\prime$ also decreases according to Eq.~\eqref{eq:xfoprime}, since $\lambda_{\chi_1}(x)$ is generally a decreasing function of $x$.    
Then, $Y_{{\rm WIMP},\chi_{1}}\left(x_{{\rm fo}}^{\prime}\right)\rightarrow0$ by the same reasoning, and the final yield of $\chi_1$ is just the standard WIMP yield; this should be since the $Y_{\rm ast.}(x)\rightarrow 0$ limit corresponds to the standard freeze-out.

\section{Temperature evolution of $\chi_1$ in the case of $p$-wave annihilation of $\chi_0$}
\label{section:pwaveann}

In this appendix, we comment on the temperature evolution of $\chi_1$ when $\chi_0$-annihilation is $p$-wave.
Let us examine the asymptotic behavior of $T_{\chi_1}$ from the evolution equation Eq.~\eqref{eq:Teqnonrel}.
We take an ansatz that $T_{\chi_1}\propto 1/a^N$, where $N\leq2$ is some positive number.
As in the main text, we denote the non-relativistic limit of $\left\langle\sigma_0 v_{{\rm rel}}\right\rangle_{T_{\chi_0}}$ as $\left(\sigma_0 v_{{\rm rel}}\right)$.
The ansatz would mean $\dot{T}_{\chi_1}+NHT_{\chi_1}=0$.
Furthermore, since the kinetic decoupling of $\chi_1$ takes place after the freeze-out of $\chi_0$ and $\chi_1$, their yields $Y_{\chi_i}=n_{\chi_i}/s$ are virtually conserved.
For our ansatz to be consistent with Eq.~\eqref{eq:Teqnonrel}, following condition must hold:
\begin{equation}
\begin{aligned}
(2-N)HT_{\chi_{1}}&\simeq\frac{2}{3n_{\chi_{1}}}\delta m\left(\sigma_0 v_{{\rm rel}}\right)n_{\chi_{0}}^{2}\\
&\sim\frac{2}{3}\frac{1-r_1}{r_1}\frac{m_{\chi_{1}}}{m_{\chi_{0}}}\delta m\left(\frac{H}{s}\right)_{{\rm fo},0}s\,,
\label{eq:Tchi1balance}
\end{aligned}
\end{equation}
where we have used $Y_{\chi_0}=Y_{\chi_0}(\infty)\sim 1/\lambda_{\chi_0}(x_{\rm fo,0})$.
In order for our ansatz to be a solution, both sides of Eq.~\eqref{eq:Tchi1balance} should scale in the same way with respect to the scale factor $a$.
During the radiation-dominated era, from the first equality of Eq.~\eqref{eq:Tchi1balance}, we see that $N=1$ for $s$-wave annihilation of $\chi_0$, and $N=2$ for $p$-wave annihilation of $\chi_0$
For higher partial-wave annihilations, $N=2$ since the particular solution for the heating term in Eq.~\eqref{eq:Teqnonrel} decays away faster than the complementary solution, i.e., $T_{\chi_1}\propto1/a^2$.
In the matter-dominated era, one finds that $N=3/2$ for $s$-wave annihilation of $\chi_0$, while $N=2$ and for higher partial-wave annihilations.
Therefore, we see that $T_{\chi_1}$ redshifts slower than $1/a^2$ only when $\chi_0$ annihilation is $s$-wave.

\section{Some remarks on the temperature evolution of $\chi_1$}
\label{section:Tdmevolapp}

\begin{figure}[t!]
\centering
\includegraphics[scale=0.52]{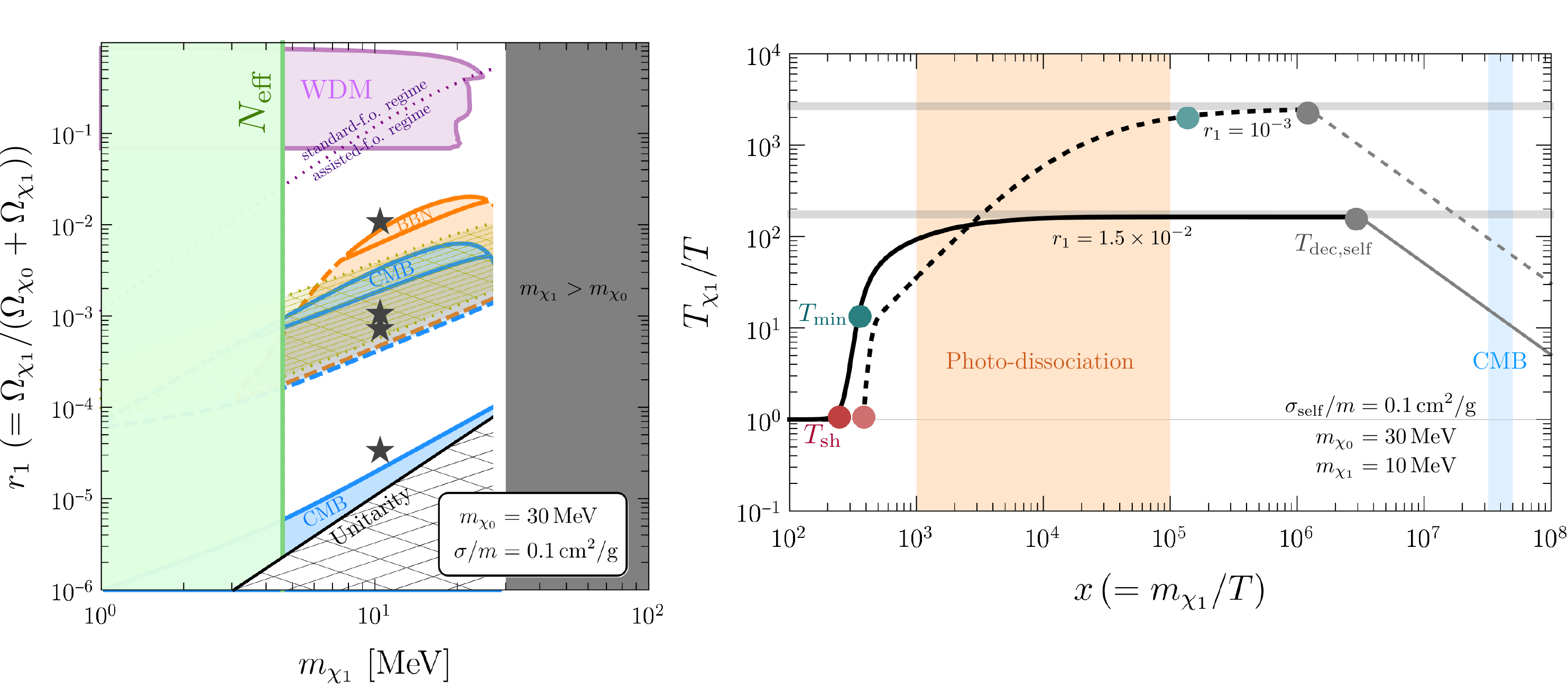}
\includegraphics[scale=0.6]{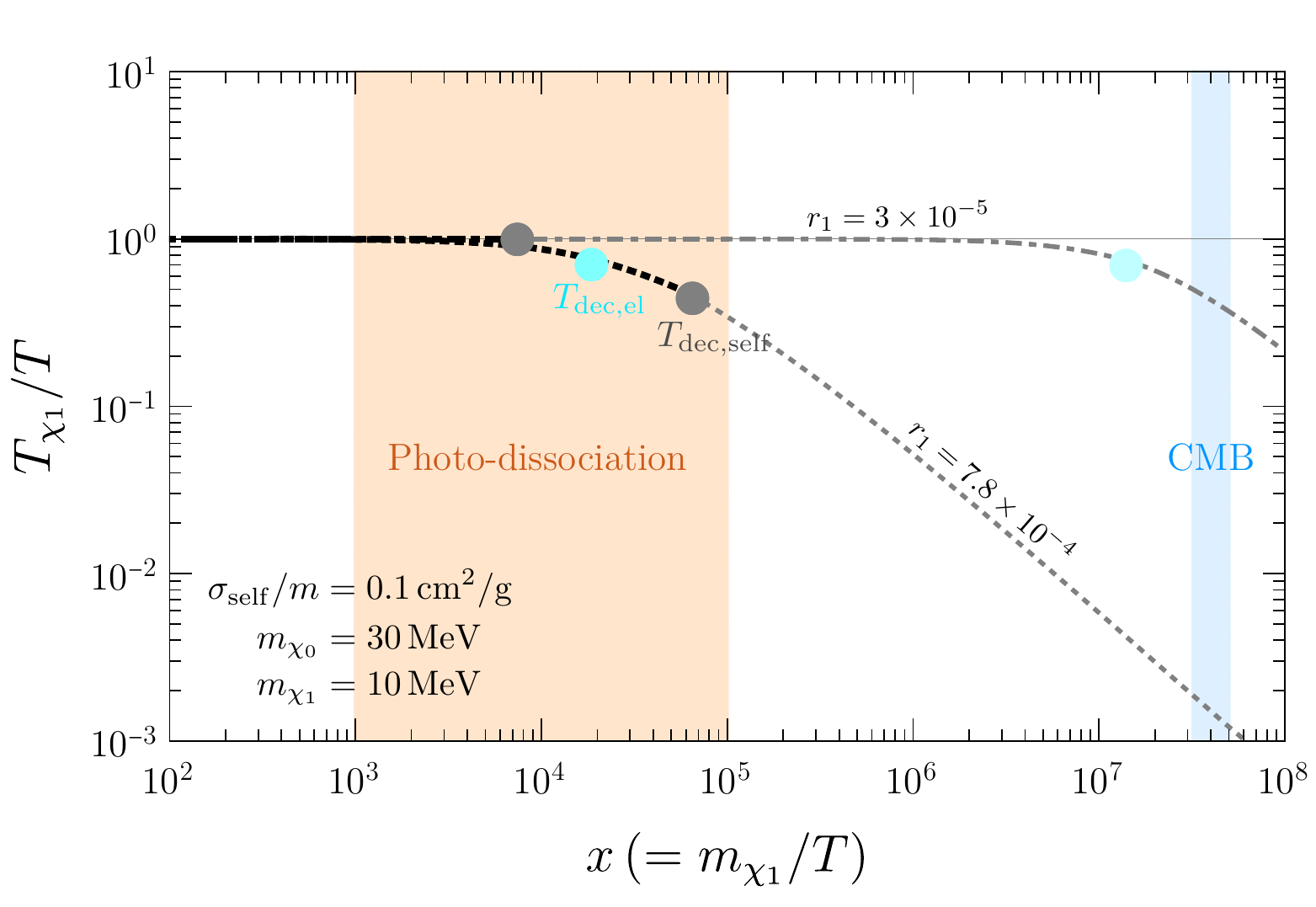}
\caption{
({\it Right}) - Same as Figure~\ref{fig:Tdmevol}, but for the abundance fractions depicted parameters (as stars) in the left panel;
$r_1=1.5\times10^{-2}$ (solid), $r_1=10^{-3}$ (dashed), $r_1=7.8\times10^{-4}$ (dotted), and $r_1=3\times10^5$ (dot-dashed).
The gray, red, and cyan circles represent the true values of $T_{\rm dec,self}$, $T_{\rm sh}$, and $T_{\rm dec,el}$ according to the presented evolution of $T_{\chi_1}$.
The green circles represent the underestimated $T_{\rm min}$~(Eq.~\eqref{eq:Tmin}), after which we may apply the naive estimation for $T_{\chi_1}$ given in Eq.~\eqref{eq:naiveTchi1}.
}
\label{fig:Tdmevolapp}
\end{figure}

In Figure~\ref{fig:paramSH} of the main text, we have discussed the parameter regions that are constrained by the cosmological observations, e.g., photo-dissociation and CMB constraints on $\chi_1$-annihilation.
Since the constraints depend on the temperature evolution of $\chi_1$, it is in principle best to scope out the evolution of $T_{\chi_1}$ by following the temperature evolution equation Eq.~\eqref{eq:Tchi1evolre}.
For practical reasons, we do not attempt to realize the exact evolutions of $T_{\chi_1}$ for every parameter points.
Furthermore, investigating the photo-dissociation constraints not only requires the exact evolution of $T_{\chi_1}$, but also requires a dedicated analysis on the evolution of yields of light nuclei in the presence of DM self-heating.
Instead, we have mapped out the regions where we can robustly put constraints on DM annihilation without scoping out the exact evolutions of $T_{\chi_1}$.
We have also depicted parameter regions that could be potentially constrained by a more dedicated analysis.
In this section, we elaborate more on the evolution of $T_{\chi_1}$ in the potentially constrained regions.
We assume fixed DM yield and numerically follow the evolution of $T_{\chi_1}$ via Eq.~\eqref{eq:Tchi1evolre} as a first approximation.~\footnote{The freeze-out of $\chi_1$ could overlap with the self-heating epoch, i.e., in the yellow-hatched regions in Figure~\ref{fig:paramSH}. In such a case, Eq.~\eqref{eq:Tchi1evolre} would need to be followed alongside with Eqs.~\eqref{eq:yieldevol}.}
By presenting the approximate evolutions of $T_{\chi_1}$ for a set of parameter points, we elaborate more on the conditions that separate the robustly-constrained and potentially-constrained regions and demonstrate how the potentially constrained regions may be subject to more dedicated analyses.

The depicted parameter points (by stars) in the left panel of Figure~\ref{fig:Tdmevolapp} represents the benchmark cases of distinct evolutions of $T_{\chi_1}$.
We discuss them one by one in the descending order in $r_1$:
\begin{itemize}

\item {\bf $T_{\rm min}>10\,{\rm keV}$ and $T_{\rm min}>T_{\rm dec,self}$} :
The solid curve in the right panel of Figure~\ref{fig:Tdmevolapp} correspond to this case.
The mentioned conditions are the requirements for putting robust photo-dissociation and CMB constraints on DM annihilation, as discussed in more detail in Section~\ref{section:SHDM} of the main text.
Since $T_{\rm min}>10\,{\rm keV}$, we may apply the approximation Eq.~\eqref{eq:naiveTchi1} throughout the photo-dissociation epoch.
Since $T_{\rm min}>T_{\rm dec,self}$, $T_{\chi_1}$ reaches the asymptotic solution Eq.~\eqref{eq:Tratioasy} before the decoupling of self-scattering, and we may robustly estimate $T_{\chi_1}$ around the last scattering.
We emphasize that for robustly-constrained regions, we require the conditions, $T_{\rm min}>10\,{\rm keV}$ and $T_{\rm min}>T_{\rm dec,self}$, to be held for any possible realizations of $T_{\chi_1}$-evolution.
For example, we underestimate $T_{\rm min}$ by taking the highest possible value for $T_{\chi_1}$, i.e., the asymptotic solution Eq.~\eqref{eq:Tratioasy}.
At the same time, we overestimate $T_{\rm dec,self}$ by taking the lowest possible value, i.e., the evolution in the absence of DM self-heating Eq.~\eqref{eq:Tchi1noSH}.
Note the the gray circles in the right panel of Figure~\ref{fig:Tdmevolapp} are the true decoupling points of self-scattering according to the numerical solution of $T_{\chi_1}$.
In the current case, the true $T_{\rm dec,self}$ is well approximated by Eq.~\eqref{eq:Tself} since we can assume the asymptotic solution around the decoupling point.

\item {$T_{\rm min}<10\,{\rm keV}$ and $T_{\rm min}<T_{\rm dec,self}$} :
The dashed and the dotted curve in the right panel of Figure~\ref{fig:Tdmevolapp} correspond to this case.
Neither the photo-dissociation or the CMB constraints can be robustly put.
Since $T_{\rm min}<10\,{\rm keV}$, $T_{\chi_1}$ may not follow the asymptotic solution throughout the photodissociation epoch;
further dedicated analyses on both $T_{\chi_1}$-evolution and photo-dissociation processes are required to put a robust constraint.
Since $T_{\rm min}<T_{\rm dec,self}$, $T_{\chi_1}$ may not reach the asymptotic solution by the decoupling of self-scattering;
further dedicated analysis on $T_{\chi_1}$-evolution is needed to estimate $T_{\chi_1}$ at the last scattering.
We remark that the robustness conditions we required are at the most conservative level, and it does not mean that the constraints are vanishing when the conditions are not satisfied. 
There may still be parameter region that can be constrained.
The dashed curve represents this case.
We find that $T_{\chi_1}$ undergoes self-heating during the photo-dissociation epoch, and $T_{\chi_1}$ reaches the asymptotic solution before the true decoupling point of self-scattering.
Note that the point depicted by red circles represent $T_{\rm sh}$ (Eq.~\eqref{eq:Tsh}), which is the point where the heating term starts to dominate as a source and the DM self-heating starts afterwards.
For the dashed curve, $T_{\rm sh}=T_{\rm stop}$;
on the contrary, $T_{\rm sh}=T_{\rm heat}$ for the solid curve.
The red circle is also the deviation point from $T_{\chi_1}=T$, which is estimated by $T_{\rm max}$:
\begin{equation}
T_{\rm max}=\max\left[T_{\rm dec,el},T_{\rm sh}\right]\,,
\label{eq:Tmax}
\end{equation}
where $T_{\rm dec,el}$ and $T_{\rm sh}$ is determined by assuming $T_{\chi_1}=T$.
If $T_{\rm dec,el}>T_{\rm sh}$, the heating term is still ignorable for $T_{\rm sh} \lesssim T \lesssim T_{\rm dec,el}$ and $T_{\chi_1}\propto1/a^2$ in such interval.
We find that for most of the parameter region, we may estimate $T_{\rm max}=T_{\rm sh}$.
We require $T_{\rm sh}>100\,{\rm eV}$ for the potentially-constrained region by the photo-dissociation constraints;
the DM self-heating overlaps with the relevant epoch $100\,{\rm eV}\lesssim T\lesssim10\,{\rm keV}$.
On the other hand, the dotted curve represents the case of no DM self-heating;
the heating term never becomes important as a source in the Boltzmann equation before the decoupling of self-scattering, i.e., $T_{\rm sh}<T_{\rm dec,self}$.
In this case, $T_{\chi_1}=T$ until the decoupling of $\chi_1$-sm elastic scattering $T_{\rm dec,el}$ (depicted by cyan circle) and redshift as $T_{\chi_1}\propto1/a^2$ afterwards.
We require $T_{\rm sh}>T_{\rm dec,self}$ for the potentially-constrained region by the CMB constraints;
there may be an epoch where $T_{\chi_1}$ is enhanced through the DM self-heating before the last scattering.
The estimations of $T_{\rm sh}$ and $T_{\rm dec,self}$ for this condition are discussed in the next section.

\item {$T_{\rm sh}<T_{\rm dec,self}$} :
The dot-dashed curve in the right panel of Figure~\ref{fig:Tdmevolapp} corresponds to this case.
If $T_{\rm sh}<T_{\rm dec,self}$, the heating term never becomes relevant for the evolution of $T_{\chi_1}$, and the temperature evolution is well approximated by Eq.~\eqref{eq:Tchi1noSH};
$T_{\chi_1}=T$ until the decoupling of $\chi_1$-sm elastic scattering $T_{\rm dec,el}$ (depicted by cyan circle) and $T_{\chi_1}\propto1/a^2$ afterwards.
We require the condition at the most conservative level, i.e., we overestimate (underestimate) $T_{\rm sh}$ ($T_{\rm dec,self}$) using Eq.~\eqref{eq:Tchi1noSH} (Eq.~\eqref{eq:naiveTchi1}) for $T_{\chi_1}$.
Note that if $T_{\rm sh}>T_{\rm dec,self}$, it means there may exist DM self-heating epoch, and hence may be potentially probed by CMB.
Although there is no self-heating in this case, the $T_{\rm dec,el}$ is significantly delayed due to strong $\chi_1$-sm interaction, i.e. $\sigma_{\chi_1 e}$ is enhanced as $1/r_1^3$.
The CMB constraints on DM annihilation appears in the small $r_1$ limit because of the enhanced annihilation cross section and the delayed $T_{\rm dec,el}$.

\end{itemize}

\bibliographystyle{utphys}
\bibliography{ref}

\providecommand{\href}[2]{#2}\begingroup\raggedright\begin{thebibliography}{100}

\bibitem{Arcadi:2017kky}
G.~Arcadi, M.~Dutra, P.~Ghosh, M.~Lindner, Y.~Mambrini, M.~Pierre, S.~Profumo,
  and F.~S. Queiroz, ``{The waning of the WIMP? A review of models, searches,
  and constraints},''
  \href{http://dx.doi.org/10.1140/epjc/s10052-018-5662-y}{{\em Eur. Phys. J. C}
  {\bfseries 78} no.~3, (2018) 203},
  \href{http://arxiv.org/abs/1703.07364}{{\ttfamily arXiv:1703.07364
  [hep-ph]}}.

\bibitem{Roszkowski:2017nbc}
L.~Roszkowski, E.~M. Sessolo, and S.~Trojanowski, ``{WIMP dark matter
  candidates and searches\textemdash{}current status and future prospects},''
  \href{http://dx.doi.org/10.1088/1361-6633/aab913}{{\em Rept. Prog. Phys.}
  {\bfseries 81} no.~6, (2018) 066201},
  \href{http://arxiv.org/abs/1707.06277}{{\ttfamily arXiv:1707.06277
  [hep-ph]}}.

\bibitem{Tucker-Smith:2001myb}
D.~Tucker-Smith and N.~Weiner, ``{Inelastic dark matter},''
  \href{http://dx.doi.org/10.1103/PhysRevD.64.043502}{{\em Phys. Rev. D}
  {\bfseries 64} (2001) 043502},
  \href{http://arxiv.org/abs/hep-ph/0101138}{{\ttfamily arXiv:hep-ph/0101138}}.

\bibitem{Loeb:2010gj}
A.~Loeb and N.~Weiner, ``{Cores in Dwarf Galaxies from Dark Matter with a
  Yukawa Potential},''
  \href{http://dx.doi.org/10.1103/PhysRevLett.106.171302}{{\em Phys. Rev.
  Lett.} {\bfseries 106} (2011) 171302},
  \href{http://arxiv.org/abs/1011.6374}{{\ttfamily arXiv:1011.6374
  [astro-ph.CO]}}.

\bibitem{Schutz:2014nka}
K.~Schutz and T.~R. Slatyer, ``{Self-Scattering for Dark Matter with an Excited
  State},'' \href{http://dx.doi.org/10.1088/1475-7516/2015/01/021}{{\em JCAP}
  {\bfseries 01} (2015) 021}, \href{http://arxiv.org/abs/1409.2867}{{\ttfamily
  arXiv:1409.2867 [hep-ph]}}.

\bibitem{McDermott:2017vyk}
S.~D. McDermott, ``{Is Self-Interacting Dark Matter Undergoing Dark Fusion?},''
  \href{http://dx.doi.org/10.1103/PhysRevLett.120.221806}{{\em Phys. Rev.
  Lett.} {\bfseries 120} no.~22, (2018) 221806},
  \href{http://arxiv.org/abs/1711.00857}{{\ttfamily arXiv:1711.00857
  [hep-ph]}}.

\bibitem{Chu:2018nki}
X.~Chu and C.~Garcia-Cely, ``{Core formation from self-heating dark matter},''
  \href{http://dx.doi.org/10.1088/1475-7516/2018/07/013}{{\em JCAP} {\bfseries
  07} (2018) 013}, \href{http://arxiv.org/abs/1803.09762}{{\ttfamily
  arXiv:1803.09762 [hep-ph]}}.

\bibitem{Vogelsberger:2018bok}
M.~Vogelsberger, J.~Zavala, K.~Schutz, and T.~R. Slatyer, ``{Evaporating the
  Milky Way halo and its satellites with inelastic self-interacting dark
  matter},'' \href{http://arxiv.org/abs/1805.03203}{{\ttfamily arXiv:1805.03203
  [astro-ph.GA]}}.

\bibitem{Kamada:2019wjo}
A.~Kamada and H.~J. Kim, ``{Escalating core formation with dark matter
  self-heating},'' \href{http://dx.doi.org/10.1103/PhysRevD.102.043009}{{\em
  Phys. Rev. D} {\bfseries 102} no.~4, (2020) 043009},
  \href{http://arxiv.org/abs/1911.09717}{{\ttfamily arXiv:1911.09717
  [hep-ph]}}.

\bibitem{Chua:2020svq}
K.~T.~E. Chua, K.~Dibert, M.~Vogelsberger, and J.~Zavala, ``{The impact of
  inelastic self-interacting dark matter on the dark matter structure of a
  Milky Way halo},'' \href{http://dx.doi.org/10.1093/mnras/staa3315}{{\em Mon.
  Not. Roy. Astron. Soc.} {\bfseries 500} no.~1, (2020) 1531--1546},
  \href{http://arxiv.org/abs/2010.08562}{{\ttfamily arXiv:2010.08562
  [astro-ph.GA]}}.

\bibitem{Pollack:2014rja}
J.~Pollack, D.~N. Spergel, and P.~J. Steinhardt, ``{Supermassive Black Holes
  from Ultra-Strongly Self-Interacting Dark Matter},''
  \href{http://dx.doi.org/10.1088/0004-637X/804/2/131}{{\em Astrophys. J.}
  {\bfseries 804} no.~2, (2015) 131},
  \href{http://arxiv.org/abs/1501.00017}{{\ttfamily arXiv:1501.00017
  [astro-ph.CO]}}.

\bibitem{Choquette:2018lvq}
J.~Choquette, J.~M. Cline, and J.~M. Cornell, ``{Early formation of
  supermassive black holes via dark matter self-interactions},''
  \href{http://dx.doi.org/10.1088/1475-7516/2019/07/036}{{\em JCAP} {\bfseries
  07} (2019) 036}, \href{http://arxiv.org/abs/1812.05088}{{\ttfamily
  arXiv:1812.05088 [astro-ph.CO]}}.

\bibitem{Jo:2020ggs}
B.~Jo, H.~Kim, H.~D. Kim, and C.~S. Shin, ``{Exploring the Universe with dark
  light scalars},'' \href{http://dx.doi.org/10.1103/PhysRevD.103.083528}{{\em
  Phys. Rev. D} {\bfseries 103} no.~8, (2021) 083528},
  \href{http://arxiv.org/abs/2010.10880}{{\ttfamily arXiv:2010.10880
  [hep-ph]}}.

\bibitem{Agashe:2014yua}
K.~Agashe, Y.~Cui, L.~Necib, and J.~Thaler, ``{(In)direct Detection of Boosted
  Dark Matter},'' \href{http://dx.doi.org/10.1088/1475-7516/2014/10/062}{{\em
  JCAP} {\bfseries 10} (2014) 062},
  \href{http://arxiv.org/abs/1405.7370}{{\ttfamily arXiv:1405.7370 [hep-ph]}}.

\bibitem{Bhattacharya:2014yha}
A.~Bhattacharya, R.~Gandhi, and A.~Gupta, ``{The Direct Detection of Boosted
  Dark Matter at High Energies and PeV events at IceCube},''
  \href{http://dx.doi.org/10.1088/1475-7516/2015/03/027}{{\em JCAP} {\bfseries
  03} (2015) 027}, \href{http://arxiv.org/abs/1407.3280}{{\ttfamily
  arXiv:1407.3280 [hep-ph]}}.

\bibitem{Kong:2014mia}
K.~Kong, G.~Mohlabeng, and J.-C. Park, ``{Boosted dark matter signals uplifted
  with self-interaction},''
  \href{http://dx.doi.org/10.1016/j.physletb.2015.02.057}{{\em Phys. Lett. B}
  {\bfseries 743} (2015) 256--266},
  \href{http://arxiv.org/abs/1411.6632}{{\ttfamily arXiv:1411.6632 [hep-ph]}}.

\bibitem{Necib:2016aez}
L.~Necib, J.~Moon, T.~Wongjirad, and J.~M. Conrad, ``{Boosted Dark Matter at
  Neutrino Experiments},''
  \href{http://dx.doi.org/10.1103/PhysRevD.95.075018}{{\em Phys. Rev. D}
  {\bfseries 95} no.~7, (2017) 075018},
  \href{http://arxiv.org/abs/1610.03486}{{\ttfamily arXiv:1610.03486
  [hep-ph]}}.

\bibitem{Alhazmi:2016qcs}
H.~Alhazmi, K.~Kong, G.~Mohlabeng, and J.-C. Park, ``{Boosted Dark Matter at
  the Deep Underground Neutrino Experiment},''
  \href{http://dx.doi.org/10.1007/JHEP04(2017)158}{{\em JHEP} {\bfseries 04}
  (2017) 158}, \href{http://arxiv.org/abs/1611.09866}{{\ttfamily
  arXiv:1611.09866 [hep-ph]}}.

\bibitem{Kim:2016zjx}
D.~Kim, J.-C. Park, and S.~Shin, ``{Dark Matter
  \textquotedblleft{}Collider\textquotedblright{} from Inelastic Boosted Dark
  Matter},'' \href{http://dx.doi.org/10.1103/PhysRevLett.119.161801}{{\em Phys.
  Rev. Lett.} {\bfseries 119} no.~16, (2017) 161801},
  \href{http://arxiv.org/abs/1612.06867}{{\ttfamily arXiv:1612.06867
  [hep-ph]}}.

\bibitem{Giudice:2017zke}
G.~F. Giudice, D.~Kim, J.-C. Park, and S.~Shin, ``{Inelastic Boosted Dark
  Matter at Direct Detection Experiments},''
  \href{http://dx.doi.org/10.1016/j.physletb.2018.03.043}{{\em Phys. Lett. B}
  {\bfseries 780} (2018) 543--552},
  \href{http://arxiv.org/abs/1712.07126}{{\ttfamily arXiv:1712.07126
  [hep-ph]}}.

\bibitem{Chatterjee:2018mej}
A.~Chatterjee, A.~De~Roeck, D.~Kim, Z.~G. Moghaddam, J.-C. Park, S.~Shin, L.~H.
  Whitehead, and J.~Yu, ``{Searching for boosted dark matter at ProtoDUNE},''
  \href{http://dx.doi.org/10.1103/PhysRevD.98.075027}{{\em Phys. Rev. D}
  {\bfseries 98} no.~7, (2018) 075027},
  \href{http://arxiv.org/abs/1803.03264}{{\ttfamily arXiv:1803.03264
  [hep-ph]}}.

\bibitem{Kim:2018veo}
D.~Kim, K.~Kong, J.-C. Park, and S.~Shin, ``{Boosted Dark Matter Quarrying at
  Surface Neutrino Detectors},''
  \href{http://dx.doi.org/10.1007/JHEP08(2018)155}{{\em JHEP} {\bfseries 08}
  (2018) 155}, \href{http://arxiv.org/abs/1804.07302}{{\ttfamily
  arXiv:1804.07302 [hep-ph]}}.

\bibitem{Kim:2019had}
D.~Kim, J.-C. Park, and S.~Shin, ``{Searching for boosted dark matter via
  dark-photon bremsstrahlung},''
  \href{http://dx.doi.org/10.1103/PhysRevD.100.035033}{{\em Phys. Rev. D}
  {\bfseries 100} no.~3, (2019) 035033},
  \href{http://arxiv.org/abs/1903.05087}{{\ttfamily arXiv:1903.05087
  [hep-ph]}}.

\bibitem{Heurtier:2019rkz}
L.~Heurtier, D.~Kim, J.-C. Park, and S.~Shin, ``{Explaining the ANITA Anomaly
  with Inelastic Boosted Dark Matter},''
  \href{http://dx.doi.org/10.1103/PhysRevD.100.055004}{{\em Phys. Rev. D}
  {\bfseries 100} no.~5, (2019) 055004},
  \href{http://arxiv.org/abs/1905.13223}{{\ttfamily arXiv:1905.13223
  [hep-ph]}}.

\bibitem{Kim:2020ipj}
D.~Kim, P.~A.~N. Machado, J.-C. Park, and S.~Shin, ``{Optimizing Energetic
  Light Dark Matter Searches in Dark Matter and Neutrino Experiments},''
  \href{http://dx.doi.org/10.1007/JHEP07(2020)057}{{\em JHEP} {\bfseries 07}
  (2020) 057}, \href{http://arxiv.org/abs/2003.07369}{{\ttfamily
  arXiv:2003.07369 [hep-ph]}}.

\bibitem{DeRoeck:2020ntj}
A.~De~Roeck, D.~Kim, Z.~G. Moghaddam, J.-C. Park, S.~Shin, and L.~H. Whitehead,
  ``{Probing Energetic Light Dark Matter with Multi-Particle Tracks Signatures
  at DUNE},'' \href{http://dx.doi.org/10.1007/JHEP11(2020)043}{{\em JHEP}
  {\bfseries 11} (2020) 043}, \href{http://arxiv.org/abs/2005.08979}{{\ttfamily
  arXiv:2005.08979 [hep-ph]}}.

\bibitem{Alhazmi:2020fju}
H.~Alhazmi, D.~Kim, K.~Kong, G.~Mohlabeng, J.-C. Park, and S.~Shin,
  ``{Implications of the XENON1T Excess on the Dark Matter Interpretation},''
  \href{http://dx.doi.org/10.1007/JHEP05(2021)055}{{\em JHEP} {\bfseries 05}
  (2021) 055}, \href{http://arxiv.org/abs/2006.16252}{{\ttfamily
  arXiv:2006.16252 [hep-ph]}}.

\bibitem{Cline:2013gha}
J.~M. Cline, K.~Kainulainen, P.~Scott, and C.~Weniger, ``{Update on scalar
  singlet dark matter},''
  \href{http://dx.doi.org/10.1103/PhysRevD.88.055025}{{\em Phys. Rev. D}
  {\bfseries 88} (2013) 055025},
  \href{http://arxiv.org/abs/1306.4710}{{\ttfamily arXiv:1306.4710 [hep-ph]}}.
  [Erratum: Phys.Rev.D 92, 039906 (2015)].

\bibitem{Athron:2017kgt}
{\bfseries GAMBIT} Collaboration, P.~Athron {\em et~al.}, ``{Status of the
  scalar singlet dark matter model},''
  \href{http://dx.doi.org/10.1140/epjc/s10052-017-5113-1}{{\em Eur. Phys. J. C}
  {\bfseries 77} no.~8, (2017) 568},
  \href{http://arxiv.org/abs/1705.07931}{{\ttfamily arXiv:1705.07931
  [hep-ph]}}.

\bibitem{Athron:2018hpc}
{\bfseries GAMBIT} Collaboration, P.~Athron {\em et~al.}, ``{Global analyses of
  Higgs portal singlet dark matter models using GAMBIT},''
  \href{http://dx.doi.org/10.1140/epjc/s10052-018-6513-6}{{\em Eur. Phys. J. C}
  {\bfseries 79} no.~1, (2019) 38},
  \href{http://arxiv.org/abs/1808.10465}{{\ttfamily arXiv:1808.10465
  [hep-ph]}}.

\bibitem{Arcadi:2019lka}
G.~Arcadi, A.~Djouadi, and M.~Raidal, ``{Dark Matter through the Higgs
  portal},'' \href{http://dx.doi.org/10.1016/j.physrep.2019.11.003}{{\em Phys.
  Rept.} {\bfseries 842} (2020) 1--180},
  \href{http://arxiv.org/abs/1903.03616}{{\ttfamily arXiv:1903.03616
  [hep-ph]}}.

\bibitem{Belanger:2011ww}
G.~Belanger and J.-C. Park, ``{Assisted freeze-out},''
  \href{http://dx.doi.org/10.1088/1475-7516/2012/03/038}{{\em JCAP} {\bfseries
  03} (2012) 038}, \href{http://arxiv.org/abs/1112.4491}{{\ttfamily
  arXiv:1112.4491 [hep-ph]}}.

\bibitem{Kamada:2017gfc}
A.~Kamada, H.~J. Kim, H.~Kim, and T.~Sekiguchi, ``{Self-Heating Dark Matter via
  Semiannihilation},''
  \href{http://dx.doi.org/10.1103/PhysRevLett.120.131802}{{\em Phys. Rev.
  Lett.} {\bfseries 120} no.~13, (2018) 131802},
  \href{http://arxiv.org/abs/1707.09238}{{\ttfamily arXiv:1707.09238
  [hep-ph]}}.

\bibitem{Kamada:2018hte}
A.~Kamada, H.~J. Kim, and H.~Kim, ``{Self-heating of Strongly Interacting
  Massive Particles},''
  \href{http://dx.doi.org/10.1103/PhysRevD.98.023509}{{\em Phys. Rev. D}
  {\bfseries 98} no.~2, (2018) 023509},
  \href{http://arxiv.org/abs/1805.05648}{{\ttfamily arXiv:1805.05648
  [hep-ph]}}.

\bibitem{Izaguirre:2013uxa}
E.~Izaguirre, G.~Krnjaic, P.~Schuster, and N.~Toro, ``{New Electron Beam-Dump
  Experiments to Search for MeV to few-GeV Dark Matter},''
  \href{http://dx.doi.org/10.1103/PhysRevD.88.114015}{{\em Phys. Rev. D}
  {\bfseries 88} (2013) 114015},
  \href{http://arxiv.org/abs/1307.6554}{{\ttfamily arXiv:1307.6554 [hep-ph]}}.

\bibitem{Depta:2019lbe}
P.~F. Depta, M.~Hufnagel, K.~Schmidt-Hoberg, and S.~Wild, ``{BBN constraints on
  the annihilation of MeV-scale dark matter},''
  \href{http://dx.doi.org/10.1088/1475-7516/2019/04/029}{{\em JCAP} {\bfseries
  04} (2019) 029}, \href{http://arxiv.org/abs/1901.06944}{{\ttfamily
  arXiv:1901.06944 [hep-ph]}}.

\bibitem{Kolb:1990vq}
E.~W. Kolb and M.~S. Turner, {\em {The Early Universe}}, vol.~69.
\newblock 1990.

\bibitem{Maity:2019hre}
T.~N. Maity and T.~S. Ray, ``{Exchange driven freeze out of dark matter},''
  \href{http://dx.doi.org/10.1103/PhysRevD.101.103013}{{\em Phys. Rev. D}
  {\bfseries 101} no.~10, (2020) 103013},
  \href{http://arxiv.org/abs/1908.10343}{{\ttfamily arXiv:1908.10343
  [hep-ph]}}.

\bibitem{Saez:2021oxl}
B.~D. S\'aez, K.~M\"ohling, and D.~St\"ockinger, ``{Two Real Scalar WIMP Model
  in the Assisted Freeze-Out Scenario},''
  \href{http://arxiv.org/abs/2103.17064}{{\ttfamily arXiv:2103.17064
  [hep-ph]}}.

\bibitem{Sabti:2019mhn}
N.~Sabti, J.~Alvey, M.~Escudero, M.~Fairbairn, and D.~Blas, ``{Refined Bounds
  on MeV-scale Thermal Dark Sectors from BBN and the CMB},''
  \href{http://dx.doi.org/10.1088/1475-7516/2020/01/004}{{\em JCAP} {\bfseries
  01} (2020) 004}, \href{http://arxiv.org/abs/1910.01649}{{\ttfamily
  arXiv:1910.01649 [hep-ph]}}.

\bibitem{Aghanim:2018eyx}
{\bfseries Planck} Collaboration, N.~Aghanim {\em et~al.}, ``{Planck 2018
  results. VI. Cosmological parameters},''
  \href{http://dx.doi.org/10.1051/0004-6361/201833910}{{\em Astron. Astrophys.}
  {\bfseries 641} (2020) A6}, \href{http://arxiv.org/abs/1807.06209}{{\ttfamily
  arXiv:1807.06209 [astro-ph.CO]}}.

\bibitem{Riess:2019cxk}
A.~G. Riess, S.~Casertano, W.~Yuan, L.~M. Macri, and D.~Scolnic, ``{Large
  Magellanic Cloud Cepheid Standards Provide a 1\% Foundation for the
  Determination of the Hubble Constant and Stronger Evidence for Physics beyond
  $\Lambda$CDM},'' \href{http://dx.doi.org/10.3847/1538-4357/ab1422}{{\em
  Astrophys. J.} {\bfseries 876} no.~1, (2019) 85},
  \href{http://arxiv.org/abs/1903.07603}{{\ttfamily arXiv:1903.07603
  [astro-ph.CO]}}.

\bibitem{Tanabashi:2018oca}
{\bfseries Particle Data Group} Collaboration, M.~Tanabashi {\em et~al.},
  ``{Review of Particle Physics},''
  \href{http://dx.doi.org/10.1103/PhysRevD.98.030001}{{\em Phys. Rev. D}
  {\bfseries 98} no.~3, (2018) 030001}.

\bibitem{Pitrou:2018cgg}
C.~Pitrou, A.~Coc, J.-P. Uzan, and E.~Vangioni, ``{Precision big bang
  nucleosynthesis with improved Helium-4 predictions},''
  \href{http://dx.doi.org/10.1016/j.physrep.2018.04.005}{{\em Phys. Rept.}
  {\bfseries 754} (2018) 1--66},
  \href{http://arxiv.org/abs/1801.08023}{{\ttfamily arXiv:1801.08023
  [astro-ph.CO]}}.

\bibitem{Kawasaki:2015yya}
M.~Kawasaki, K.~Kohri, T.~Moroi, and Y.~Takaesu, ``{Revisiting Big-Bang
  Nucleosynthesis Constraints on Dark-Matter Annihilation},''
  \href{http://dx.doi.org/10.1016/j.physletb.2015.10.048}{{\em Phys. Lett. B}
  {\bfseries 751} (2015) 246--250},
  \href{http://arxiv.org/abs/1509.03665}{{\ttfamily arXiv:1509.03665
  [hep-ph]}}.

\bibitem{Protheroe:1994dt}
R.~J. Protheroe, T.~Stanev, and V.~S. Berezinsky, ``{Electromagnetic cascades
  and cascade nucleosynthesis in the early universe},''
  \href{http://dx.doi.org/10.1103/PhysRevD.51.4134}{{\em Phys. Rev. D}
  {\bfseries 51} (1995) 4134--4144},
  \href{http://arxiv.org/abs/astro-ph/9409004}{{\ttfamily
  arXiv:astro-ph/9409004}}.

\bibitem{Kawasaki:1994af}
M.~Kawasaki and T.~Moroi, ``{Gravitino production in the inflationary universe
  and the effects on big bang nucleosynthesis},''
  \href{http://dx.doi.org/10.1143/PTP.93.879}{{\em Prog. Theor. Phys.}
  {\bfseries 93} (1995) 879--900},
  \href{http://arxiv.org/abs/hep-ph/9403364}{{\ttfamily arXiv:hep-ph/9403364}}.

\bibitem{Cyburt:2002uv}
R.~H. Cyburt, J.~R. Ellis, B.~D. Fields, and K.~A. Olive, ``{Updated
  nucleosynthesis constraints on unstable relic particles},''
  \href{http://dx.doi.org/10.1103/PhysRevD.67.103521}{{\em Phys. Rev. D}
  {\bfseries 67} (2003) 103521},
  \href{http://arxiv.org/abs/astro-ph/0211258}{{\ttfamily
  arXiv:astro-ph/0211258}}.

\bibitem{Hufnagel:2018bjp}
M.~Hufnagel, K.~Schmidt-Hoberg, and S.~Wild, ``{BBN constraints on MeV-scale
  dark sectors. Part II. Electromagnetic decays},''
  \href{http://dx.doi.org/10.1088/1475-7516/2018/11/032}{{\em JCAP} {\bfseries
  11} (2018) 032}, \href{http://arxiv.org/abs/1808.09324}{{\ttfamily
  arXiv:1808.09324 [hep-ph]}}.

\bibitem{Poulin:2015woa}
V.~Poulin and P.~D. Serpico, ``{Loophole to the Universal Photon Spectrum in
  Electromagnetic Cascades and Application to the Cosmological Lithium
  Problem},'' \href{http://dx.doi.org/10.1103/PhysRevLett.114.091101}{{\em
  Phys. Rev. Lett.} {\bfseries 114} no.~9, (2015) 091101},
  \href{http://arxiv.org/abs/1502.01250}{{\ttfamily arXiv:1502.01250
  [astro-ph.CO]}}.

\bibitem{Padmanabhan:2005es}
N.~Padmanabhan and D.~P. Finkbeiner, ``{Detecting dark matter annihilation with
  CMB polarization: Signatures and experimental prospects},''
  \href{http://dx.doi.org/10.1103/PhysRevD.72.023508}{{\em Phys. Rev. D}
  {\bfseries 72} (2005) 023508},
  \href{http://arxiv.org/abs/astro-ph/0503486}{{\ttfamily
  arXiv:astro-ph/0503486}}.

\bibitem{Green:2018pmd}
D.~Green, P.~D. Meerburg, and J.~Meyers, ``{Aspects of Dark Matter Annihilation
  in Cosmology},'' \href{http://dx.doi.org/10.1088/1475-7516/2019/04/025}{{\em
  JCAP} {\bfseries 04} (2019) 025},
  \href{http://arxiv.org/abs/1804.01055}{{\ttfamily arXiv:1804.01055
  [astro-ph.CO]}}.

\bibitem{Essig:2013goa}
R.~Essig, E.~Kuflik, S.~D. McDermott, T.~Volansky, and K.~M. Zurek,
  ``{Constraining Light Dark Matter with Diffuse X-Ray and Gamma-Ray
  Observations},'' \href{http://dx.doi.org/10.1007/JHEP11(2013)193}{{\em JHEP}
  {\bfseries 11} (2013) 193}, \href{http://arxiv.org/abs/1309.4091}{{\ttfamily
  arXiv:1309.4091 [hep-ph]}}.

\bibitem{Cirelli:2020bpc}
M.~Cirelli, N.~Fornengo, B.~J. Kavanagh, and E.~Pinetti, ``{Integral X-ray
  constraints on sub-GeV Dark Matter},''
  \href{http://dx.doi.org/10.1103/PhysRevD.103.063022}{{\em Phys. Rev. D}
  {\bfseries 103} no.~6, (2021) 063022},
  \href{http://arxiv.org/abs/2007.11493}{{\ttfamily arXiv:2007.11493
  [hep-ph]}}.

\bibitem{Angle:2011th}
{\bfseries XENON10} Collaboration, J.~Angle {\em et~al.}, ``{A search for light
  dark matter in XENON10 data},''
  \href{http://dx.doi.org/10.1103/PhysRevLett.107.051301}{{\em Phys. Rev.
  Lett.} {\bfseries 107} (2011) 051301},
  \href{http://arxiv.org/abs/1104.3088}{{\ttfamily arXiv:1104.3088
  [astro-ph.CO]}}. [Erratum: Phys.Rev.Lett. 110, 249901 (2013)].

\bibitem{Aprile:2016wwo}
{\bfseries XENON} Collaboration, E.~Aprile {\em et~al.}, ``{Low-mass dark
  matter search using ionization signals in XENON100},''
  \href{http://dx.doi.org/10.1103/PhysRevD.94.092001}{{\em Phys. Rev. D}
  {\bfseries 94} no.~9, (2016) 092001},
  \href{http://arxiv.org/abs/1605.06262}{{\ttfamily arXiv:1605.06262
  [astro-ph.CO]}}. [Erratum: Phys.Rev.D 95, 059901 (2017)].

\bibitem{Essig:2017kqs}
R.~Essig, T.~Volansky, and T.-T. Yu, ``{New Constraints and Prospects for
  sub-GeV Dark Matter Scattering off Electrons in Xenon},''
  \href{http://dx.doi.org/10.1103/PhysRevD.96.043017}{{\em Phys. Rev. D}
  {\bfseries 96} no.~4, (2017) 043017},
  \href{http://arxiv.org/abs/1703.00910}{{\ttfamily arXiv:1703.00910
  [hep-ph]}}.

\bibitem{Agnes:2018oej}
{\bfseries DarkSide} Collaboration, P.~Agnes {\em et~al.}, ``{Constraints on
  Sub-GeV Dark-Matter\textendash{}Electron Scattering from the DarkSide-50
  Experiment},'' \href{http://dx.doi.org/10.1103/PhysRevLett.121.111303}{{\em
  Phys. Rev. Lett.} {\bfseries 121} no.~11, (2018) 111303},
  \href{http://arxiv.org/abs/1802.06998}{{\ttfamily arXiv:1802.06998
  [astro-ph.CO]}}.

\bibitem{Agnese:2018col}
{\bfseries SuperCDMS} Collaboration, R.~Agnese {\em et~al.}, ``{First Dark
  Matter Constraints from a SuperCDMS Single-Charge Sensitive Detector},''
  \href{http://dx.doi.org/10.1103/PhysRevLett.121.051301}{{\em Phys. Rev.
  Lett.} {\bfseries 121} no.~5, (2018) 051301},
  \href{http://arxiv.org/abs/1804.10697}{{\ttfamily arXiv:1804.10697
  [hep-ex]}}. [Erratum: Phys.Rev.Lett. 122, 069901 (2019)].

\bibitem{Crisler:2018gci}
{\bfseries SENSEI} Collaboration, M.~Crisler, R.~Essig, J.~Estrada,
  G.~Fernandez, J.~Tiffenberg, M.~Sofo~haro, T.~Volansky, and T.-T. Yu,
  ``{SENSEI: First Direct-Detection Constraints on sub-GeV Dark Matter from a
  Surface Run},'' \href{http://dx.doi.org/10.1103/PhysRevLett.121.061803}{{\em
  Phys. Rev. Lett.} {\bfseries 121} no.~6, (2018) 061803},
  \href{http://arxiv.org/abs/1804.00088}{{\ttfamily arXiv:1804.00088
  [hep-ex]}}.

\bibitem{Abramoff:2019dfb}
{\bfseries SENSEI} Collaboration, O.~Abramoff {\em et~al.}, ``{SENSEI:
  Direct-Detection Constraints on Sub-GeV Dark Matter from a Shallow
  Underground Run Using a Prototype Skipper-CCD},''
  \href{http://dx.doi.org/10.1103/PhysRevLett.122.161801}{{\em Phys. Rev.
  Lett.} {\bfseries 122} no.~16, (2019) 161801},
  \href{http://arxiv.org/abs/1901.10478}{{\ttfamily arXiv:1901.10478
  [hep-ex]}}.

\bibitem{Emken:2019tni}
T.~Emken, R.~Essig, C.~Kouvaris, and M.~Sholapurkar, ``{Direct Detection of
  Strongly Interacting Sub-GeV Dark Matter via Electron Recoils},''
  \href{http://dx.doi.org/10.1088/1475-7516/2019/09/070}{{\em JCAP} {\bfseries
  09} (2019) 070}, \href{http://arxiv.org/abs/1905.06348}{{\ttfamily
  arXiv:1905.06348 [hep-ph]}}.

\bibitem{Dvorkin:2013cea}
C.~Dvorkin, K.~Blum, and M.~Kamionkowski, ``{Constraining Dark Matter-Baryon
  Scattering with Linear Cosmology},''
  \href{http://dx.doi.org/10.1103/PhysRevD.89.023519}{{\em Phys. Rev. D}
  {\bfseries 89} no.~2, (2014) 023519},
  \href{http://arxiv.org/abs/1311.2937}{{\ttfamily arXiv:1311.2937
  [astro-ph.CO]}}.

\bibitem{Binder:2016pnr}
T.~Binder, L.~Covi, A.~Kamada, H.~Murayama, T.~Takahashi, and N.~Yoshida,
  ``{Matter Power Spectrum in Hidden Neutrino Interacting Dark Matter Models: A
  Closer Look at the Collision Term},''
  \href{http://dx.doi.org/10.1088/1475-7516/2016/11/043}{{\em JCAP} {\bfseries
  11} (2016) 043}, \href{http://arxiv.org/abs/1602.07624}{{\ttfamily
  arXiv:1602.07624 [hep-ph]}}.

\bibitem{Boddy:2018wzy}
K.~K. Boddy, V.~Gluscevic, V.~Poulin, E.~D. Kovetz, M.~Kamionkowski, and
  R.~Barkana, ``{Critical assessment of CMB limits on dark matter-baryon
  scattering: New treatment of the relative bulk velocity},''
  \href{http://dx.doi.org/10.1103/PhysRevD.98.123506}{{\em Phys. Rev. D}
  {\bfseries 98} no.~12, (2018) 123506},
  \href{http://arxiv.org/abs/1808.00001}{{\ttfamily arXiv:1808.00001
  [astro-ph.CO]}}.

\bibitem{Binder:2017rgn}
T.~Binder, T.~Bringmann, M.~Gustafsson, and A.~Hryczuk, ``{Early kinetic
  decoupling of dark matter: when the standard way of calculating the thermal
  relic density fails},''
  \href{http://dx.doi.org/10.1103/PhysRevD.96.115010}{{\em Phys. Rev. D}
  {\bfseries 96} no.~11, (2017) 115010},
  \href{http://arxiv.org/abs/1706.07433}{{\ttfamily arXiv:1706.07433
  [astro-ph.CO]}}. [Erratum: Phys.Rev.D 101, 099901 (2020)].

\bibitem{Harada:2014lma}
A.~Harada and A.~Kamada, ``{Structure formation in a mixed dark matter model
  with decaying sterile neutrino: the 3.5 keV X-ray line and the Galactic
  substructure},'' \href{http://dx.doi.org/10.1088/1475-7516/2016/01/031}{{\em
  JCAP} {\bfseries 01} (2016) 031},
  \href{http://arxiv.org/abs/1412.1592}{{\ttfamily arXiv:1412.1592
  [astro-ph.CO]}}.

\bibitem{Dienes:2020bmn}
K.~R. Dienes, F.~Huang, J.~Kost, S.~Su, and B.~Thomas, ``{Deciphering the
  archaeological record: Cosmological imprints of nonminimal dark sectors},''
  \href{http://dx.doi.org/10.1103/PhysRevD.101.123511}{{\em Phys. Rev. D}
  {\bfseries 101} no.~12, (2020) 123511},
  \href{http://arxiv.org/abs/2001.02193}{{\ttfamily arXiv:2001.02193
  [astro-ph.CO]}}.

\bibitem{Press:1973iz}
W.~H. Press and P.~Schechter, ``{Formation of galaxies and clusters of galaxies
  by selfsimilar gravitational condensation},''
  \href{http://dx.doi.org/10.1086/152650}{{\em Astrophys. J.} {\bfseries 187}
  (1974) 425--438}.

\bibitem{Baur:2017stq}
J.~Baur, N.~Palanque-Delabrouille, C.~Yeche, A.~Boyarsky, O.~Ruchayskiy,
  E.~Armengaud, and J.~Lesgourgues, ``{Constraints from Ly-$\alpha$ forests on
  non-thermal dark matter including resonantly-produced sterile neutrinos},''
  \href{http://dx.doi.org/10.1088/1475-7516/2017/12/013}{{\em JCAP} {\bfseries
  12} (2017) 013}, \href{http://arxiv.org/abs/1706.03118}{{\ttfamily
  arXiv:1706.03118 [astro-ph.CO]}}.

\bibitem{Pagels:1981ke}
H.~Pagels and J.~R. Primack, ``{Supersymmetry, Cosmology and New TeV
  Physics},'' \href{http://dx.doi.org/10.1103/PhysRevLett.48.223}{{\em Phys.
  Rev. Lett.} {\bfseries 48} (1982) 223}.

\bibitem{Bond:1982uy}
J.~R. Bond, A.~S. Szalay, and M.~S. Turner, ``{Formation of Galaxies in a
  Gravitino Dominated Universe},''
  \href{http://dx.doi.org/10.1103/PhysRevLett.48.1636}{{\em Phys. Rev. Lett.}
  {\bfseries 48} (1982) 1636}.

\bibitem{Kamada:2013sya}
A.~Kamada, M.~Shirasaki, and N.~Yoshida, ``{Weighing the Light Gravitino Mass
  with Weak Lensing Surveys},''
  \href{http://dx.doi.org/10.1007/JHEP06(2014)162}{{\em JHEP} {\bfseries 06}
  (2014) 162}, \href{http://arxiv.org/abs/1311.4323}{{\ttfamily arXiv:1311.4323
  [hep-ph]}}.

\bibitem{Osato:2016ixc}
K.~Osato, T.~Sekiguchi, M.~Shirasaki, A.~Kamada, and N.~Yoshida,
  ``{Cosmological Constraint on the Light Gravitino Mass from CMB Lensing and
  Cosmic Shear},'' \href{http://dx.doi.org/10.1088/1475-7516/2016/06/004}{{\em
  JCAP} {\bfseries 06} (2016) 004},
  \href{http://arxiv.org/abs/1601.07386}{{\ttfamily arXiv:1601.07386
  [astro-ph.CO]}}.

\bibitem{Boyarsky:2008xj}
A.~Boyarsky, J.~Lesgourgues, O.~Ruchayskiy, and M.~Viel, ``{Lyman-alpha
  constraints on warm and on warm-plus-cold dark matter models},''
  \href{http://dx.doi.org/10.1088/1475-7516/2009/05/012}{{\em JCAP} {\bfseries
  05} (2009) 012}, \href{http://arxiv.org/abs/0812.0010}{{\ttfamily
  arXiv:0812.0010 [astro-ph]}}.

\bibitem{Diamanti:2017xfo}
R.~Diamanti, S.~Ando, S.~Gariazzo, O.~Mena, and C.~Weniger, ``{Cold dark matter
  plus not-so-clumpy dark relics},''
  \href{http://dx.doi.org/10.1088/1475-7516/2017/06/008}{{\em JCAP} {\bfseries
  06} (2017) 008}, \href{http://arxiv.org/abs/1701.03128}{{\ttfamily
  arXiv:1701.03128 [astro-ph.CO]}}.

\bibitem{Dodelson:1993je}
S.~Dodelson and L.~M. Widrow, ``{Sterile-neutrinos as dark matter},''
  \href{http://dx.doi.org/10.1103/PhysRevLett.72.17}{{\em Phys. Rev. Lett.}
  {\bfseries 72} (1994) 17--20},
  \href{http://arxiv.org/abs/hep-ph/9303287}{{\ttfamily arXiv:hep-ph/9303287}}.

\bibitem{Weinberg:2008zzc}
S.~Weinberg, {\em {Cosmology}}.
\newblock 2008.

\bibitem{Batell:2009di}
B.~Batell, M.~Pospelov, and A.~Ritz, ``{Exploring Portals to a Hidden Sector
  Through Fixed Targets},''
  \href{http://dx.doi.org/10.1103/PhysRevD.80.095024}{{\em Phys. Rev. D}
  {\bfseries 80} (2009) 095024},
  \href{http://arxiv.org/abs/0906.5614}{{\ttfamily arXiv:0906.5614 [hep-ph]}}.

\bibitem{Battaglieri:2017aum}
M.~Battaglieri {\em et~al.}, ``{US Cosmic Visions: New Ideas in Dark Matter
  2017: Community Report},'' in {\em {U.S. Cosmic Visions: New Ideas in Dark
  Matter}}.
\newblock 7, 2017.
\newblock \href{http://arxiv.org/abs/1707.04591}{{\ttfamily arXiv:1707.04591
  [hep-ph]}}.

\bibitem{Dutta:2019nbn}
B.~Dutta, D.~Kim, S.~Liao, J.-C. Park, S.~Shin, and L.~E. Strigari, ``{Dark
  matter signals from timing spectra at neutrino experiments},''
  \href{http://dx.doi.org/10.1103/PhysRevLett.124.121802}{{\em Phys. Rev.
  Lett.} {\bfseries 124} no.~12, (2020) 121802},
  \href{http://arxiv.org/abs/1906.10745}{{\ttfamily arXiv:1906.10745
  [hep-ph]}}.

\bibitem{DUNE:2020fgq}
{\bfseries DUNE} Collaboration, B.~Abi {\em et~al.}, ``{Prospects for beyond
  the Standard Model physics searches at the Deep Underground Neutrino
  Experiment},'' \href{http://dx.doi.org/10.1140/epjc/s10052-021-09007-w}{{\em
  Eur. Phys. J. C} {\bfseries 81} no.~4, (2021) 322},
  \href{http://arxiv.org/abs/2008.12769}{{\ttfamily arXiv:2008.12769
  [hep-ex]}}.

\bibitem{Fitzpatrick:2020vba}
P.~J. Fitzpatrick, H.~Liu, T.~R. Slatyer, and Y.-D. Tsai, ``{New Pathways to
  the Relic Abundance of Vector-Portal Dark Matter},''
  \href{http://arxiv.org/abs/2011.01240}{{\ttfamily arXiv:2011.01240
  [hep-ph]}}.

\bibitem{NA64:2019imj}
D.~Banerjee {\em et~al.}, ``{Dark matter search in missing energy events with
  NA64},'' \href{http://dx.doi.org/10.1103/PhysRevLett.123.121801}{{\em Phys.
  Rev. Lett.} {\bfseries 123} no.~12, (2019) 121801},
  \href{http://arxiv.org/abs/1906.00176}{{\ttfamily arXiv:1906.00176
  [hep-ex]}}.

\bibitem{Battaglieri:2016ggd}
{\bfseries BDX} Collaboration, M.~Battaglieri {\em et~al.}, ``{Dark Matter
  Search in a Beam-Dump eXperiment (BDX) at Jefferson Lab},''
  \href{http://arxiv.org/abs/1607.01390}{{\ttfamily arXiv:1607.01390
  [hep-ex]}}.

\bibitem{Battaglieri:2020lds}
M.~Battaglieri {\em et~al.}, ``{The BDX-MINI detector for Light Dark Matter
  search at JLab},''
  \href{http://dx.doi.org/10.1140/epjc/s10052-021-08957-5}{{\em Eur. Phys. J.
  C} {\bfseries 81} no.~2, (2021) 164},
  \href{http://arxiv.org/abs/2011.10532}{{\ttfamily arXiv:2011.10532
  [physics.ins-det]}}.

\bibitem{Lees:2017lec}
{\bfseries BaBar} Collaboration, J.~P. Lees {\em et~al.}, ``{Search for
  Invisible Decays of a Dark Photon Produced in ${e}^{+}{e}^{-}$ Collisions at
  BaBar},'' \href{http://dx.doi.org/10.1103/PhysRevLett.119.131804}{{\em Phys.
  Rev. Lett.} {\bfseries 119} no.~13, (2017) 131804},
  \href{http://arxiv.org/abs/1702.03327}{{\ttfamily arXiv:1702.03327
  [hep-ex]}}.

\bibitem{Binder:2021bmg}
T.~Binder, T.~Bringmann, M.~Gustafsson, and A.~Hryczuk, ``{DRAKE: Dark matter
  Relic Abundance beyond Kinetic Equilibrium},''
  \href{http://arxiv.org/abs/2103.01944}{{\ttfamily arXiv:2103.01944
  [hep-ph]}}.

\bibitem{Randall:2007ph}
S.~W. Randall, M.~Markevitch, D.~Clowe, A.~H. Gonzalez, and M.~Bradac,
  ``{Constraints on the Self-Interaction Cross-Section of Dark Matter from
  Numerical Simulations of the Merging Galaxy Cluster 1E 0657-56},''
  \href{http://dx.doi.org/10.1086/587859}{{\em Astrophys. J.} {\bfseries 679}
  (2008) 1173--1180}, \href{http://arxiv.org/abs/0704.0261}{{\ttfamily
  arXiv:0704.0261 [astro-ph]}}.

\bibitem{Harvey:2018uwf}
D.~Harvey, A.~Robertson, R.~Massey, and I.~G. McCarthy, ``{Observable tests of
  self-interacting dark matter in galaxy clusters: BCG wobbles in a constant
  density core},'' \href{http://dx.doi.org/10.1093/mnras/stz1816}{{\em Mon.
  Not. Roy. Astron. Soc.} {\bfseries 488} no.~2, (2019) 1572--1579},
  \href{http://arxiv.org/abs/1812.06981}{{\ttfamily arXiv:1812.06981
  [astro-ph.CO]}}.

\bibitem{Sagunski:2020spe}
L.~Sagunski, S.~Gad-Nasr, B.~Colquhoun, A.~Robertson, and S.~Tulin,
  ``{Velocity-dependent Self-interacting Dark Matter from Groups and Clusters
  of Galaxies},'' \href{http://dx.doi.org/10.1088/1475-7516/2021/01/024}{{\em
  JCAP} {\bfseries 01} (2021) 024},
  \href{http://arxiv.org/abs/2006.12515}{{\ttfamily arXiv:2006.12515
  [astro-ph.CO]}}.

\bibitem{Inoue:2014jka}
K.~T. Inoue, R.~Takahashi, T.~Takahashi, and T.~Ishiyama, ``{Constraints on
  warm dark matter from weak lensing in anomalous quadruple lenses},''
  \href{http://dx.doi.org/10.1093/mnras/stv194}{{\em Mon. Not. Roy. Astron.
  Soc.} {\bfseries 448} no.~3, (2015) 2704--2716},
  \href{http://arxiv.org/abs/1409.1326}{{\ttfamily arXiv:1409.1326
  [astro-ph.CO]}}.

\bibitem{Kamada:2016vsc}
A.~Kamada, K.~T. Inoue, and T.~Takahashi, ``{Constraints on mixed dark matter
  from anomalous strong lens systems},''
  \href{http://dx.doi.org/10.1103/PhysRevD.94.023522}{{\em Phys. Rev. D}
  {\bfseries 94} no.~2, (2016) 023522},
  \href{http://arxiv.org/abs/1604.01489}{{\ttfamily arXiv:1604.01489
  [astro-ph.CO]}}.

\bibitem{Kamada:2017icv}
A.~Kamada, K.~T. Inoue, K.~Kohri, and T.~Takahashi, ``{Constraints on
  long-lived electrically charged massive particles from anomalous strong lens
  systems},'' \href{http://dx.doi.org/10.1088/1475-7516/2017/11/008}{{\em JCAP}
  {\bfseries 11} (2017) 008}, \href{http://arxiv.org/abs/1703.05145}{{\ttfamily
  arXiv:1703.05145 [astro-ph.CO]}}.

\bibitem{Birrer:2017rpp}
S.~Birrer, A.~Amara, and A.~Refregier, ``{Lensing substructure quantification
  in RXJ1131-1231: A 2 keV lower bound on dark matter thermal relic mass},''
  \href{http://dx.doi.org/10.1088/1475-7516/2017/05/037}{{\em JCAP} {\bfseries
  05} (2017) 037}, \href{http://arxiv.org/abs/1702.00009}{{\ttfamily
  arXiv:1702.00009 [astro-ph.CO]}}.

\bibitem{Gilman:2017voy}
D.~Gilman, S.~Birrer, T.~Treu, C.~R. Keeton, and A.~Nierenberg, ``{Probing the
  nature of dark matter by forward modelling flux ratios in strong
  gravitational lenses},'' \href{http://dx.doi.org/10.1093/mnras/sty2261}{{\em
  Mon. Not. Roy. Astron. Soc.} {\bfseries 481} no.~1, (2018) 819--834},
  \href{http://arxiv.org/abs/1712.04945}{{\ttfamily arXiv:1712.04945
  [astro-ph.CO]}}.

\bibitem{Vegetti:2018dly}
S.~Vegetti, G.~Despali, M.~R. Lovell, and W.~Enzi, ``{Constraining sterile
  neutrino cosmologies with strong gravitational lensing observations at
  redshift z \ensuremath{\sim} 0.2},''
  \href{http://dx.doi.org/10.1093/mnras/sty2393}{{\em Mon. Not. Roy. Astron.
  Soc.} {\bfseries 481} no.~3, (2018) 3661--3669},
  \href{http://arxiv.org/abs/1801.01505}{{\ttfamily arXiv:1801.01505
  [astro-ph.CO]}}.

\bibitem{Rivero:2018bcd}
A.~D\'\i{}az~Rivero, C.~Dvorkin, F.-Y. Cyr-Racine, J.~Zavala, and
  M.~Vogelsberger, ``{Gravitational Lensing and the Power Spectrum of Dark
  Matter Substructure: Insights from the ETHOS N-body Simulations},''
  \href{http://dx.doi.org/10.1103/PhysRevD.98.103517}{{\em Phys. Rev. D}
  {\bfseries 98} no.~10, (2018) 103517},
  \href{http://arxiv.org/abs/1809.00004}{{\ttfamily arXiv:1809.00004
  [astro-ph.CO]}}.

\bibitem{Gilman:2019nap}
D.~Gilman, S.~Birrer, A.~Nierenberg, T.~Treu, X.~Du, and A.~Benson, ``{Warm
  dark matter chills out: constraints on the halo mass function and the
  free-streaming length of dark matter with eight quadruple-image strong
  gravitational lenses},'' \href{http://dx.doi.org/10.1093/mnras/stz3480}{{\em
  Mon. Not. Roy. Astron. Soc.} {\bfseries 491} no.~4, (2020) 6077--6101},
  \href{http://arxiv.org/abs/1908.06983}{{\ttfamily arXiv:1908.06983
  [astro-ph.CO]}}.

\bibitem{Sitwell:2013fpa}
M.~Sitwell, A.~Mesinger, Y.-Z. Ma, and K.~Sigurdson, ``{The Imprint of Warm
  Dark Matter on the Cosmological 21-cm Signal},''
  \href{http://dx.doi.org/10.1093/mnras/stt2392}{{\em Mon. Not. Roy. Astron.
  Soc.} {\bfseries 438} no.~3, (2014) 2664--2671},
  \href{http://arxiv.org/abs/1310.0029}{{\ttfamily arXiv:1310.0029
  [astro-ph.CO]}}.

\bibitem{Sekiguchi:2014wfa}
T.~Sekiguchi and H.~Tashiro, ``{Constraining warm dark matter with 21 cm line
  fluctuations due to minihalos},''
  \href{http://dx.doi.org/10.1088/1475-7516/2014/08/007}{{\em JCAP} {\bfseries
  08} (2014) 007}, \href{http://arxiv.org/abs/1401.5563}{{\ttfamily
  arXiv:1401.5563 [astro-ph.CO]}}.

\bibitem{Safarzadeh:2018hhg}
M.~Safarzadeh, E.~Scannapieco, and A.~Babul, ``{A limit on the warm dark matter
  particle mass from the redshifted 21 cm absorption line},''
  \href{http://dx.doi.org/10.3847/2041-8213/aac5e0}{{\em Astrophys. J. Lett.}
  {\bfseries 859} no.~2, (2018) L18},
  \href{http://arxiv.org/abs/1803.08039}{{\ttfamily arXiv:1803.08039
  [astro-ph.CO]}}.

\bibitem{Schneider:2018xba}
A.~Schneider, ``{Constraining noncold dark matter models with the global 21-cm
  signal},'' \href{http://dx.doi.org/10.1103/PhysRevD.98.063021}{{\em Phys.
  Rev. D} {\bfseries 98} no.~6, (2018) 063021},
  \href{http://arxiv.org/abs/1805.00021}{{\ttfamily arXiv:1805.00021
  [astro-ph.CO]}}.

\bibitem{Lidz:2018fqo}
A.~Lidz and L.~Hui, ``{Implications of a prereionization 21-cm absorption
  signal for fuzzy dark matter},''
  \href{http://dx.doi.org/10.1103/PhysRevD.98.023011}{{\em Phys. Rev. D}
  {\bfseries 98} no.~2, (2018) 023011},
  \href{http://arxiv.org/abs/1805.01253}{{\ttfamily arXiv:1805.01253
  [astro-ph.CO]}}.

\bibitem{Lopez-Honorez:2018ipk}
L.~Lopez-Honorez, O.~Mena, and P.~Villanueva-Domingo, ``{Dark matter
  microphysics and 21 cm observations},''
  \href{http://dx.doi.org/10.1103/PhysRevD.99.023522}{{\em Phys. Rev. D}
  {\bfseries 99} no.~2, (2019) 023522},
  \href{http://arxiv.org/abs/1811.02716}{{\ttfamily arXiv:1811.02716
  [astro-ph.CO]}}.

\bibitem{Nebrin:2018vqt}
O.~Nebrin, R.~Ghara, and G.~Mellema, ``{Fuzzy Dark Matter at Cosmic Dawn: New
  21-cm Constraints},''
  \href{http://dx.doi.org/10.1088/1475-7516/2019/04/051}{{\em JCAP} {\bfseries
  04} (2019) 051}, \href{http://arxiv.org/abs/1812.09760}{{\ttfamily
  arXiv:1812.09760 [astro-ph.CO]}}.

\bibitem{Chatterjee:2019jts}
A.~Chatterjee, P.~Dayal, T.~R. Choudhury, and A.~Hutter, ``{Ruling out 3 keV
  warm dark matter using 21 cm EDGES data},''
  \href{http://dx.doi.org/10.1093/mnras/stz1444}{{\em Mon. Not. Roy. Astron.
  Soc.} {\bfseries 487} no.~3, (2019) 3560--3567},
  \href{http://arxiv.org/abs/1902.09562}{{\ttfamily arXiv:1902.09562
  [astro-ph.CO]}}.

\end{thebibliography}\endgroup

\end{document}